\documentclass[11pt]{article}

\usepackage{graphicx}
\usepackage{newtxtext}
\usepackage{newtxmath}
\usepackage{natbib}
\usepackage{hyperref}
\usepackage[a4paper, total={6.5in, 9.5in}]{geometry}
\hypersetup{
    colorlinks = true,
    urlcolor   = blue,
    citecolor  = black,
}
\usepackage{amsmath}
\usepackage{natbib}

\newcommand{\refe}[1]{(\ref{#1})}
\newcommand{\red}[1]{\textcolor{black}{#1}}

\title{Dynamics of particle aggregation in \red{dewetting} films of complex liquids}

\author{J.~Zhang, D.~N.~Sibley, D. Tseluiko and A.~J.~Archer\\
Department of Mathematical Sciences, Loughborough University,\\Loughborough LE11 3TU, UK}

\begin{document}
\maketitle

\begin{abstract}
We consider the dynamic wetting and \red{dewetting} processes of films and droplets of complex liquids on planar surfaces, focusing on the case of colloidal suspensions, where the particle interactions can be sufficiently attractive to cause agglomeration of the colloids within the film. This leads to \red{an interesting array of dynamic behaviours} within the liquid and of the liquid-air interface. Incorporating concepts from thermodynamics and using the thin-film approximation, we construct a model consisting of a pair of coupled partial differential equations that represent the evolution of the liquid film and the effective colloidal height profiles. We determine the relevant phase behaviour of the uniform system, including finding associated binodal and spinodal curves, helping to uncover how the emerging behaviour depends on the particle interactions. Performing a linear stability analysis of our system enables us to identify parameter regimes where agglomerates form, which we independently confirm through numerical simulations and continuation of steady states, to construct bifurcation diagrams. We obtain various dynamics such as uniform colloidal profiles in an unstable situation evolving into agglomerates and thus elucidate the interplay between \red{dewetting} and particle aggregation in complex liquids on surfaces. 
\end{abstract}

\section{Introduction}
\label{sec:intro}

The \red{dynamics and equilibration} of films and droplets of colloidal suspensions over surfaces is an everyday process \citep{kalliadasis2007thin}. Paints and coatings are a classic example of such liquids. For example, dispersions containing polymer particles are routinely used as paints \citep{keddie2010fundamentals}. These are formulated so that the pigment and other suspended particles remain well dispersed throughout the solvent liquid. However, this is not always the case; sufficiently strong attractive interactions between the colloids can lead to agglomeration \citep{hansen2013theory}. Such particle ordering within liquid films on surfaces is of interest due to the resulting pattern-formation having potential uses for structure-creation after the solvent liquid has evaporated, to leave a dried-on structure. This includes the well-known coffee stain effect discussed by \citet{deegan1997capillary} and also a wide variety of other ways that colloidal particles in thin liquid films can organise themselves \citep{thiele2014patterned}.

Particle suspensions are also used in the manufacture of structures on surfaces. For example, inks containing conductive copper particles are used to form electrical connections via ink-jet printing \citep{chalmers2017modelling}. Another application is to use colloids which assemble naturally to form silicon photonic bandgap crystals \citep{vlasov2001chip}. In nature, bacterial colonies on surfaces are another example of particles (the bacteria) in a liquid film \citep{mimura2000reaction, trinschek2018modelling}, although in this case the particles are also active particles. 

There is much previous work on droplets wetting, spreading or dewetting from surfaces, discussion of theories for contact line motion and the dynamical equations used to describe liquids on surfaces, perhaps the most notable being the thin-film equation, obtained via the lubrication approximation. There are several excellent reviews on this broad area, including by \citet{de1985wetting}, \citet{oron1997long}, \citet{bonn2009wetting}\red{, \citet{craster2009dynamics}, \citet{lohse2022fundamental} and \citet{WilsonRev}.} What is of interest here is the dewetting behaviour of liquid films and in particular on how this is influenced by (and coupled to) the aggregation/demixing of colloidal particles suspended in the liquid. Dewetting is the process by which an initially uniform film on a surface breaks up into droplets, leaving the surface bare in places. For pure liquids, this has been extensively studied both in experiments\red{, see e.g.\ \citet{reiter1992dewetting, seemann2001gaining}, and in theory; for example, see  \citet{thiele2001dewetting, becker2003complex}.}

Studies of the influence of suspended colloidal particles on the dynamics of liquid films on surfaces include those of \citet{parisse1996shape, parisse1997drying} who observed that the colloids can alter significantly the shape of the surface of droplets. They also developed a simple model to explain the changes during drying. One way to model the influence of colloids on droplet dynamics is via particle-based models. Typically, these have a stochastic dynamics, generally referred to as kinetic Monte Carlo (KMC) models. This effective dynamics arises because these are coarse-grained models: simulating over the relevant timescales the true molecular dynamics of even a micron sized droplet containing just a few colloids is just not feasible even with modern computers, due to the huge numbers of solvent molecules to be simulated. Examples of these sorts of KMC models include \red{effective two-dimensional (2D) models, such as those of \citet{rabani2003drying, martin2004nanoparticle, vancea2008front}, and also fully three-dimensional (3D) models, such as the models of \citet{sztrum2005self, kim2011computational, chalmers2017modelling, areshi2019kinetic}; see also} references therein. 

Such considerations highlight the importance and need for coarse-grained continuum models. Given the success of thin-film hydrodynamic models, it is natural to seek to incorporate the influences of suspended colloids into such models. \citet{thiele2009modelling} gives a review of some of these approaches. Most directly connected with the KMC models are those that use dynamical density functional theory (DDFT) to construct coupled partial differential equations (PDEs) for the dynamics of the liquid and the colloids over the surface  \citep{archer2010dynamical, robbins2011modelling, chalmers2017dynamical}. The real value of the DDFT approach is that it is based on thermodynamics, building this into the resulting models correctly. The dynamical equations are based on a free energy functional which incorporates the correct physics. Here, we do not use a DDFT, but we do enforce that our theory be based on a free energy functional having the necessary terms.

Hydrodynamic models of the thin-film-equation-type can also be constructed based on a free energy \red{functional \citep{thiele2018recent, mitlin1993dewetting, kalliadasis2007thin, thiele2009modelling}. Here, we should also mention the work of \citet{naraigh2010nonlinear}, who derived a long-wave model, starting from the full model-H hydrodynamics.} Improvements and applications to various different problems can then generally be made by either adding additional terms (additional physics) to the free energy functional or by modifying the dynamical coefficients to take account of any additional kinetic mechanisms the system may have. These include e.g.\ the work of
\citet{warner2003surface, frastia2011dynamical, frastia2012modelling, zigelman2019analysis} on the formation of periodic line-deposits by drying colloidal films, or the work in \citet{thiele2016gradient} describing liquid films containing surfactant molecules. In \citet{naraigh2010nonlinear} and \citet{thiele2011note}, models consisting of a pair of coupled equations for the dynamics of the film height and the local height averaged concentration of the colloids were derived. These works, and also \citet{thiele2013gradient} showed that dewetting can be triggered by the coupling between the film height and concentration fluctuations. Such fluctuations are typically thermal in origin; the recent work of \citet{zhang2019molecular} and \citet{zhao2022fluctuation} explain well how the microscopic (molecular) scale properties of liquid films connect to the mesoscopic scales at which the thin-film equation operates. The thermal fluctuations lead to a stochastic term in the equations \citep{grun2006thin}, that becomes important in certain regimes, such as when there is the possibility of holes in the film to be nucleated.

The thin-film hydrodynamic model that we develop here builds on the results in \citet{thiele2011note, thiele2013gradient, todorova2013modelling}. \red{As derived in these works, the free energy can be rather general, but then in their subsequent analysis} the authors assume that the interactions between the colloids are such that there is no aggregation and that when the film thickness becomes large the colloids remain well dispersed in the (bulk) liquid. One can go beyond this by \red{including} additional terms to the free energy to incorporate the effect of the interactions between the colloids. This was the approach taken in \citet{naraigh2010nonlinear} and also in the recent study conducted by \citet{diez2021simultaneous}, where a thin-film model for the decomposition and dewetting of nanoscale alloys was developed. Their theory is based on a simple Cahn-Hilliard-type \citep{hilliard1958nature, cahn1965phase, langer1992introduction} double-well free energy, appropriate to the binary alloys they consider. But for the colloidal suspensions of interest here, we must go beyond this by including in the free energy both the logarithmic ideal-gas term (in order to correctly describe the low density limit), as well as terms describing the effect of the attractions and steric repulsions between the colloids.

It should be mentioned that the thin-film models discussed above, together with the theory we develop here, all assume that the distribution of the colloids over the surface can be described by a height-averaged field, i.e.\ averaging the local concentration of the colloids over the direction perpendicular to the substrate. This assumption is valid in many cases, but it is not always the case. To include the effect of variations in the local density of the colloids perpendicular to the substrate one must e.g.\ take the approach of \citet{maki2011fast} who use the thin-film approximation to describe the solvent liquid, but then describe the transport of the colloids within the film with the full convection-diffusion equation. Alternatively, one can use DDFT models, such as that of \citet{chalmers2017dynamical}, although this neglects some aspects of hydrodynamic flow over surfaces.

This paper is organised as follows: In \S~\ref{sec:TFmodel}, we discuss how to extend the pair of coupled thin-film equations from \citet{thiele2011note}, to incorporate colloidal interactions and thus allow for the possibility of particle agglomeration, if the attractive interactions are strong enough. A linear stability analysis of the resulting equations is conducted in \S~\ref{sec:LSA}, together with a brief explanation of the bulk (uniform film) phase diagram. In \S~\ref{sec:nandr} we present numerical results from solving the coupled equations for the film height and concentration profiles over time in one spatial dimension, i.e.\ assuming that the film is two-dimensional. Then, in \S~\ref{sec:bif}, we present results from numerical continuation of stationary states as the system size is varied, together with corresponding bifurcation diagrams showing how the various different solutions originate and are connected. In \S~\ref{sec:2D} we present results from solving the coupled equations over time in two spatial dimension, i.e.\ assuming that the film is fully three-dimensional. Finally, in \S~\ref{sec:conc} we present a few concluding remarks.

\section{The thin-film model}
\label{sec:TFmodel}

\begin{figure}
  \centerline{\includegraphics{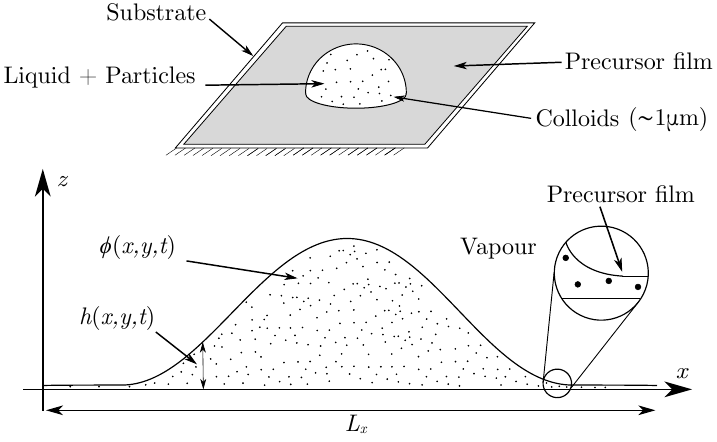}}
  \caption{Illustration of the system we consider. On the \red{top} is a sketch of a droplet of a colloidal suspension deposited on a surface. On the \red{bottom} is the cross-section sketch with system size $L_x$. $h(x,y,t)$ is the film height and $\phi(x,y,t)$ is the effective local concentration.}
\label{fig:geoillu}
\end{figure}

Consider the dynamics of a film or droplet of a partially wetting liquid on a flat substrate, which contains colloidal particles suspended inside the liquid. We introduce a cartesian coordinate system $(x,y,z)$ with the $x$- and $y$-axes pointing along the substrate and the $z$-axis perpendicular to the substrate -- see Fig.~\ref{fig:geoillu}. Particles that can be classified as colloids, i.e.\ that remain suspended in the solvent liquid, typically have radii $R$ in the range of $1nm < R < 10\mu m$ \citep{mewis2012colloidal}. In our study we assume that the liquid is incompressible and has constant surface tension $\gamma$ and constant viscosity $\eta$. The variables that we use to describe the system are the liquid film height $h(x,y,t)$ and the effective colloid height $\psi(x,y,t)= h(x,y,t)\phi(x,y,t)$, which both change in the space and time domain. The effective height $\psi$ is the product of the film height, $h$ and \red{the dimensionless} local height averaged colloid concentration, $\phi$.

The governing equations for the dynamics of the coupled fields $h$ and $\psi$ can be written as the following gradient dynamics system \citep{thiele2011note}:
\begin{eqnarray}
\partial_t h & = &
\nabla\cdot\left[Q_{hh}\nabla\frac{\delta F}{\delta h}\,+\,Q_{h\psi}\nabla\frac{\delta F}{\delta \psi}\right],
\label{eq:1}
\\[3pt]
\partial_t \psi & = &
\nabla\cdot\left[Q_{\psi h}\nabla\frac{\delta F}{\delta h}\,+\,Q_{\psi \psi}\nabla\frac{\delta F}{\delta \psi}\right],
\label{eq:solsusp-coup-grad-evap}
\end{eqnarray}
where $F[h,\psi]$ is the free energy of the system and $Q_{ij}$, with $i,j = h, \psi$, are the elements of the film mobility matrix. When there is no slip at the surface, this takes the form \citep{thiele2011note}:
\begin{equation}
\mathbf{Q}\,=\,\left( 
\begin{array}{cc}  
Q_{hh} & Q_{h\psi} \\[.3ex]
Q_{\psi h} &Q_{\psi \psi}
\end{array}
\right)
\,=\,\frac{1}{3\eta}\left( 
\begin{array}{cc}  
h^3 & h^2\psi \\[.3ex]
h^2\psi \phantom{x}& h\psi^2+ 3 \widetilde D \psi
\end{array}
\right).
\label{eq:solsusp-mob}
\end{equation}
Note that in the limit $\psi\to0$ (i.e.\ no colloids), Eqs.~\eqref{eq:1} and \eqref{eq:solsusp-coup-grad-evap} reduce to the usual thin-film equation for a pure liquid, with mobility coefficent $Q_{hh}=h^3/3\eta$. Note also that $\phi Q_{hh} = Q_{h\psi} = Q_{\psi h}$ and $\widetilde{D} = a^2/6\pi$, where $a$ is a molecular length scale, is a dynamical coefficient related to the diffusion coefficient of the colloids in the liquid. We say more on this below.
$\delta F/\delta h$ and $\delta F/\delta \psi$ are functional derivatives of the \red{free energy functional
\begin{equation}
F[h,\psi] = \iint \left[g(h)+\frac{\gamma}{2}|\nabla h|^2+hf\left(\frac{\psi}{h}\right)+\frac{\epsilon h}{2}\left|\nabla \left(\frac{\psi}{h}\right) \right|^2\right]dx dy.
\label{freeenergyfunc}
\end{equation}
The} first term in the integrand, $g(h)$, is the binding potential, which is the effective interaction between the liquid-air interface and the solid-liquid interface below it. It results from the molecular interactions in the liquid and so is short ranged, influencing the system largely in the vicinity of the contact line. The derivative $\Pi(h) \equiv -\partial g/\partial h$ is the disjoining pressure. In this paper we use \citep{kalliadasis2007thin, craster2009dynamics, bonn2009wetting}
\begin{equation}
g(h)=\frac{B}{h^3}-\frac{A}{h^2},
\label{binding}
\end{equation}
where here we take $A$ and $B$ to be constants. \red{This commonly used approximation may be obtained as the leading-order terms arising from an expansion of the full binding potential in powers of $1/h$ \citep{dietrich88, schick1990introduction, hughes2015liquid, hughes2017}. In particular, the term $-A/h^2$, which dominates for large $h$ and determines whether the liquid wets the surface or not, originates from integrating over the van der Waals interactions between molecules. A variety of different approximations for $g(h)$ are used in the literature. The main influence of the particular form of $g(h)$ is to determine the contact angle droplets make with the surface and whether the wetting transition which occurs as the value of $A$ is varied is either continuous or is a first-order transition \citep{dietrich88}. However, even subtle details such as the nature of the molecular ordering in the liquid in the vicinity of the surface can influence the form of the biding potential \citep{hughes2017, yin2017films, macdowell2019surface, llombart2020rounded}.} In \citet{thiele2013gradient} the case where $A$ \red{in Eq.~\eqref{binding}} is treated as a function of the colloid concentration $\phi$ is considered. Note that Eq.~\eqref{binding} can be written as $g(h)={B}(1-{Ah}/{B})/h^3$. In other words, the ratio $\widetilde{h}\equiv B/A$ is one of the relevant length scales in our system and is the lengthscale we use in our non-dimensionalisation below. This lengthscale is related to the thickness of the equilibrium precursor film, $h_{eq}=1.5\widetilde{h}$, which corresponds to the minimum of $g(h)$, i.e.\ where $g'(h_{eq})=0$.

The second term in Eq.~\eqref{freeenergyfunc} describes the surface tension contribution and gives a contribution to the free energy that is proportional to the area of the liquid-air interface in the long-wave limit. Just the first two terms in Eq.~\eqref{freeenergyfunc} alone are what one would use to describe a liquid film with no colloids involved and the dynamical equation that follows from this is commonly used to describe the dewetting of liquid films.

The third and fourth terms in Eq.~(\ref{freeenergyfunc}) incorporate the contribution to the free energy of the suspended colloids. Note that since $\phi=\psi/h$, the \red{third and fourth terms} in the integrand of Eq.~(\ref{freeenergyfunc}) \red{together are simply $h(f(\phi)+(\epsilon/2)|\nabla\phi|^2)$}, where $f(\phi)$ is a Helmholtz free energy density that depends on the local height-averaged colloid concentration $\phi$ \citep{thiele2011note, thiele2013gradient}. \red{In other words, we just use a simple square-gradient approximation for the free energy of the colloids \citep{hansen2013theory}.} We seek to model colloids that can agglomerate, i.e.\ that can exhibit colloidal-demixing into a colloid-rich phase that coexists with a colloid-poor phase, due to sufficiently strong attractive interactions between the colloids. For example, depletion interactions due to the presence of non-adsorbing polymers in the system can drive such phase separation \citep{mao1995depletion, likos2001effective, hansen2013theory}. In the limit where the film thickness $h$ remains constant, our aim is to make these terms lead to a generalised Cahn-Hilliard equation for the local concentration field $\phi$. Another way to say this is that when $h=$ constant, the colloids are described by a simple DDFT with a square-gradient approximation for the free energy \citep{hansen2013theory}. Of course, because these terms are coupled to the film height, in general $h$ is not a constant and in particular, when $\psi$ is spatially varying, then so is $h$ because of this. When the concentration of the colloids is small, then the colloidal free energy density is just the ideal-gas contribution \citep{hansen2013theory, thiele2011note, thiele2013gradient, thiele2016gradient}
\begin{equation}
f \approx f_{id} = \frac{k_B T}{a^3} \phi \ln(\phi),
\label{bulkfreeen}
\end{equation}
where $k_B$ is Boltzmann's constant, $T$ is the temperature and $a$ is a molecular length scale. Note that this term, combined with the diffusive term involving $\widetilde{D}$ in the mobility matrix~(\ref{eq:solsusp-mob}), results in the Einstein-Stokes relation for the diffusion coefficient of the colloids
\begin{equation}
D=\red{ \frac{\widetilde{D}}{\eta}\frac{k_B T}{a^3}= } \frac{k_B T}{6 \pi R \eta}\red{,}
\label{Einstein}
\end{equation}
\red{when $a=R$.}
This construction means that in the limit when $h=$ constant and $\psi$ is small, Eqs.~\eqref{eq:1} and \eqref{eq:solsusp-coup-grad-evap} yield the diffusion equation for the local colloid concentration. However, in general we cannot assume that the colloids act as an ideal-gas. This means the Helmholtz free energy density in Eq.~(\ref{bulkfreeen}) must be modified to take the colloid interactions into consideration. In order to do this, we recall that a virial expansion for the pressure $p$ (strictly, the osmotic pressure of the colloids) is
\begin{equation}
p=k_B T\left[\rho + B_2 \rho^2 +B_3 \rho^3 +B_4 \rho^4 +\cdots\right],
\label{virial}
\end{equation}
where $\rho$ is number density of the colloidal particles (i.e.\ $\rho \propto \phi$) and the virial coefficients $B_n$ are coefficients that depend on temperature. From the thermodynamic identity
\begin{equation}
 p=-f+\rho\mu,
\label{pressurevir}
\end{equation}
where the chemical potential of the colloids is \citep{hansen2013theory}
\begin{equation}
\mu=\frac{\partial f}{\partial \rho},
\label{eq:mu}
\end{equation}
we can then integrate Eq.~(\ref{virial}) to obtain 
\begin{equation}
f=k_B T\rho \left[\ln (\Lambda^3\rho) -1\right] +B_2\rho^2 +\frac{1}{2}B_3 \rho^3 +\frac{1}{3}B_4\rho^4+\cdots,
\label{freeenergyperV}
\end{equation}
where $\Lambda$ is the thermal de Broglie wavelength. Recall that if the colloid particles interact via the pair potential $u(r)$, which includes an effective solvent-mediated contribution \citep{likos2001effective}, then the second virial coefficient is \citep{hansen2013theory}
\begin{equation}
B_2(T) = -2\pi \int_{0}^{\infty} \left[\exp\left(-\frac{u(r)}{k_B T}\right)-1\right]r^2\,dr.
\label{2ndvircoef}
\end{equation}
So, as the attractive interactions between the colloids increases in strength, then $B_2$ becomes increasingly negative. At some point, it becomes sufficiently negative to drive aggregation of the colloids, leading to phase separation into a colloid poor `gas' phase and a colloid rich `liquid' phase. The density in the aggregated phase is determined by the size of the colloids (i.e.\ the steric repulsions) and packing effects which are controlled by the cubic and higher order terms in Eq.~\eqref{freeenergyperV}. For simplicity, here we just include one higher-order term \red{(the one with coefficient $B_4$)} and assume the following expression for the free energy of the colloids (recalling $\phi \propto \rho$):
\begin{equation}
f(\phi)=\frac{k_B T}{a^3} \phi \ln \phi -\frac{\alpha}{2} \phi ^2+\frac{\beta}{4} \phi ^4.
\label{freeenergynogra}
\end{equation}
\red{Note that if we instead retain the cubic $B_3$ term, this does not lead to any significant qualitative changes to the model. The key feature that the model must have, and we enforce, is that the free energy is bounded from below for $\phi>0$ and has two minima, corresponding to the colloidal gas and liquid phases, respectively. Our main reason for retaining instead the quartic term is that the resulting model has some useful connections to the Cahn-Hilliard equation \citep{cahn1965phase}.} Thus, the parameter $\alpha \propto -B_2$ models the influence of the attractive interactions between the colloids and the term with coefficient $\beta$ \red{incorporating} the effect of the steric repulsions between the colloids.

Since there is also an interfacial tension between the colloidal-liquid and colloidal-gas phase, we must have the fourth term in Eq.~\eqref{freeenergyfunc}, where the parameter $\epsilon$ determines the value of this interfacial tension. It turns out that $\epsilon$ can often be related to $-B_2$ \citep{robbins2011modelling, hansen2013theory}, but here we treat it as an independent parameter in the model.

Recalling that $\phi=\psi/h$, inserting Eqs.~(\ref{binding}) and (\ref{freeenergynogra}) into (\ref{freeenergyfunc}), and scaling all lengths with the length-scale $\widetilde{h}\equiv B/A$, as follows
\begin{equation}
h = \widetilde{h} h^*,\hspace{0.3cm}\psi = \widetilde{h} \psi^*, \hspace{0.3cm}x = \widetilde{h} x^*,\hspace{0.3cm}y = \widetilde{h} y^*,
\label{nondpara}
\end{equation}
we can non-dimensionalise the free energy in Eq.~\eqref{freeenergyfunc} to obtain:
\begin{eqnarray}
F^*[h^*,\psi^*]\equiv\frac{F}{\gamma \widetilde{h}^2} & = & \iint \bigg[\frac{1}{2}|\nabla^* h^*|^2+A'\left(\frac{1}{h^{*3}}-\frac{1}{h^{*2}}\right) \nonumber\\
&& + K'\psi^*\ln\frac{\psi^*}{h^*}-\frac{\alpha'}{2}\frac{\psi^{*2}}{h^*}+\frac{\beta'}{4}\frac{\psi^{*4}}{h^{*3}}+\frac{\epsilon'}{2}h^*\left|\nabla \frac{\psi^*}{h^*} \right|^2\bigg]dx^* dy^*.
\label{freeenergyfuncphinond2}
\end{eqnarray}
\red{More generally, this can be written as
\begin{eqnarray}
F^*[h^*,\psi^*]  =  \iint \bigg[\frac{1}{2}|\nabla^* h^*|^2+{g^*}(h^*)+ h^*{f^*}\left(\frac{\psi^*}{h^*}\right)+\frac{\epsilon'}{2}h^*\left|\nabla \frac{\psi^*}{h^*} \right|^2\bigg]dx^*dy^*,
\label{freeenergyfuncphinond_general}
\end{eqnarray}
where $g^*$ and $f^*$ are the dimensionless binding potential and colloid free energy, respectively. Here, these take the specific form
\begin{eqnarray}
g^*(h^*) = A'\left(\frac{1}{h^{*3}}-\frac{1}{h^{*2}}\right),
\label{eq:dim_g1}
\end{eqnarray}
and
\begin{eqnarray}
f^*(\phi^*) = K'\phi^*\ln\phi^*-\frac{\alpha'}{2}\phi^{*2}+\frac{\beta'}{4}\phi^{*4}.
\label{eq:dim_g2}
\end{eqnarray}
}
As a result, Eq.~(\ref{freeenergyfuncphinond2}) depends on five dimentionless parameters, namely 
\begin{equation}
A'=\frac{A}{\gamma \widetilde{h}^2}=\frac{B}{\gamma \widetilde{h}^3}, \hspace{0.3cm}
K'=\frac{k_B T\widetilde{h}}{\gamma a^3},\hspace{0.3cm}
\alpha' = \frac{\alpha \widetilde{h}}{\gamma},\hspace{0.3cm}
\beta'=\frac{\beta \widetilde{h}}{\gamma},\hspace{0.3cm}
\epsilon' = \frac{\epsilon}{\gamma \widetilde{h}}.
\label{dimentionlessnumbers}
\end{equation}
From this, we can calculate the functional derivatives of the free energy with respect to film height $h$ and effective height $\psi$, which are required for the dynamical equations~(\ref{eq:solsusp-coup-grad-evap}). For reference, in Appendix~\ref{appA}, we write down in full these functional derivatives \red{and their} gradients. Scaling time as $t=\tau t^*$, where
\begin{equation}
\tau = \frac{3\eta \widetilde{h}}{\gamma},
\end{equation}
we can write the dynamical equations~(\ref{eq:solsusp-coup-grad-evap}) as the following non-dimensional equations
\begin{eqnarray}
\label{eq:21}
\frac{\partial h^*}{\partial t^*} =
\nabla^*\cdot\left[h^{*3}\nabla^*\frac{\delta F^*}{\delta h^*}+h^{*2}\psi^*\nabla^*\frac{\delta F^*}{\delta \psi^*}\right],\\
\frac{\partial \psi^*}{\partial t^*} =
\nabla^*\cdot\left[h^{*2}\psi^*\nabla^*\frac{\delta F^*}{\delta h^*}+\left(h^{*} \psi^{*2}+\frac{a^{*2}}{2\pi}\psi^*\right)\nabla^*\frac{\delta F^*}{\delta \psi^*}\right],
\label{eq:solsusp-coup-grad}
\end{eqnarray}
with dimensionless molecular length scale $a^*=a/\widetilde{h} $.
It is worth noting that the typical (precursor-film) velocity in our system, $U\equiv\widetilde{h}/\tau$, corresponds to a Capillary number, $\mathit{Ca} \equiv \eta U/\gamma\sim{\cal O}(1)$. Henceforth, we abandon the superscript `$*$' on all quantities and deal solely with the non-dimensional variables and parameters.

\section{Linear stability analysis, bulk phase diagram and dispersion relation}
\label{sec:LSA}
\subsection{Linear stability analysis}

To determine the parameter values for which a uniform film with $h=h_i$ and $\psi=\psi_i$ is stable, where $h_i$ and $\psi_i$ are constants, we perform a linear stability analysis. To do this, we substitute the following ansatz
\begin{eqnarray}
\label{eq:23}
h=h_i+\kappa e^{i(k_xx+k_yy)+\omega t},\\
\psi=\psi_i+\chi e^{i(k_xx+k_yy)+\omega t},
\label{linvar}
\end{eqnarray}
into the coupled pair of equations~(\ref{eq:solsusp-coup-grad}) and then linearise in the amplitudes $\kappa$ and $\chi$, which are assumed to be small. This corresponds to a sinusoidal perturbation with dimensionless wavevector ${\bf k}\equiv(k_x,k_y)$ and dimensionless dispersion relation (growth rate) $\omega$. The aim is to obtain the dependence of $\omega$ on the wavenumber $k=|{\bf k}|$ and in particular the sign of $\omega(k)$. If $\omega(k)>0$ for any $k$, then the uniform film is unstable, since this indicates the perturbations in Eq.~\eqref{linvar} grow over time. Taking together Eqs.~\eqref{eq:21}, (\ref{eq:solsusp-coup-grad}), \eqref{eq:23} and \eqref{linvar}, Taylor expanding, and then linearising in $\kappa$ and $\chi$, we obtain the system
\begin{equation}
\omega 
\begin{bmatrix} 
\kappa\\
\chi
\end{bmatrix}
=
\begin{bmatrix} 
C_1 & C_2\\
C_3 & C_4
\end{bmatrix}
\begin{bmatrix} 
\kappa\\
\chi
\end{bmatrix}
\red{
=
{\mathbf{C}}\begin{bmatrix} 
\kappa\\
\chi
\end{bmatrix}
},
\label{LSAoverall}
\end{equation}
\red{where the matrix $\bar{\mathbf{C}}$ can be written as the product of two matrices, $\bar{\mathbf{C}}=\bar{\mathbf{Q}}\bar{\mathbf{J}}$, where the first is the linearised mobility matrix [c.f.\ Eq.~\eqref{eq:solsusp-mob}]
\begin{equation}
\bar{\mathbf{Q}}
=h_i^3
\begin{bmatrix} 
1 & \phi_i\\
\phi_i & \phi_i^2 + \frac{a^2\phi_i}{2\pi h_i^2}
\end{bmatrix},
\label{eq:bar_Q}
\end{equation}
and the second is the following (largely thermodynamic in origin) matrix
\begin{align}
\bar{\mathbf{J}}
=-k^2
\begin{bmatrix} 
k^2 + g''(h_i) + \phi_i \bar{F} & -\bar{F}\\
-\bar{F} & \frac{1}{\phi_i}\bar{F}
\end{bmatrix},
\label{eq:bar_J}
\end{align}
where $\bar{F} = (f''(\phi_i)+ \epsilon'\phi_i k^2)/h_i=\left(K' - \alpha'\phi_i + 3\beta'\phi_i^3 + \epsilon'\phi_i k^2\right)/h_i$, and where we use} $\partial{h}/\partial t =  \omega \kappa e^{i(k_xx+k_yy)+\omega t}$ and $\nabla{h} = i{\bf k}\kappa e^{i(k_xx+k_yy)+\omega t}$, together with corresponding derivatives of $\psi$\red{. The} Jacobian matrix coefficients \red{for our specific $g$ and $f$} are
\begin{equation}
C_1 = -h_i^{3}k^{2}\left[k^{2} + A'\left(-\frac{6}{h_i^{4}}+\frac{12}{h_i^{5}}\right) \right],
\label{C1}
\end{equation}
\begin{equation}
C_2 = 0,
\label{C2}
\end{equation}
\begin{equation}
C_3 = -h_i^{3}k^{2}\phi_i\left[k^{2} + A'\left(-\frac{6}{h_i^{4}}+\frac{12}{h_i^{5}}\right) \right] + \frac{a^{2} \phi_i}{2 \pi} k^{2}\left[K'-\alpha'\phi_i + 3\beta'\phi_i^3 + \epsilon'\phi_i k^{2}\right],
\label{C3}
\end{equation}
and
\begin{equation}
C_4 = -\frac{a^{2}}{2 \pi} k^{2}\left[K'-\alpha'\phi_i + 3\beta'\phi_i^3 + \epsilon'\phi_i k^{2}\right].
\label{C4}
\end{equation}
Note that $C_3 = \phi_i (C_1 - C_4)$. We also observe that $C_2 = 0$, which originates from the multiple cancellations of terms that occur when the functional derivates of the free energy are substituted into the dynamical equations (see Appendix~\ref{appA})\red{, and it is an expression of the fact that bulk concentration gradients do not drive osmotic flow.} In order to find the relationship between $\omega$ and $k$, we simply find the eigenvalues of the matrix:
\begin{equation}
\begin{bmatrix} 
C_1-\omega & C_2\\
C_3 & C_4-\omega
\end{bmatrix}
.
\end{equation}
The first eigenvalue is:
\begin{equation}
\omega_h = -h_i^{3}k^{2}\left[k^{2} + A'\left(-\frac{6}{h_i^{4}}+\frac{12}{h_i^{5}}\right) \right],
\label{trace2dimless}
\end{equation}
and the second is  
\begin{equation}
\omega_\psi = -\frac{a^{2}}{2 \pi} k^{2}\left[K'-\alpha'\phi_i + 3\beta'\phi_i^3 + \epsilon'\phi_i k^{2}\right].
\label{omega_TFCS}
\end{equation}
It is because $C_2=0$ and $C_3 = \phi_i(\omega_h - \omega_\psi)$, that we obtain the simple results $\omega_h=C_1$ and $\omega_\psi=C_4$.

\red{By writing the Jacobian matrix as the product of two symmetric matrices (shown in Eqs.~\eqref{eq:bar_Q}--\eqref{eq:bar_J}) and noting also that $\bar{\mathbf{Q}}$ is positive definite allows one to write the linear problem as a generalized eigenvalue problem and therefore to directly deduce that (i) all eigenvalues are real, and that (ii) the stability thresholds do not depend at all on the mobility coefficients in $\bar{\mathbf{Q}}$, which is what one must expect, given the overall gradient dynamics structure in Eq.~\eqref{eq:solsusp-coup-grad-evap}.}

\red{Note also} that $\omega_h$ is the dispersion relation that is obtained for a pure liquid film, in the limit $\psi\to0$. Moreover, $\omega_\psi$ is the dispersion relation that is obtained when considering just the colloids evolving in a liquid film that remains constant in thickness for all time. In other words, there is no coupling between film height fluctuations and colloidal concentration fluctuations at the linear level. It is only at the nonlinear level that they are coupled. \red{If we were to replace the binding potential in Eq.~\eqref{binding} with one including a coupling between $\psi$ and $h$, then our linear decoupling would no longer occur, as in the models of \citet{naraigh2010nonlinear} and \citet{thiele2013gradient}. Such a coupling may even push an otherwise stable systems over the instability threshold.} 

Because of the decoupling just mentioned, the standard pure-liquid-film condition for $\omega_h(k)>0$ also applies here, i.e.\ one must have
\begin{equation}
\frac{\partial^2g}{\partial h^2}\bigg|_{h_i}=A'\left(-\frac{6}{h_i^{4}}+\frac{12}{h_i^{5}}\right)>0.
\label{eq:h_stab_cond}
\end{equation}
The limit of linear stability for the liquid film is thus the locus in the phase diagram of the equation
\begin{equation}
\frac{\partial^2g}{\partial h^2}\bigg|_{h_i}=0.
\end{equation}
From Eq.~\eqref{eq:h_stab_cond} we see that this linear stability threshold is simply $h_i=2$, i.e.\ for $h_i>2$ the film is linearly unstable. Note however that when $h_i$ becomes large, the growth rate becomes very small and the unstable wavenumbers also become very small, i.e.\ corresponding to large length-scale instabilities.

Similarly, for the colloids to be stable, with $\omega_\psi(k)\leq0$ for all $k$, we must have
\begin{equation}
\frac{\partial^2f}{\partial \phi^2}\bigg|_{h_i,\psi_i}=K'-\alpha'\phi_i + 3\beta'\phi_i^3>0.
\label{eq:stab_cond}
\end{equation}
The corresponding limit of linear stability is just the locus in the phase diagram of the equation
\begin{equation}
\frac{\partial^2 f}{\partial \phi^2}\bigg|_{h_i,\psi_i}=0.
\label{spin2}
\end{equation}
Owing to the fact that we derive our dynamics from the free energy functional, as we show in the following subsection, this condition for the spinodal (onset of linear instability) for colloidal phase separation, is precisely the same as that we obtain from the standard thermodynamic definition of the spinodal.

\subsection{Bulk phase diagram for the colloids}

\begin{figure}
 \centerline{\includegraphics[width=250pt]{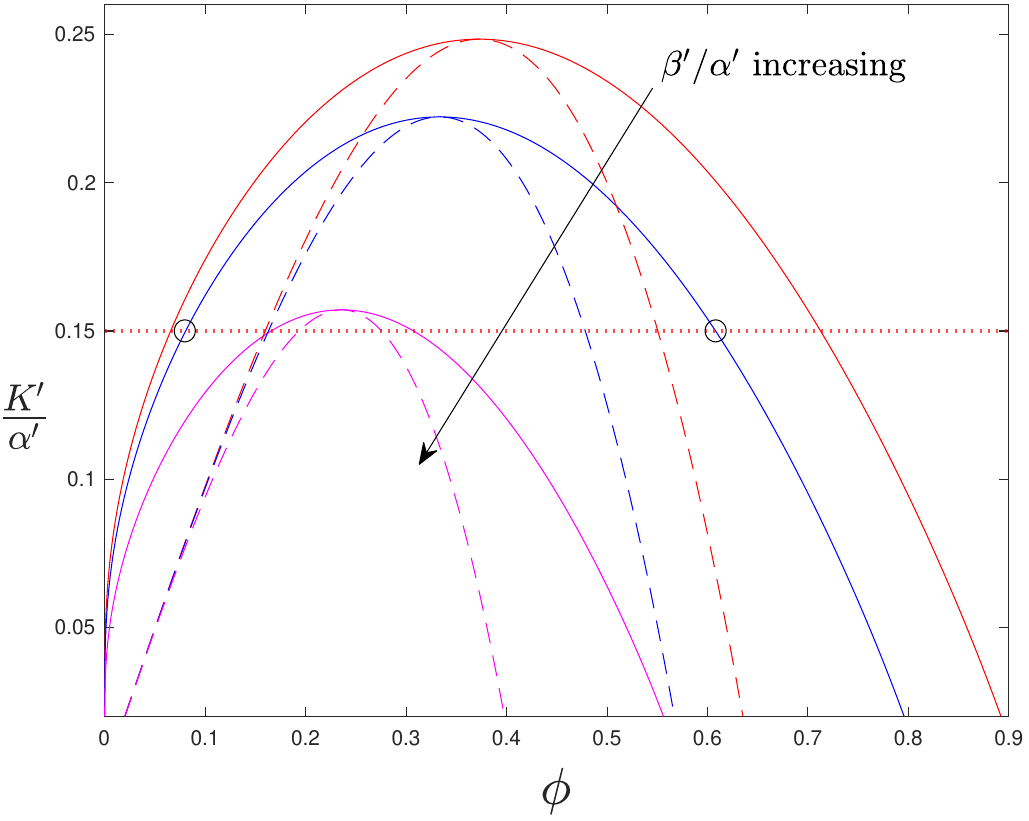}}
\caption{Bulk colloid phase diagram in the plane of dimensionless temperature $K'/\alpha'=k_BT/(\alpha a^3)$ versus colloid concentration $\phi$, for the three values of $\beta'/\alpha'=\beta/\alpha = 0.8$, 1 and 2. The solid lines are the binodals and the dashed lines are the corresponding spinodals. The circles identify coexisting colloid concentrations for the particular temperature $K'/\alpha'=0.15$ and $\beta'/\alpha'=1$, that are referred to in \S~\ref{sec:case1}.}
\label{Bulkphasediagram}
\end{figure}

When the colloids agglomerate, the system separates into two phases, one with colloid concentration $\phi_a$ and the other with concentration $\phi_b$. These two phases are in thermodynamic coexistence. The values of $\phi_a$ and $\phi_b$ vary as a function of the temperature and the other parameters in the model (i.e.\ $K'$, $\alpha'$ and $\beta'$). A plot of $\phi_a$ and $\phi_b$ in the temperature versus concentration plane yields the binodal curve. In the same diagram, we can also plot the spinodal curve from Eq.~\eqref{spin2}. These two curves meet at a single point in the phase diagram, namely the critical point \citep{hansen2013theory}. The binodal curve separates the single-phase and two-phase-coexistence regions of the phase diagram, while the spinodal curve is the linear-stability boundary: i.e.\ the system is linearly unstable if we choose the parameters so that the system is initially uniform and inside the spinodal curve. \red{The phase diagram for our model is shown in Fig.~\ref{Bulkphasediagram}, and here we discuss how it is constructed.}

The binodal curve (coexisting states) can be determined by considering the thermodynamic requirements for two different phases to coexist: namely, they must be in thermal, mechanical and chemical equilibrium. This means that for phases $a$ and $b$ the temperature, pressure and chemical potential must be simultaneously equal
\begin{equation}
T_a = T_b, \hspace{0.5cm}p_a=p_b, \hspace{0.5cm}\mu_a=\mu_b.
\end{equation}
The first condition is equivalent to the requirement that the value of $K'$ be the same in both phases -- see Eq.~\eqref{dimentionlessnumbers}. Taking the second and third conditions, using the expressions in Eqs.~\eqref{pressurevir} and \eqref{eq:mu} for $p$ and $\mu$, and remembering that the density of the colloids $\rho\propto\phi$, we obtain the following pair of simultaneous equations to be solved for the two coexisting concentrations $\phi_a$ and $\phi_b$:
\begin{equation}
   \frac{\partial f}{\partial \phi}\bigg|_{\phi = \phi_a}=\frac{\partial f}{\partial \phi}\bigg|_{\phi = \phi_b},
\label{equatechempot}
\end{equation}
and
\begin{equation}
   \left[\phi\left(\frac{\partial f}{\partial \phi}\right)-f\right]\bigg|_{\phi = \phi_a} = \left[\phi\left(\frac{\partial f}{\partial \phi}\right)-f\right]\bigg|_{\phi = \phi_b}.
\label{equatepressure}
\end{equation}
Solving these for fixed $\alpha'$ and $\beta'$ and for a range of values of $K'$ yields the binodal curve. The corresponding spinodal is obtained from Eq.~\eqref{spin2}.
These are shown in Fig.~\ref{Bulkphasediagram} for three different values of $\beta'/\alpha'=\beta/\alpha$. We plot the binodal and spinodal curves in the dimensionless temperature $K'/\alpha'=k_BT/(\alpha a^3)$ versus concentration $\phi$ plane. One can see that $K'/\alpha'$ and $\beta'/\alpha'$ are the relevant ratios to consider, by simply taking the spinodal condition obtained from Eq.~\eqref{spin2} [see also Eq.~\eqref{eq:stab_cond}] and dividing through by $\alpha'$ to obtain the following expression for the spinodal
\begin{equation}
\frac{K'}{\alpha'}=\phi_i - 3\frac{\beta'}{\alpha'}\phi_i^3.
\label{eq:spinodal}
\end{equation}
From our dynamical equations, we identify the spinodal as a linear-stability threshold. However, from the thermodynamic point of view, it is the line in the phase diagram at which the isothermal compressibility $\chi_T$ diverges. This compressibility can be evaluated as \citep{hansen2013theory}
\begin{equation}
\chi_T=\frac{1}{\phi}\frac{\partial\phi}{\partial p}.
\end{equation}
From this expression, together with Eqs.~\eqref{pressurevir} and \eqref{eq:mu}, one can easily show that that the spinodal condition in Eq.~\eqref{spin2} is precisely the line in the phase diagram where $\chi_T\to\infty$.

Inspecting Fig.~\ref{Bulkphasediagram}, we see that increasing the ratio $\beta'/\alpha'$ moves the critical point to the left and downwards in the phase diagram. Recall that $\beta'$ is the parameter originating from the steric repulsions between the colloids and $\alpha'$ is related to the strength of the attraction between the colloids. Thus, adjusting the ratio $\beta'/\alpha'$ for fixed $K'/\alpha'$ varies the value of $\phi$ in the dense colloidal phase, while at the same time also shifting the critical point. As long as $\beta'/\alpha'\sim{\cal O}(1)$, the precise value makes very little difference qualitatively to the behaviour of our model. Thus, henceforth we fix $\beta'/\alpha'=1$ \red{\cite[so that our phase diagram roughly matches some of the typical colloid phase diagrams displayed in chapter 12 of][and references therein]{hansen2013theory}} and vary just the ratio $K'/\alpha'=k_BT/(\alpha a^3)$, which is equivalent to either varying the temperature or the strength of the attraction between the colloids. On the other hand, if modelling a particular experimental system, then adjusting the values of $\beta'/\alpha'$ and $K'/\alpha'$ to values different to those used here may be more appropriate.

\subsection{Dispersion relations}

Having obtained the dispersion relations (growth rates) in Eqs.~\eqref{trace2dimless} and \eqref{omega_TFCS}, we now discuss the different possible stability regimes. As mentioned already, the liquid film is unstable for $h_i>2$ and for all $A'>0$, where we have $\omega_h(k)>0$ for a band of wavenumbers $0<k<\sqrt{2}k_{h}$, where
\begin{equation}
k_{h}=\frac{\sqrt{3A'h_i(h_i - 2)}}{h_i^3}
\label{eq:k_h}
\end{equation}
is the fastest growing mode, i.e.\ $\omega_h(k_h)$ is the maximum growth rate. The largest possible value of $k_h$ is $k_h^{max}=2\sqrt{60A'}/125\approx0.124\sqrt{A'}$, which occurs when $h_i=2.5$. 

\begin{figure}
        \includegraphics[width=0.48\linewidth]{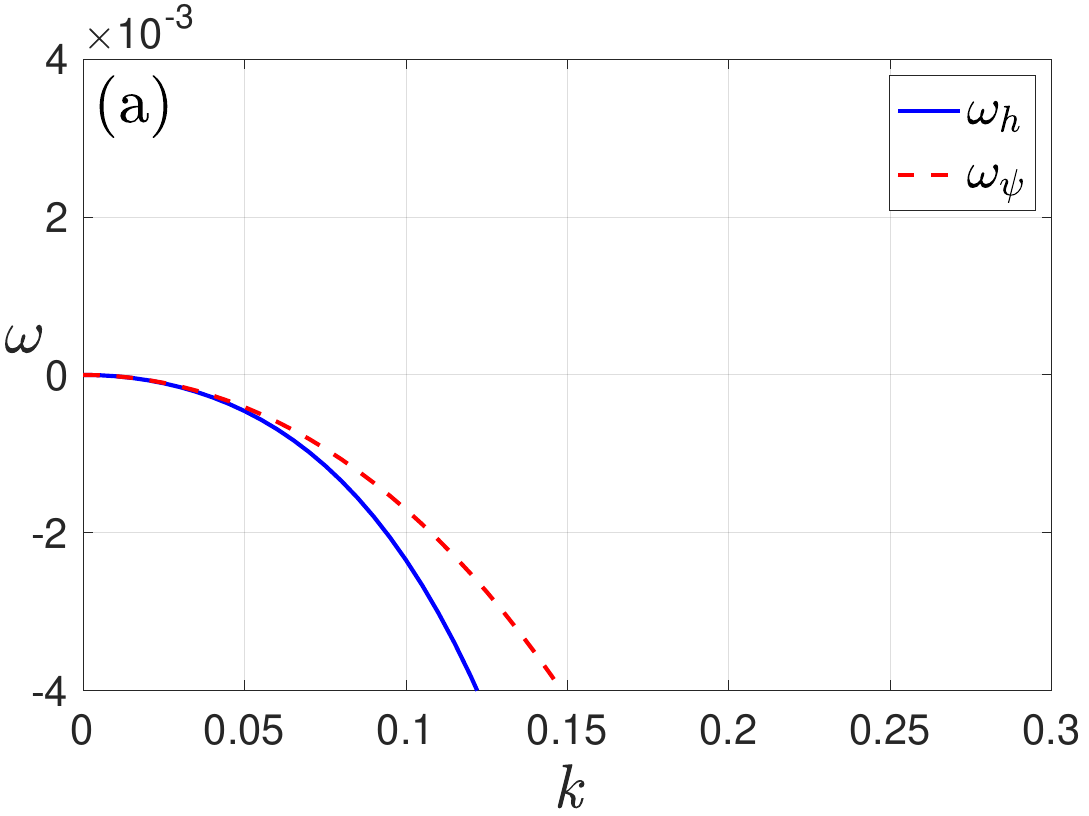}
        \includegraphics[width=0.48\linewidth]{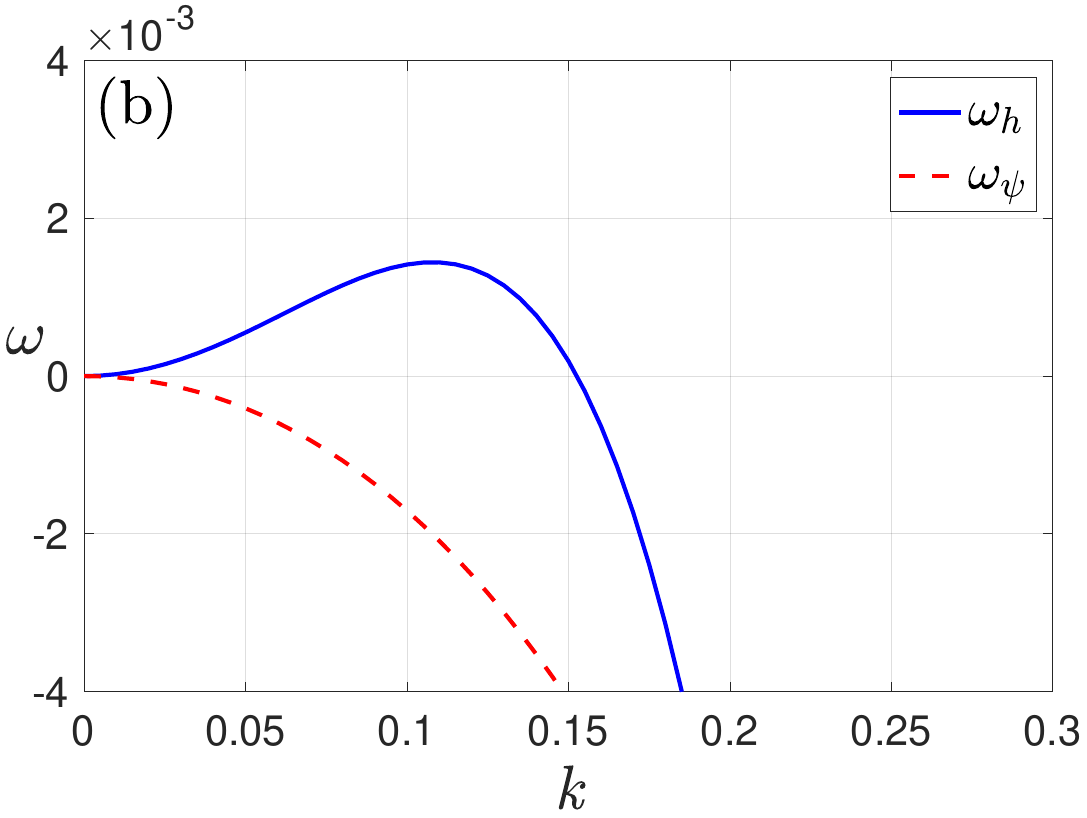}
        \includegraphics[width=0.48\linewidth]{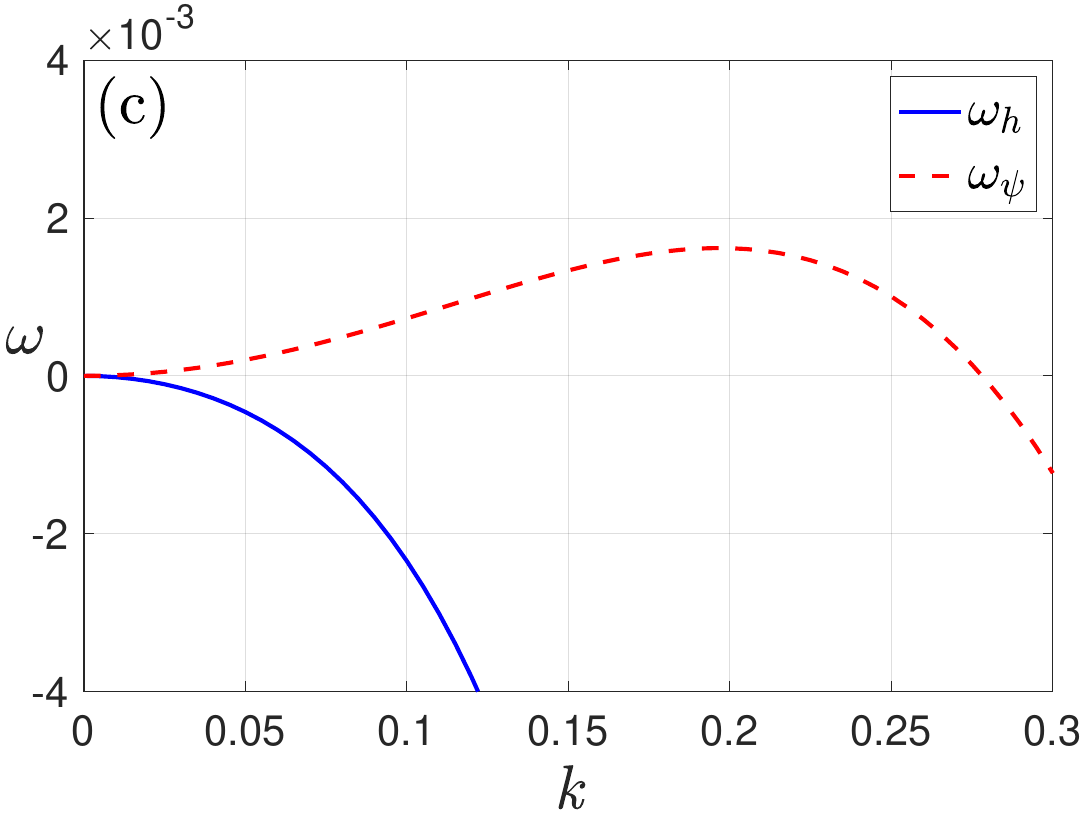}
        \hspace{0.45cm}
        \includegraphics[width=0.48\linewidth]{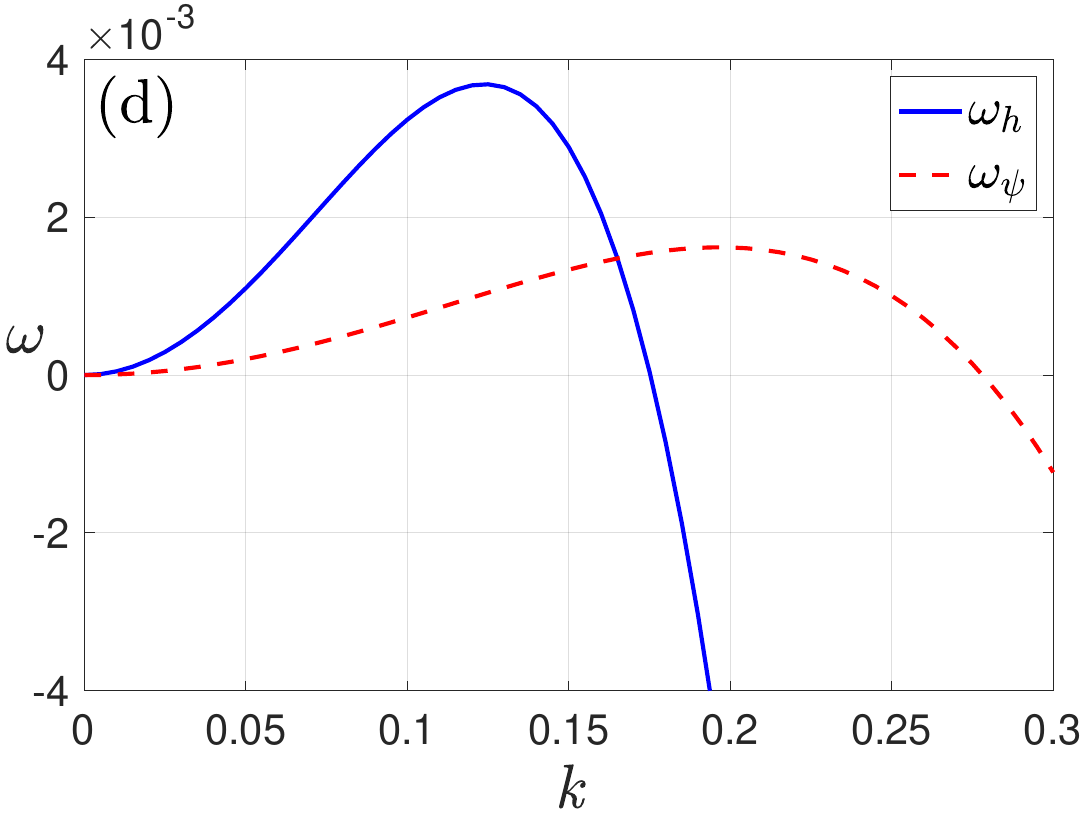}
\caption{Dispersion relation for four cases with $A' = 1$, $K' = 0.15$, $\epsilon' = 0.4$ and \red{$a^{2} = 100$}. In case (a), both the film height and colloids are stable ($h_i = 1.9$ and $\phi_i = 0.15$). In case (b), the film height is unstable and colloids are stable ($h_i = 2.2$ and $\phi_i = 0.15$). In (c) the film height is stable and colloids are unstable ($h_i = 1.9$ and $\phi_i = 0.17$). In (d), both the film height and colloids are unstable ($h_i = 2.5$ and $\phi_i = 0.17$).}
\label{disp1}
\end{figure}

Similarly, the colloids are unstable when the concentration takes a value inside the spinodal, where $f''(\phi)<0$ (see Eq.~\eqref{eq:stab_cond} and Fig.~\ref{Bulkphasediagram}) and also where $\omega_\psi(k)>0$ for a band of wavenumbers $0<k<\sqrt{2}k_\psi$, where
\begin{equation}
k_\psi=\sqrt{\frac{-(K'-\alpha'\phi_i+3\beta'\phi_i^3)}{2\epsilon'\phi_i}},
\label{eq:k_psi}
\end{equation}
which is the fastest growing mode, i.e.\ $\omega_\psi(k_\psi)$ is the maximum growth rate. The largest possible value of $k_\psi=k_\psi^{max}$ occurs when $\phi_i=(K'/6\beta')^{{1}/{3}}$, which corresponds to a curve in the phase diagram $\propto\phi_i^3$ that passes through both the origin and the critical point and where
\begin{equation}
k_\psi^{max}=\sqrt{\frac{{\alpha'-((\frac{9}{2})^{2}K'^2\beta')^{1/3}}}{2\epsilon'}}.
\end{equation}
Thus, we see that in general there are four possible stability scenarios, with four corresponding arrangements for the dispersion relations $\omega_h$ and $\omega_\psi$: (a) both the liquid and the colloids are stable [see Fig.~\ref{disp1}(a)]; (b) the liquid film is unstable, but the colloids are stable [see Fig.~\ref{disp1}(b)]; (c) the liquid stable, but the colloids are unstable [see Fig.~\ref{disp1}(c)]; and finally, (d) both are unstable [see Fig.~\ref{disp1}(d)]. In all cases we set $\beta'/\alpha' = 1$, corresponding to the middle of the three phase diagrams in Fig.~\ref{Bulkphasediagram}.

When both the liquid film and the colloids are unstable, then there are four possible arrangements of the dispersion relations, related to the question of whether $k_\psi>k_h$ or not and which leads to the fastest growth rate, i.e.\ whether $\omega_\psi(k_\psi)>\omega_h(k_h)$ or not. Recall that the fastest growing modes $k_h$ and $k_\psi$ each correspond to initial fastest growing wavelengths $\lambda_h = 2\pi/k_h$ and $\lambda_\psi = 2\pi/k_\psi$, respectively. Thus, suppose $\omega_\psi(k_\psi)<\omega_h(k_h)$ so that the fastest growing mode is that corresponding to perturbations of the film height. When $k_\psi>k_h$, then the wavelength of the fastest growing film-height mode is longer than the corresponding fastest mode in the colloidal demixing. In contrast, when $k_\psi<k_h$, then the wavelength of the fastest film-height mode is shorter than the fastest colloidal demixing mode. However, when $\omega_\psi(k_\psi)>\omega_h(k_h)$, then the fastest growing mode is that due to colloidal demixing and this can have a wavelength either longer or shorter than the fastest growing film-height mode, depending on whether $k_\psi>k_h$ or not. All four of these cases are possible, depending on the choice of parameters in Eqs.~\eqref{trace2dimless} and \eqref{omega_TFCS}. \red{Note that it is the presence of the two interfacial tensions $\gamma\propto1/A'$ and $\epsilon'$ in the denominators of \eqref{eq:k_h} and \eqref{eq:k_psi}, respectively, that are the main players in determining the characteristic length scales $\lambda_h$ and $\lambda_\psi$.}

\section{Numerical results}
\label{sec:nandr}

We return to the full non-dimensional system in Eqs.~\eqref{eq:solsusp-coup-grad}, which are a pair of coupled PDEs that are first-order in time and fourth-order in each spatial variable. As illustrated in Fig.~\ref{fig:geoillu}, here we consider Cartesian coordinates in both the case of a single spatial variable $x$, representing a two-dimensional colloidal droplet, and the case of spatial variables $(x,y)$, representing the three-dimensional situation. To complete the PDE system, we must supply appropriate initial conditions for the dependent variables $h$ and $\psi$, and for simplicity in all situations we solve our system with periodic boundary conditions.

To obtain solutions of the full system, we developed independent codes for the two- and three-dimensional situations in \red{\textsc{Matlab}}. For the spatial derivatives we use central finite differences with the periodic boundary conditions included. We choose a sufficiently large number of spatial discretisation points for a converged simulation \red{-- i.e.\ we performed convergence tests (not displayed) with varying mesh spacing to check convergence}. In our two-dimensional code we solve in a spatial domain $x\in [0,L_x)$ with $500$ discretisation points, and in three-dimensions in $(x,y)\in [0,L_x)\times[0,L_y)$ with $110$ points in either direction. To integrate in time, we use the variable-step, variable-order solver \emph{ode15s} from \cite{matlabodesuite} which efficiently allows for integrating over many orders of magnitude in time whilst capturing processes on fast times-scales. Due to the adaptive nature of \emph{ode15s} a time-step is not prescribed, but absolute and relative error tolerances were set to $10^{-9}$ for all computations. \red{Note that decreasing the tolerances by an order of magnitude does not qualitatively change our results.}

In order to verify our codes and investigate the behaviour of colloidal liquids from our model, we conduct a series of simulations and compare the results with the theoretical dispersion relations from \S~\ref{sec:LSA}. For this, we set initial conditions as flat profiles with perturbations: 
\begin{align}
h(x,t=0) = h_0(x) \equiv h_i+A_i(x),\nonumber\\
\psi(x,t=0) = \psi_0(x)\equiv \psi_i+B_i(x),
\label{inicon1D}
\end{align}
where $0 \leq x < L_x$. \red{When discretised like the fields $h$ and $\psi$, at each lattice point,} the perturbations $A_i(x)$ and $B_i(x)$ are small-amplitude randomly generated numbers uniformly distributed in $(-\varepsilon_h,\varepsilon_h)$, $(-\varepsilon_\psi,\varepsilon_\psi)$, where we retain control of the orders of magnitude of the perturbations of the film height and colloid profiles $\varepsilon_h$ and $\varepsilon_\psi$, respectively. This initial condition corresponds to an initially uniform-height well-mixed liquid layer deposited on the surface, with the small-amplitude perturbations corresponding to the small variations that always exist either due to the manner in which such films are deposited on surfaces or due to the thermal fluctuations that are always present on colloidal scales.

The initial conditions \refe{inicon1D} mirror the linear stability analysis ansatz \refe{linvar}, allowing a direct comparison between analytical and numerical results. The early evolution from near-flat profiles is visually unremarkable, however the modes of growth become clear upon transformation into Fourier space. For example, even the wavenumber $k$ that corresponds to the fastest-growing mode of wavelength $\lambda_h = 2\pi/k_h$ or $\lambda_\psi = 2\pi/k_\psi$ can only be distinguished at early times from initially selected noise after Fourier transform. It is thus more convenient to directly compare the analytical and numerically-evaluated dispersion relations. To do this, we Fourier transform our variables as $\widehat{h}=\mathcal{F}(h)$ and $\widehat{\psi}=\mathcal{F}(\psi)$, where $\mathcal{F}(\cdot)$ denotes the Fourier transform, such that the linearised dynamical equations \eqref{eq:solsusp-coup-grad} [c.f.~Eq.~\eqref{LSAoverall}] become
\begin{equation}
\begin{bmatrix} 
\partial_t{\widehat{h}}\\
\partial_t{\widehat{\psi}}
\end{bmatrix}
=
\begin{bmatrix} 
\omega_h & 0\\
\phi_i(\omega_h -\omega_\psi) & \omega_\psi
\end{bmatrix}
\begin{bmatrix} 
\widehat{h}\\
\widehat{\psi}
\end{bmatrix}.
\label{LSAcompare}
\end{equation}
Noting that the $\widehat{h}$ equation decouples (as discussed in \S~\ref{sec:LSA}, due to $C_2 = 0$), we can directly find an analytical solution. We then solve for $\widehat{\psi}$ in terms of $\widehat{h}$, i.e.
\begin{eqnarray}
\widehat{h}(t) & = & \widehat{h}_0\exp(\omega_h t),
\\
\widehat{\psi}(t) & = & \left(\widehat{\psi}_0-\phi_i\widehat{h}_0\right)\exp(\omega_\psi t)+\phi_i\widehat{h}_0\exp(\omega_h t),\label{fourier2}
\end{eqnarray}
where $\widehat{h}_0$ and $\widehat{\psi}_0$ are the Fourier transformed initial conditions and $\phi_i = \psi_i/h_i$. From this result, we can rearrange (\ref{fourier2}) to obtain the computational $\omega_h$ and $\omega_\psi$ as 
\begin{eqnarray}
\label{eq:51}
\omega_h & = & \frac{1}{t}\ln\left(\frac{\widehat{h}_c(t)}{\widehat{h}_0}\right),
\\
\omega_\psi & = & \frac{1}{t}\ln\left(\frac{h_i\widehat{\psi}_c(t)-\psi_i\widehat{h}_c(t)}{h_i\widehat{\psi}_0-\widehat{h}_0\psi_i}\right),\label{eq:numdisp}
\end{eqnarray}
with $\widehat{h}_c(t)$ and $\widehat{\psi}_c(t)$ being the Fourier transforms of the computed results at any time $t>0$. The time we end the simulation must remain within the linear regime for comparison to the linear stability analysis, since at later times it departs from the linear regime, as the higher-order terms become significant and more complex evolution of the system occurs. To see when this occurs, it is instructive to consider $\ln|h - h_i|$ (equivalently $\ln|\psi - \psi_i|$) at a point in the domain, since rearrangement of Eq.~\refe{linvar} gives that
\begin{equation}
\ln|h - h_i| \sim \omega_h t+\cdots,
\label{eq:lnmax}
\end{equation}
and thus the slope of $\ln(\mbox{max}|h - h_i|)$ against $t$ gives an approximation of $\omega_h(k_h)$. We discuss these results in the context of various cases next.

\subsection[Case when the film height instability is dominant]{Case when the film height instability is dominant and $k_h<k_\psi$}
\label{sec:case1}

\begin{figure}
        \includegraphics[width=0.5\linewidth]{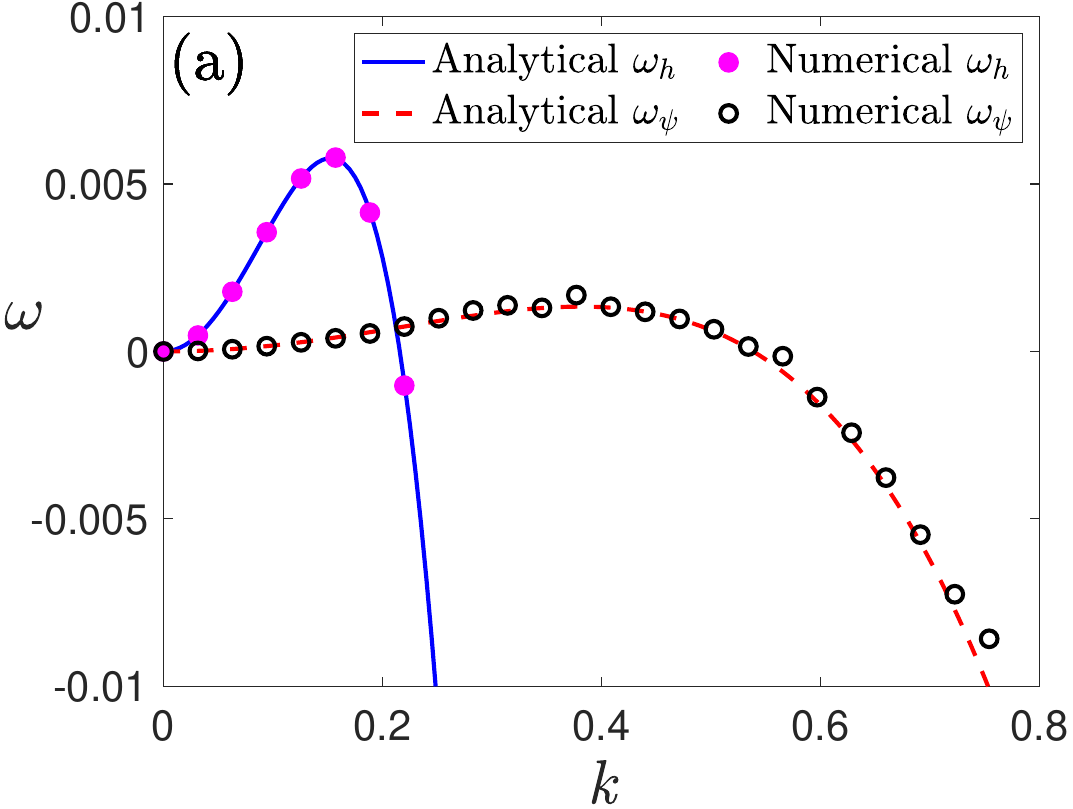}
        \includegraphics[width=0.5\linewidth]{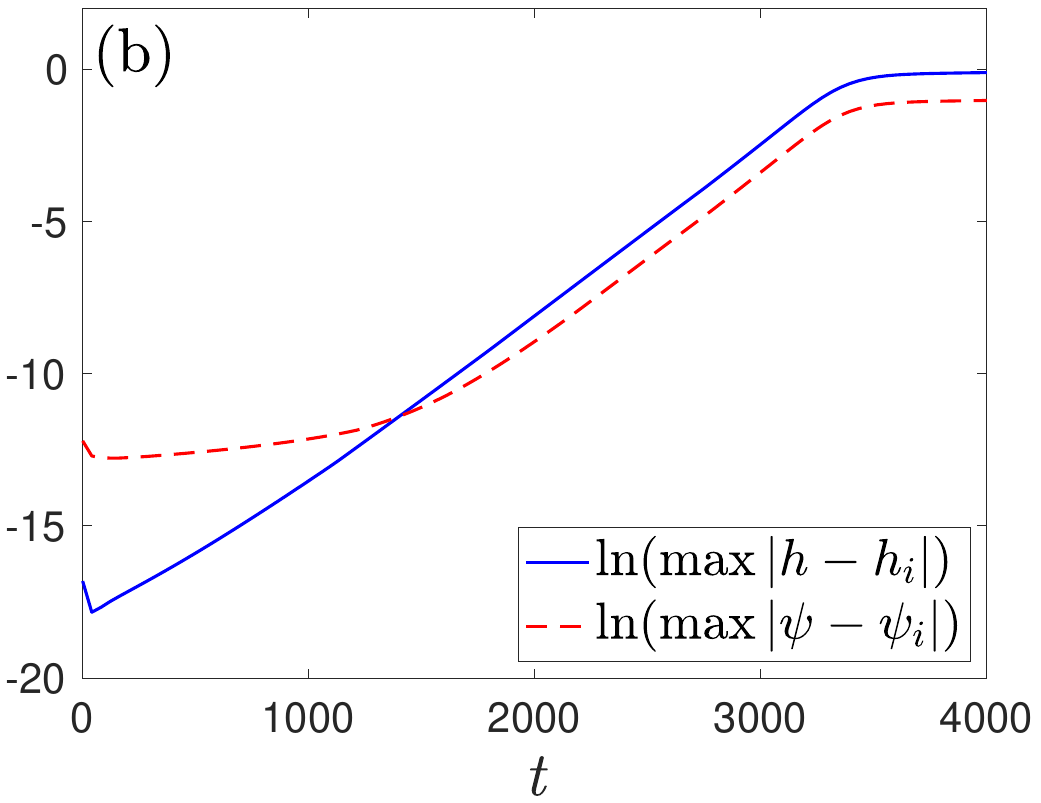}
\caption{Results for a case with $A' = 2$, $K' = 0.15$, $\alpha'=1$, $\epsilon' = 0.5$, \red{$a^{2} = 2$}, $h_i = 2.2$, and $\phi_i = 0.4$. (a) Dispersion relation calculated numerically via Eqs.~\eqref{eq:51} and \eqref{eq:numdisp} using $\varepsilon_h = 10^{-7}$, $\varepsilon_\psi = 10^{-5}$ (symbols) compared to the analytic results in Eqs.~\refe{trace2dimless} and \refe{omega_TFCS} (lines). (b) Measure of the evolution [c.f.\ Eq.~\refe{eq:lnmax}].}
\label{case1}
\end{figure}

As a first case, we consider a situation where $\omega_h(k_h) > \omega_\psi(k_\psi)$, i.e.\ the film height instability grows faster and dominates over any colloidal modes, and also where $k_h<k_\psi$, so \red{it} is at longer wavelength. The relevant dispersion relations are shown in Fig.~\ref{case1}(a). Using the results in Eqs.~\eqref{eq:51} and \refe{eq:numdisp} to numerically-evaluate the dispersion relations we can verify our numerical scheme through comparison to the analytical results in Eqs.~\refe{trace2dimless} and \refe{omega_TFCS}, which show excellent agreement.

In Fig.~\ref{case1}, we set the system length $L_x = 200$, with model parameters selected as $A' = 2$, $K' = 0.15$, $\alpha'=1$, $\epsilon' = 0.5$, and \red{$a^{2} = 2$}, with $h_i = 2.2$ and $\phi_i = 0.4$, and, as previously mentioned, we fix $\beta'/\alpha'=1$ throughout. In Fig.~\ref{case1}(a) we see that the computational result (\red{magenta dots} for $h$ and black circles for $\psi$) lie on the analytical dispersion relation curve, with little error. Errors increase with system length $L_x$, since more growing modes fit in the domain (recall that all possible Fourier modes must satisfy our periodic boundary conditions), which then couple/interfere at an earlier stage in the dynamics. Through numerical experimentation we regularly see excellent agreement between the numerical and analytical values for the film height eigenvalue $\omega_h$, whereas for the colloids there can be deviations of the numerical results from the analytically obtained curve $\omega_\psi$; they do however always follow the correct trend. This comparison makes for a very sensitive test of the numerics.

Next, we consider the evolution over time of the fastest growing mode through our result in Eq.~\refe{eq:lnmax}. In Fig.~\ref{case1}(b) we plot the logarithm of the maximum difference (as a function of position) between the initial and evolving heights. For this case, the film height instability dominates, which is demonstrated by the slope of $0.0055$, matching the value of $\omega_h(k=k_h)$ from Fig.~\ref{case1}(a) with negligible error (c.f.\ the slower maximum growth rate of the colloids is $\omega_\psi(k=k_\psi)=0.0013$). Additionally, we can predict that the system exits the linear regime as being when $\varepsilon_h e^{\omega_h t} \approx 1$, i.e.\ around $t \approx 3000$, which corresponds well with the region where linearity starts to be lost in Fig.~\ref{case1}(b).
Note that the first few points of the curves in Fig.~\ref{case1}(b) do not correspond to the fastest growing mode. This is because for simplicity we have chosen to quantify the evolution by considering the maximum change from the initial condition. The randomness of the initial condition \eqref{inicon1D} dictates that this does not initially coincide with a location demonstrating the growth of the fastest growing mode. Thus, different random seeds exhibit minor differences, until the fastest growing mode dominates over the effect of the choice of randomness. We further note that our result for $\psi$ in Fig.~\ref{case1}(b) shows that after $t\approx1500$ it does not grow with rate $\omega_\psi(k_\psi)$, due to being coupled to $h$; i.e.\ after $t\approx1500$ the fastest growing modes for both $h$ and $\psi$ are those with wavenumber $k_h$.

\begin{figure}
\centering
        \includegraphics[width=0.48\linewidth]{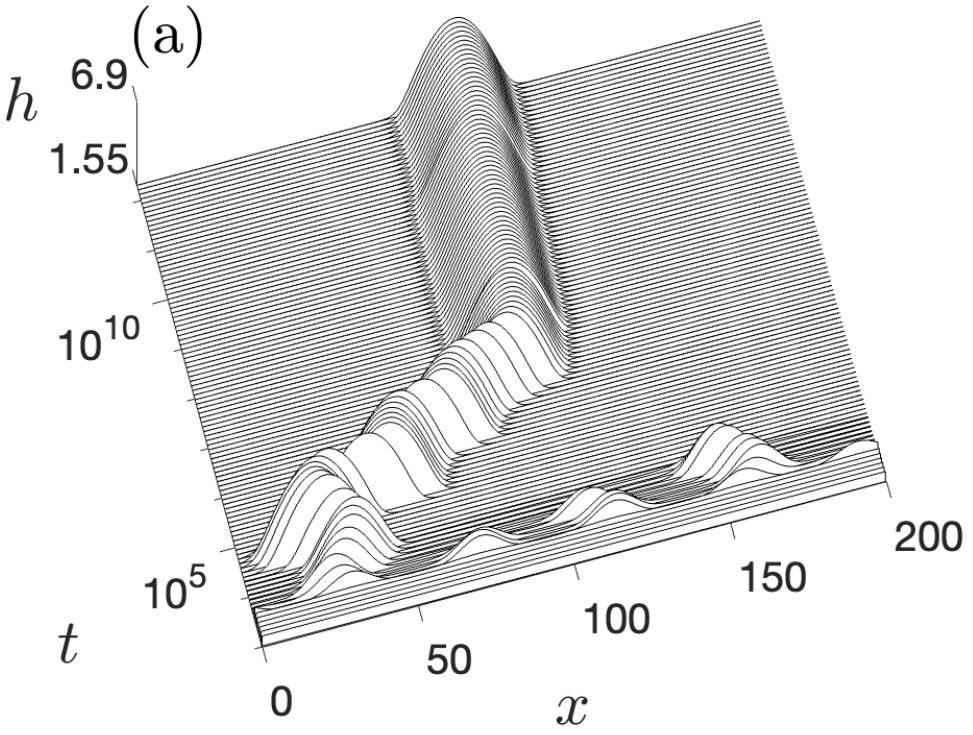}
        \hspace{0.2cm}
        \includegraphics[width=0.48\linewidth]{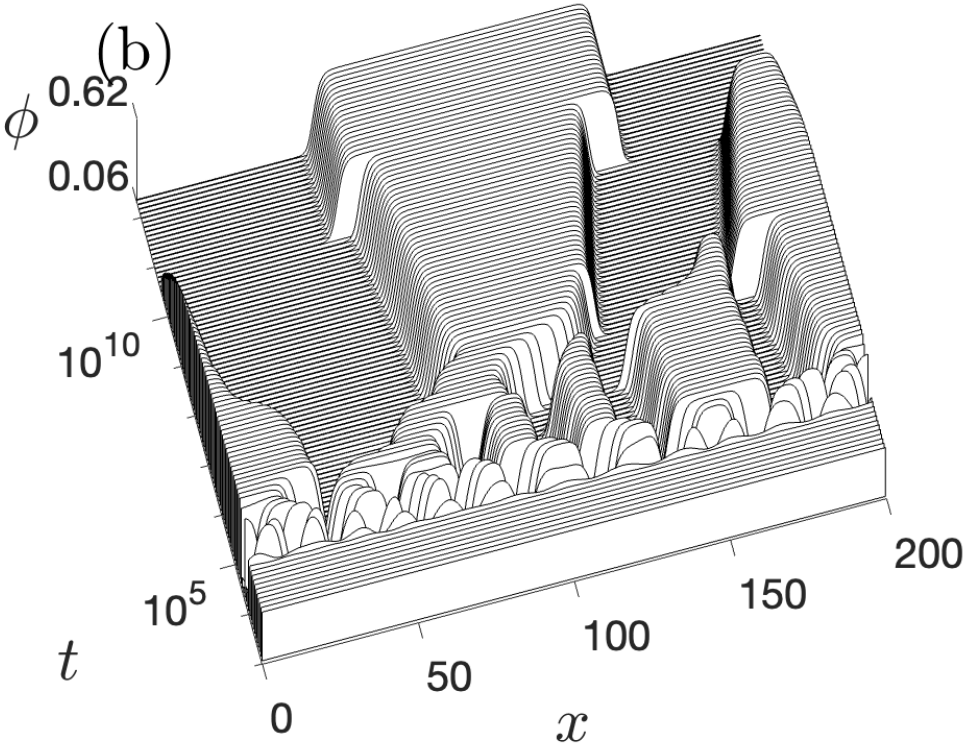}
        \includegraphics[width=0.48\linewidth]{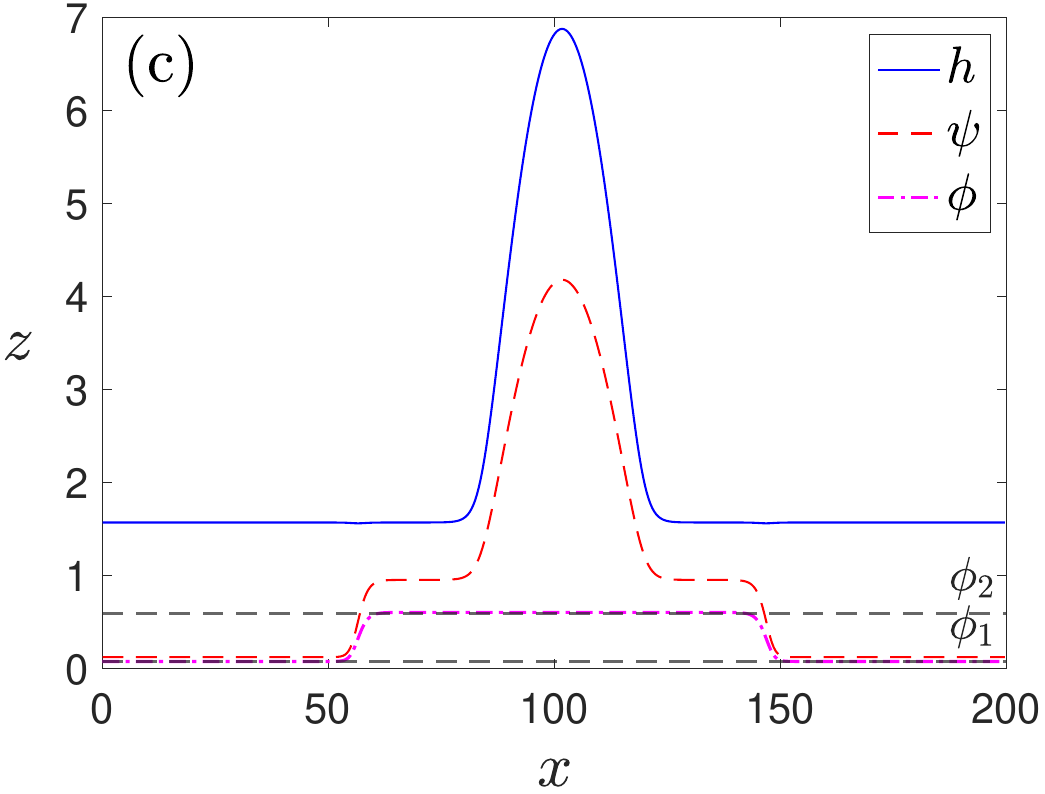}
        \includegraphics[width=0.48\linewidth]{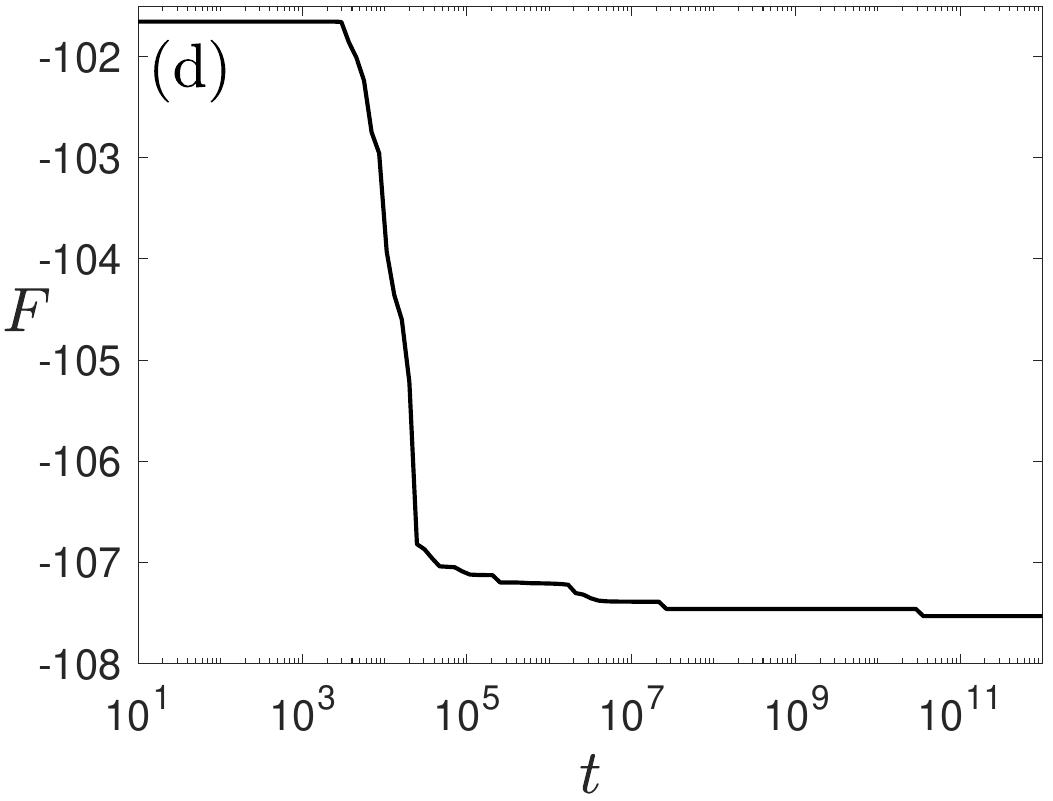}
\caption{Panel (a) shows a waterfall plot of the local film height over time and (b) shows the corresponding local concentration of colloids. Both use a logarithmic scale in $t$. Panel (c) shows the final equilibrium profiles. These are for $A' = 2$, $K' = 0.15$, $\alpha' = 1$, $\epsilon' = 0.5$, \red{$a^{2} = 2$}, $h_i = 2.2$ and $\phi_i = 0.4$. In (c) the dashed black horizontal lines denote the two coexisting $\phi$ values, indicated in Fig.~\ref{Bulkphasediagram}. \red{Panel (d) shows the free energy of the system against time.}}
\label{case1_evolve}
\end{figure}

From the wavenumbers $k_h$ and $k_\psi$ of the fastest growing modes in the system, the corresponding wavelengths $\lambda = 2\pi/k$, and the length of the system $L_x = 200$, we can predict the number of peaks that initially form in the system. Using the values of $k_h$ and $k_\psi$ in Eqs.~\eqref{eq:k_h} and \eqref{eq:k_psi}, i.e.\ at the peaks in $\omega_h$ and $\omega_\psi$, together with the Fig.~\ref{case1} parameter values, we have $\lambda_h \approx 42$ and $\lambda_\psi \approx 15$, so we expect at early times approximately 4 peaks in the film height, and 13 for the colloids. This is confirmed in Fig.~\ref{case1_evolve}, which displays the full time evolution of both $h$ and $\phi$.

Having observed the agreement between our analytical and numerical results in the linear regime, it is interesting to explore the longer-time dynamics of our system. Figure~\ref{case1_evolve} shows an example evolution, where we display the film height $h(x,t)$ in Fig.~\ref{case1_evolve}(a), the colloidal concentration $\phi(x,t)$ in Fig.~\ref{case1_evolve}(b), \red{the final profiles of the system in Fig.~\ref{case1_evolve}(c), and the free energy of the system against time in Fig.~\ref{case1_evolve}(d). We can see that the free energy of the system always decreases over time.} At early times we see the exponential growth or decay of modes as predicted by the linear theory. This corresponds to a decay of the small-wavelength modes, so the profiles initially appear to become smoother, but longer-wavelength unstable modes grow in amplitude, so that after some time the system starts forming peaks. As mentioned, the number of peaks is determined by the fastest growing modes and can be predicted by considering the dispersion relations. In Fig.~\ref{case1_evolve}(a) these peaks become visible at $t\approx2.5\times 10^{3}$ for the film height $h$, with the subdominant instability in the colloid profile taking longer to become visible, at around $t\approx10^4$, as can be seen in Fig.~\ref{case1_evolve}(b). 

The dynamics after the initial linear growth regime can be complex, with evolution due to capillarity of the film height, spinodal decomposition and coarsening of the colloids, as well as dynamics driven by the coupling between the two. In the case shown in Fig.~\ref{case1_evolve}(a), multiple small droplets initially form and then subsequently coalesce, with translation occurring as a consequence of the coalescence events. The evolution is somewhat similar to that typically observed for films of a pure liquid evolving from a uniform film to form droplets and then ultimately a single droplet, as the liquid dewets from a surface. However, the major difference in the case here is that there are two sudden translations of the droplet around $t=2.3\times 10^7$ and $t=3.2 \times 10^{10}$, both of which arise through the coupling to the colloid concentration profile, which is exhibiting coarsening via Ostwald ripening -- i.e.\ diffusion of colloids from one dense region to another through the low density regions \citep{lifshitz1961kinetics, wagner1961theorie} -- to reduce the number of colloid-rich regions from 3 to 2, and then from 2 to 1.

Consider now the colloid evolution displayed in Fig.~\ref{case1_evolve}(b): The phase diagram in Fig.~\ref{Bulkphasediagram} shows that for a given temperature ($K'/\alpha'$), there are two equilibrium values of $\phi$ that occur in a bulk system. These are indicated by the circles in Fig.~\ref{Bulkphasediagram}, with concentrations $\phi_1 = 0.08$ and $\phi_2 = 0.6$. Their respective values being the intersection points between the horizontal temperature line and the corresponding binodal curve. These are the values that the colloidal concentration wants to evolve towards, with the interfacial tension term in our model [the term with coefficient $\epsilon$ in Eq.~\eqref{freeenergyfunc}] penalising interfaces and thus driving the system to have as few regions of dense/sparse colloids as possible. This is indeed observed, with regions of high density having a clear plateau that corresponds to the dense state point, and these regions evolving in a similar way to droplets through translation and coalescence or joining through Ostwald ripening. Whether aggregation occurs through translation and joining or through the Ostwald mode depends on the state point, size of the droplets and distance between them. The situations in which each coarsening mode dominates can be understood and determined using the methods presented in \citet{pismen2004mobility, glasner2005collision, glasner2009ostwald, dai2010mean, pototsky2014coarsening, henkel2021gradient}.

For visual ease, and to highlight the relation to the phase diagram, we show the final profile of the system in Fig.~\ref{case1_evolve}(c). In this final state we observe that all colloids gather into a single region, with a single droplet too. Reaching equilibrium is slow for the film height, and even more so for the colloidal coarsening. To verify this profile as being final, we have separately considered the time evolution of the center of mass and the free energy, both of which reach a plateau. The approach to equilibrium is faster when considering a 3D situation (two spatial variables $x$ and $y$) since aggregation and droplet motion can occur in different directions, accelerating the process.

\subsection[Case when the colloidal instability is dominant]{Case when the colloidal instability is dominant and $k_h<k_\psi$}
\label{sec:case2}

\begin{figure}
        \includegraphics[width=0.48\linewidth]{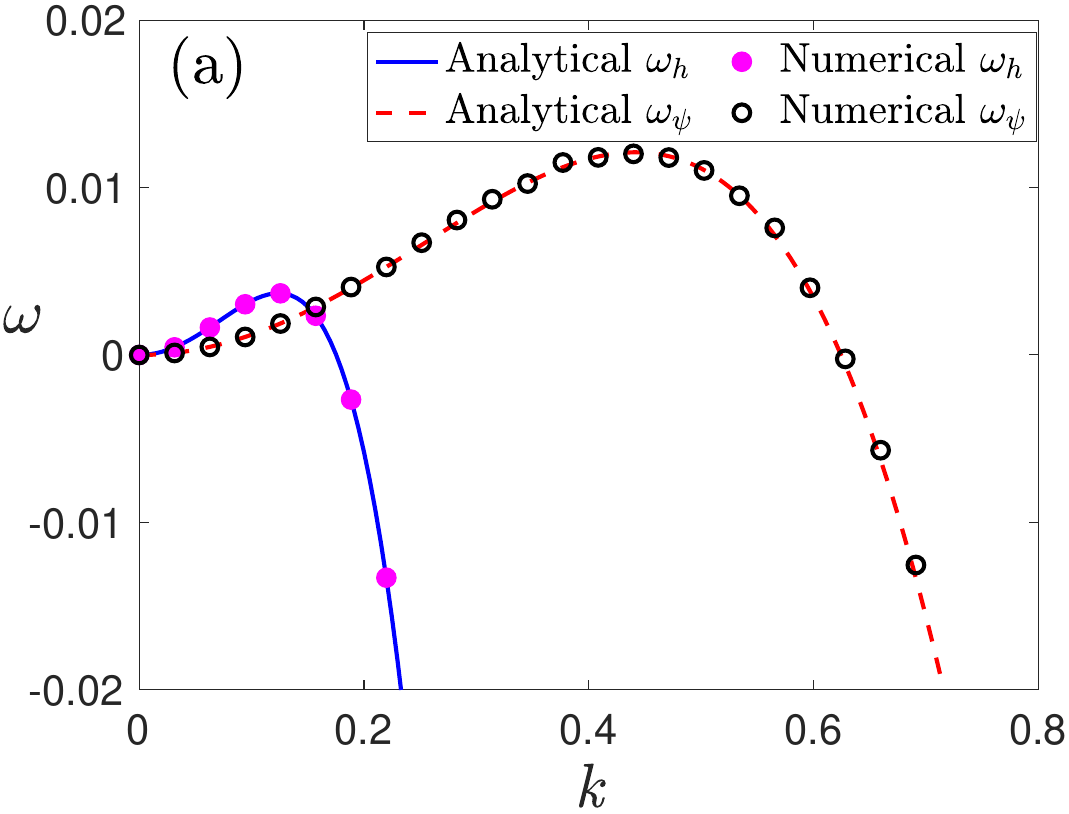}
        \includegraphics[width=0.48\linewidth]{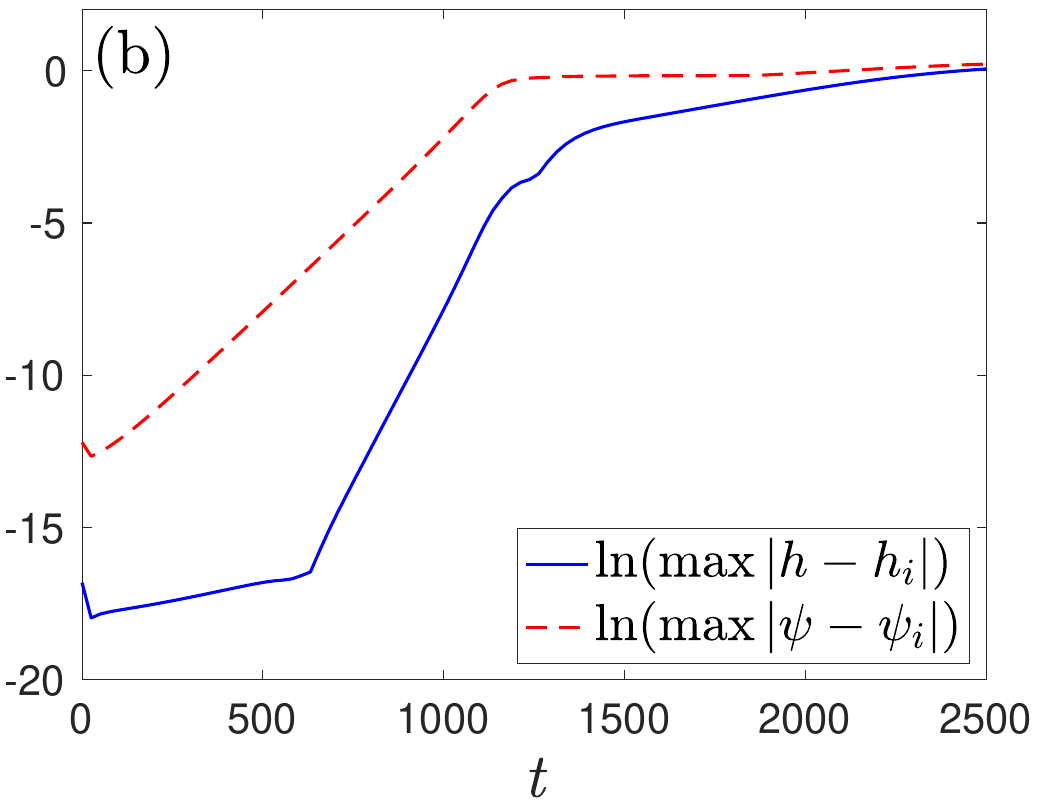}
\caption{Results for a case with $A' = 1$, $K' = 0.13$, $\alpha'=1$, $\epsilon' = 0.5$, \red{$a^{2} = 10$}, $h_i = 2.5$ and $\phi_i = 0.4$. (a) Dispersion relation calculated numerically via Eqs.~\eqref{eq:51} and \eqref{eq:numdisp} using $\varepsilon_h = 10^{-7}$, $\varepsilon_\psi = 10^{-5}$ (symbols) compared to the analytic results in Eqs.~\refe{trace2dimless} and \refe{omega_TFCS} (lines). (b) Measure of the evolution [c.f.\ Eq.~\refe{eq:lnmax}].}
\label{case2}
\end{figure}

A different case that is interesting to compare to the one in the previous subsection is a situation where both the film height and colloid profiles remain linearly unstable, with the fastest growing mode for the film height being still at a longer wavelength than that of the colloids, but we swap the relative growth rates of the instabilities, i.e.\ $\omega_\psi(k=k_\psi) > \omega_h(k=k_h)$, so that the colloid concentration fluctuations grow the fastest. We use the following parameter values to achieve this situation: $A' = 1$, $K' = 0.13$, $\alpha'=1$, $\epsilon' = 0.5$, \red{$a^{2} = 10$}, $h_i = 2.5$ and $\phi_i = 0.4$. This situation is clear from the dispersion relations displayed in Fig.~\ref{case2}(a), where once again the analytical and numerical results show excellent agreement. The evolution of the profiles (not shown) is similar to that displayed in Fig.~\ref{case1_evolve}, except the dominant early-time linear-stage dynamics\red{, which} instead corresponds to the growth of colloidal concentration fluctuations. An interesting feature of the dynamics in this case can be observed from Fig.~\ref{case2}(b), where $L_x=300$. From Eq.~\refe{eq:lnmax} we expect a linear-growth regime at early times, which we do indeed observe, with the dominant colloid instability growing with a steeper slope than in the case shown in Fig.~\ref{case1}. However, the line in Fig.~\ref{case2}(b) for the film height has a sharp turn, departing at an early stage from its corresponding slower linear-growth regime. It is clear that the coupling to the colloidal demixing drives this later growth, which occurs at a rate greater than the \red{rate predicted by linear analysis} and therefore must be due to the nonlinear coupling terms. \red{This difference to the case displayed in Fig.~\ref{case1} can be traced back to the top-right zero in the matrix in Eq.~\eqref{LSAcompare}, meaning that nonlinear couplings are required for fluctuations in $\psi$ to influence $h$.}

\subsection{Two strongly coupled cases where the long wavelength colloid instability dominates} 
\label{sec:case3}

\begin{figure}
        \includegraphics[width=0.48\linewidth]{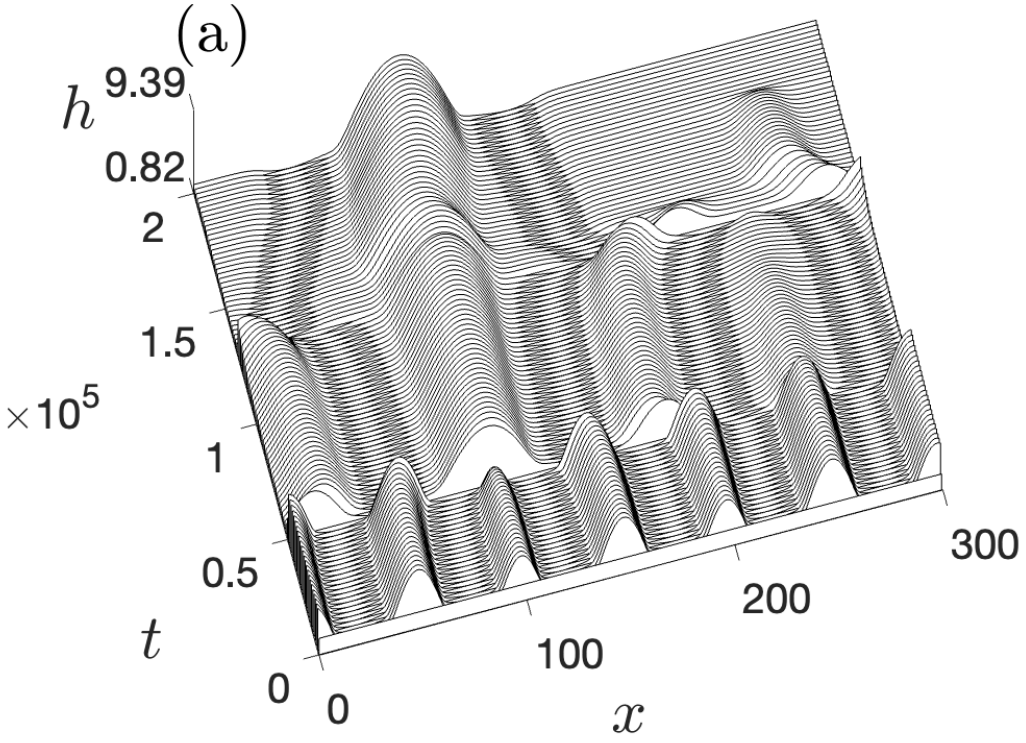}
        \includegraphics[width=0.48\linewidth]{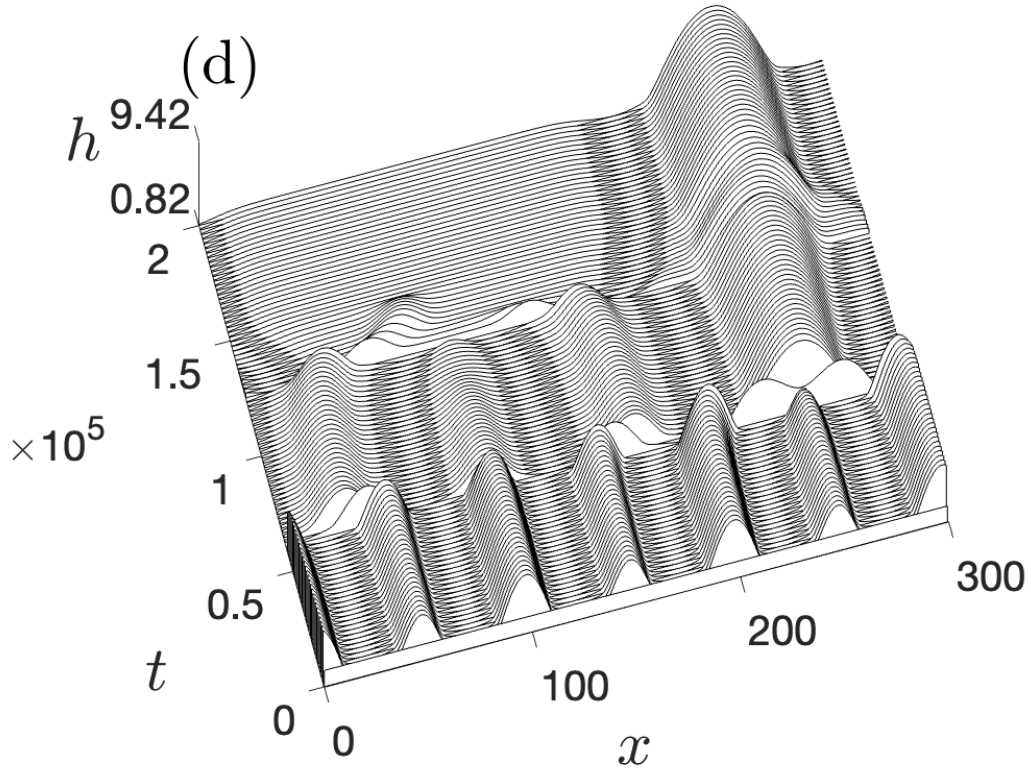}
        \includegraphics[width=0.48\linewidth]{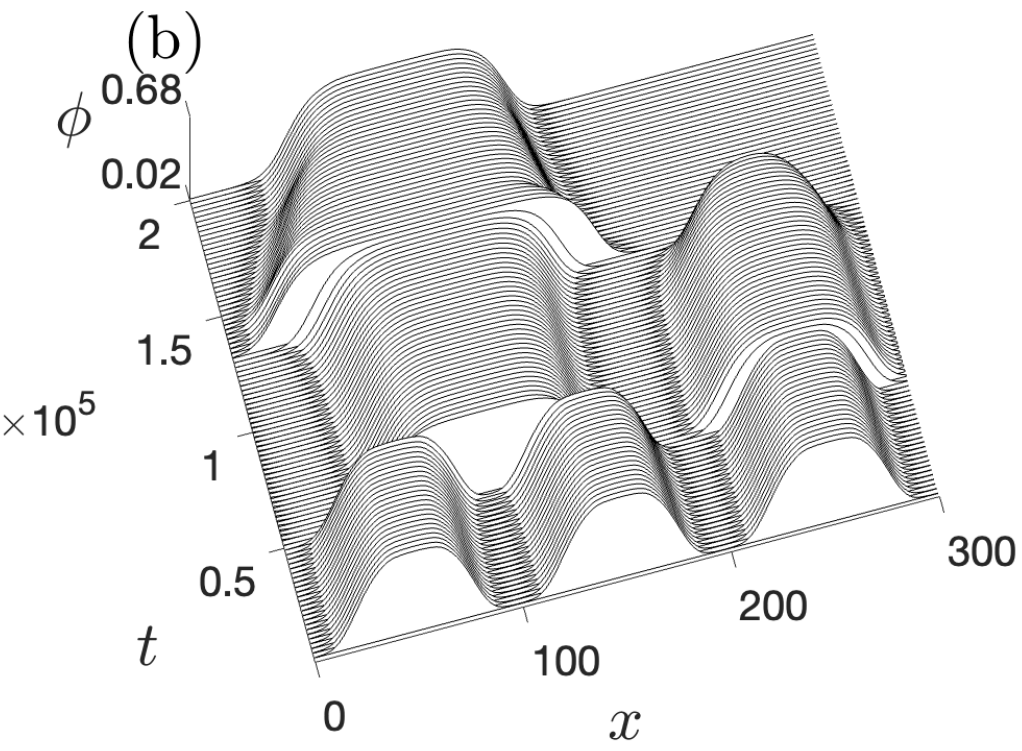}
        \includegraphics[width=0.48\linewidth]{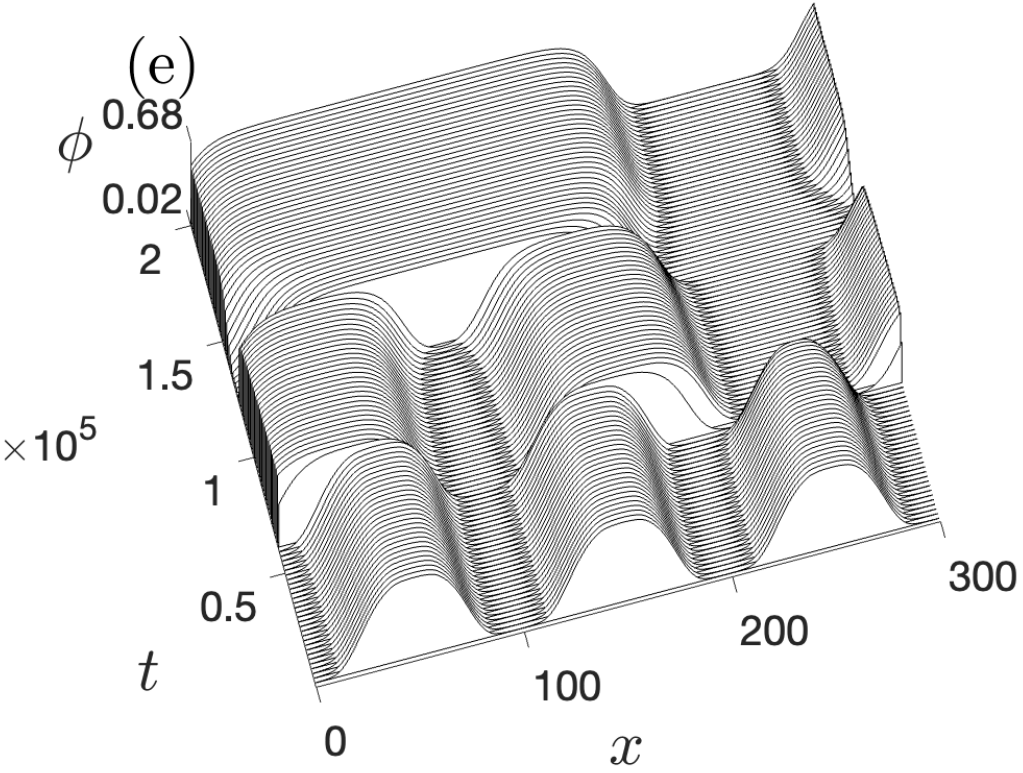}
        \includegraphics[width=0.48\linewidth]{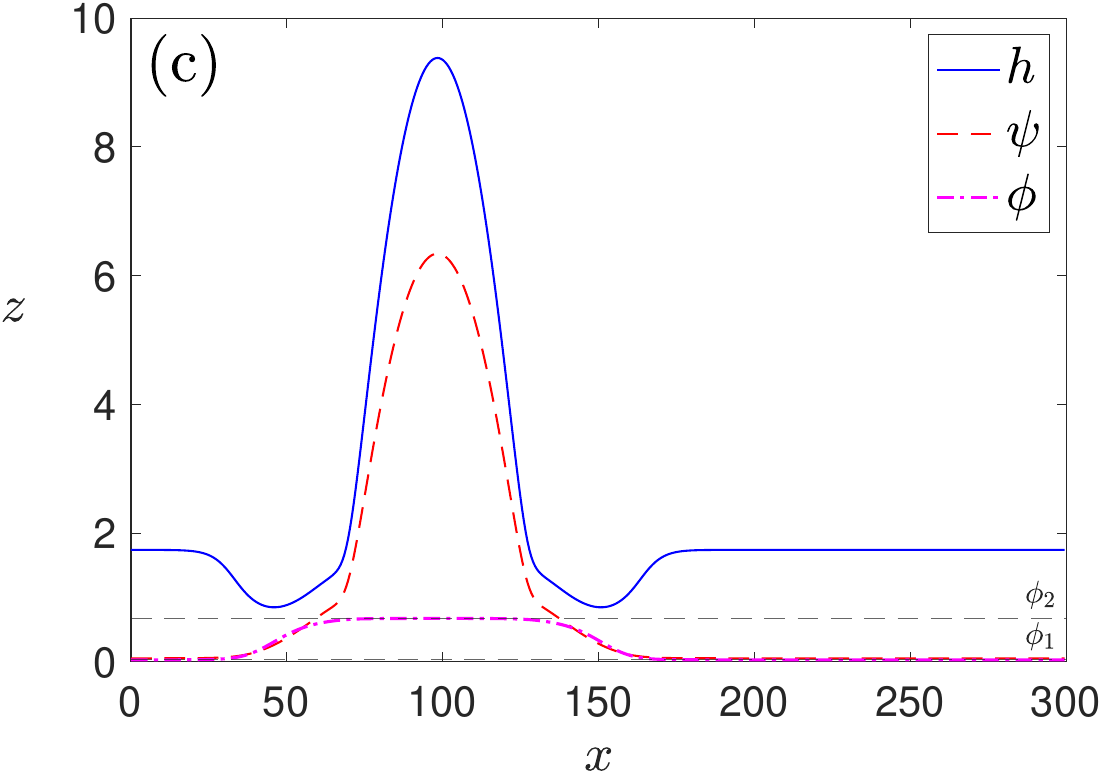}
        \includegraphics[width=0.48\linewidth]{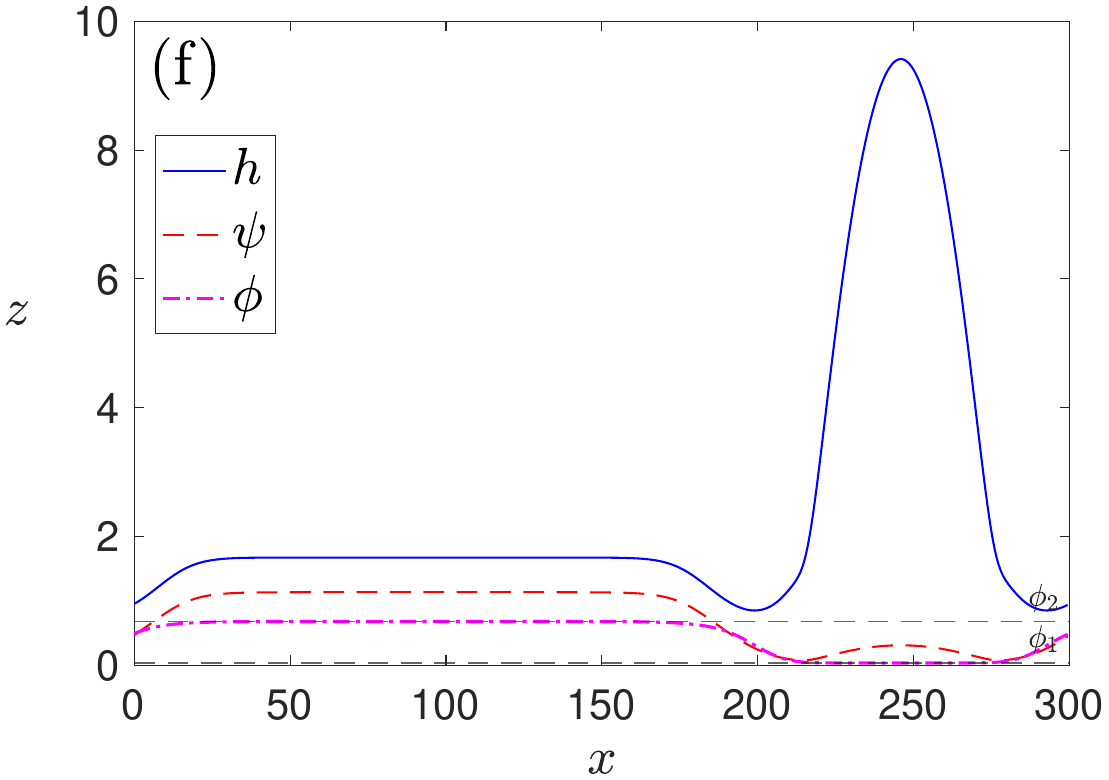}
\caption{Waterfall plots (using linear $t$) of (a) the film height, and (b) the local colloid concentration, for $A' = 1$, $K' = 11$, $\alpha' =\beta' = 100$, $\epsilon' = 4000$, \red{$a^{2} = 100$}, $h_i = 2.5$ and $\phi_i = 0.4$. Panel (c) displays the final equilibrium profiles. Panels (d)--(f) show similar results, but with $\phi_i = 0.3$, corresponding to a decrease in the total amount of colloids in the system.}
\label{case3}
\end{figure}

It is now of interest to consider cases where the coupling between the film height and the local colloid concentration is stronger. Inspecting the free energy in Eq.~\refe{freeenergyfuncphinond2}, we see that terms involving both $h$ and $\psi$ have coefficients $K'$, $\alpha'$, $\beta'$ and $\epsilon'$. To uniformly increase the coupling between the two fields without changing the colloidal bulk phase behaviour (rather than independently changing the relative strength of colloidal interactions, interfacial tension, etc.), we increase all of these parameters whilst maintaining the values of the ratios $K'/\alpha'$, $\beta'/\alpha'$ and $\epsilon'/\alpha'$ the same. The parameter values we now use are: $A' = 1$, $K' = 11$, $\alpha' = \beta' = 100$, $\epsilon' = 4000$, \red{$a^{2} = 100$} and $h_i = 2.5$. First, we consider the case $\phi_i = 0.4$. The dispersion relations (not displayed) show that the colloidal instability dominates at longer wavelength (smaller $k$) than the film height instability. From the dispersion relations in Eqs.~\eqref{trace2dimless} and \eqref{omega_TFCS} we obtain the fastest growing wavelengths to be $\lambda_h = 52$ and $\lambda_c = 115$, so in a system of length $L_x=300$ we should observe around three maxima to initially develop in the colloid concentration profile and six in the film height. This simulation is displayed in Figs.~\ref{case3}(a)--(c), where we see that the early-time dynamics agrees with this calculation.

The simulation reaches an equilibrium state, displayed in Fig.~\ref{case3}(c), where the film height and colloid concentration profiles coarsen into a single drop. The equilibration process is much faster than in the cases considered in the previous subsections as there are fewer initial wavelengths, and the approach to equilibrium of the colloids is quicker because the case we consider here in Figs.~\ref{case3}(a)-(c) has all terms in the free energy that control the colloids multiplied by a factor of $100$. The influence of the much stronger coupling can be observed in Fig.~\ref{case3}(c) where either side of of the droplet (peak in the film height), there are deep depressions in the film height that correspond in location to the interfaces between low and high density regions of the colloids. The influence of the strong coupling can also be observed during the evolution in the waterfall plot in Fig.~\ref{case3}(a), where these depressions appear as deep trenches. It is instructive to compare the equilibria of Fig.~\ref{case3}(c) with Fig.~\ref{case1_evolve}(c). In the latter, there is weak coupling, and hence the system evolves to a state where there is a droplet sitting on a precursor film at $h\approx 1.5$, corresponding to the ideal height based on the binding potential of our model, and the droplet being of a shape dictated by conservation of mass from that given by the initial condition and surface tension of the liquid. Similarly, the colloid concentration profile evolves to plateaus at high and low density phases with values given in the phase diagram in Fig.~\ref{Bulkphasediagram}, with a minimum number of interfaces. By contrast, the highly coupled situation displayed in Fig.~\ref{case3}(c) has distorted profiles. In particular, we see strong depletion regions where the liquid height is significantly below the ideal precursor film value, which occurs so as to accommodate the mass of colloidal particles in a single region within the droplet, and also the required interfaces between high and low density states. \red{However, we would caution against ascribing too much significance to details of the ordering observed in the precursor film, since the most natural interpretation of films of this thickness is as a monolayer adsorbed on the surface, where the thin-film equation is arguably not a good description \citep{yin2017films}.}

As mentioned, in Fig.~\ref{case3}(c) we observe that the colloids end up within the single final droplet on the surface, i.e.\ the film height and the colloidal concentration profiles are in-phase in the final result. This is not always the case; the outcome largely depends on the total amount of colloids within the system. In order to change the final result from being in-phase to \red{anti-phase} [an example of the latter is shown in Fig.~\ref{case3}(f)] we can control the initial colloid concentration to a different location on the phase diagram in Fig.~\ref{Bulkphasediagram}. In the case shown in Figs.~\ref{case3}(d)--(f), we decrease the initial average concentration to $\phi_i=0.3$, and achieve an \red{anti-phase} final result. We observe that the system is generally \red{anti-phase} when the total concentration of colloids is low (i.e.\ lower $\phi_i$, corresponding to locations on the left-hand side of the phase diagram within the spinodal in Fig.~\ref{Bulkphasediagram}), and in-phase for higher concentrations (i.e.\ higher $\phi_i$, corresponding to locations on the right-hand side within the spinodal). We investigate this matter further in \S~\ref{sec:bif}, with the aid of bifurcation diagrams.

\subsection{Cases including stability of either film height or colloid profiles} 
\label{sec:case4}

\begin{figure}
        \includegraphics[width=0.48\linewidth]{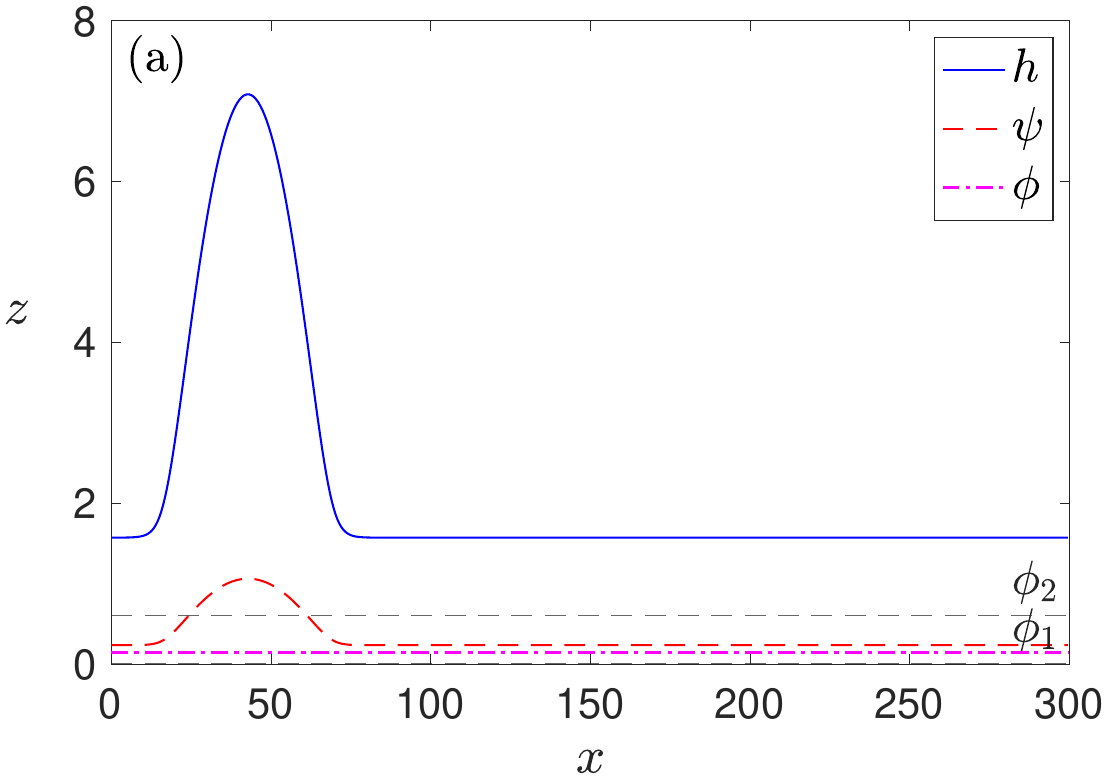}
        \includegraphics[width=0.48\linewidth]{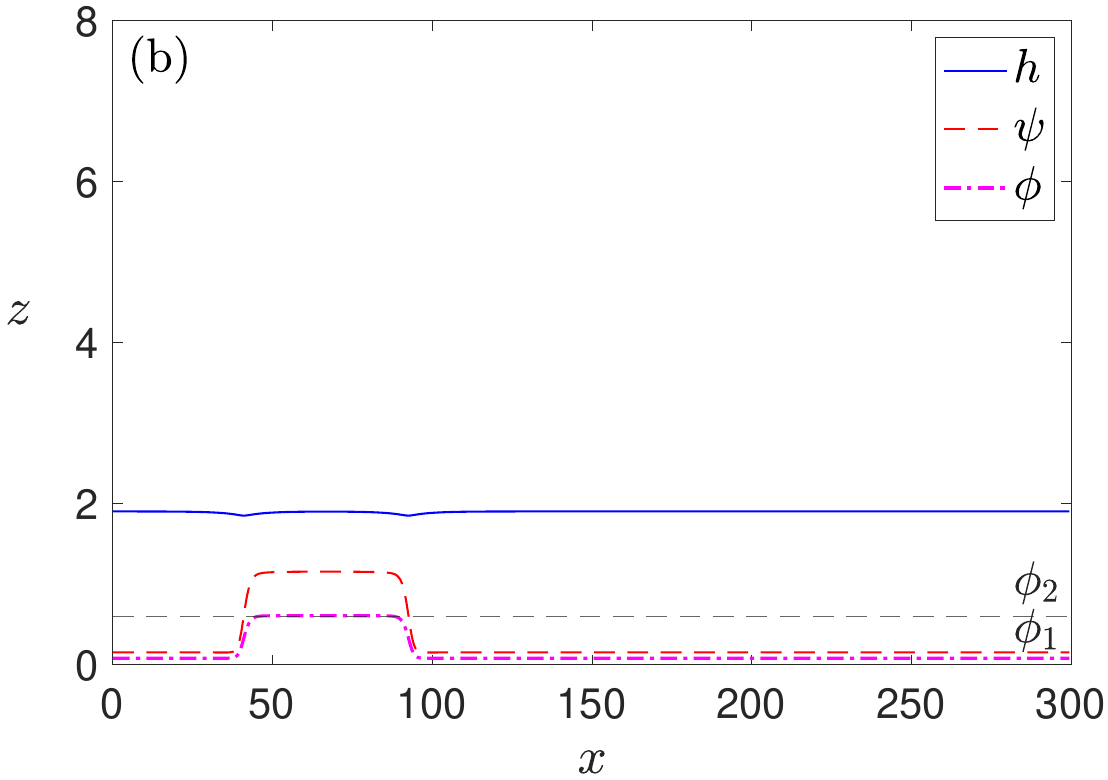}
\caption{Final equilibrium profiles corresponding to cases where (a) the colloids are stable (and the film height is unstable) and (b) the film height is stable (with unstable colloids). The parameter values are the same as those for Figs.~\ref{disp1}(b) and (c), respectively.}
\label{stable}
\end{figure}

As previously discussed, there are also situations where either a uniform film height or colloid concentration can be linearly stable. The case where both are stable is obvious, because in this situation any small perturbations in the initial profiles decay over time and the profiles both become flat. We also analysed some cases with profiles where one of the film height or colloid concentration profiles is stable, while the other unstable. Typical dispersion relations for cases like these are shown in Figs.~\ref{disp1}(b) and \ref{disp1}(c). Examples of equilibrium profiles from our dynamic simulations are shown in Fig.~\ref{stable}. 

In Fig.~\ref{stable}(a), we have stable colloids and unstable film height. The film height evolves to form a droplet, while the concentration of the colloids remains roughly the same as the initial concentration $\phi_i$. As a result, the colloids follow the overall film height, as can be seen in the field $\psi$, so that the colloid profile ends up in-phase with the film height profile. When the film height is stable, the system does not form a droplet. However, if the colloids demix, the film height is still affected by the unstable colloids. For example, in Fig.~\ref{stable}(b), we see that the colloids have separated into high and low density regions, and the interfaces between these distort the film height such that together the profiles minimise the overall free energy. We also notice that the computation time for this case is much longer. This is because when the film is stable, the colloids lose the driving from the film height evolution, so the system reaches equilibrium in a much longer time. \red{To put this another way: when the film height remains roughly constant, the only process governing the time evolution of the colloids is diffusive aggregation and this is a much slower process than the liquid dewetting dynamics.}

\subsection{Dynamics leading to asymmetrical final profiles}
\label{sec:asymm}

\begin{figure}
        \includegraphics[width=0.47\linewidth]{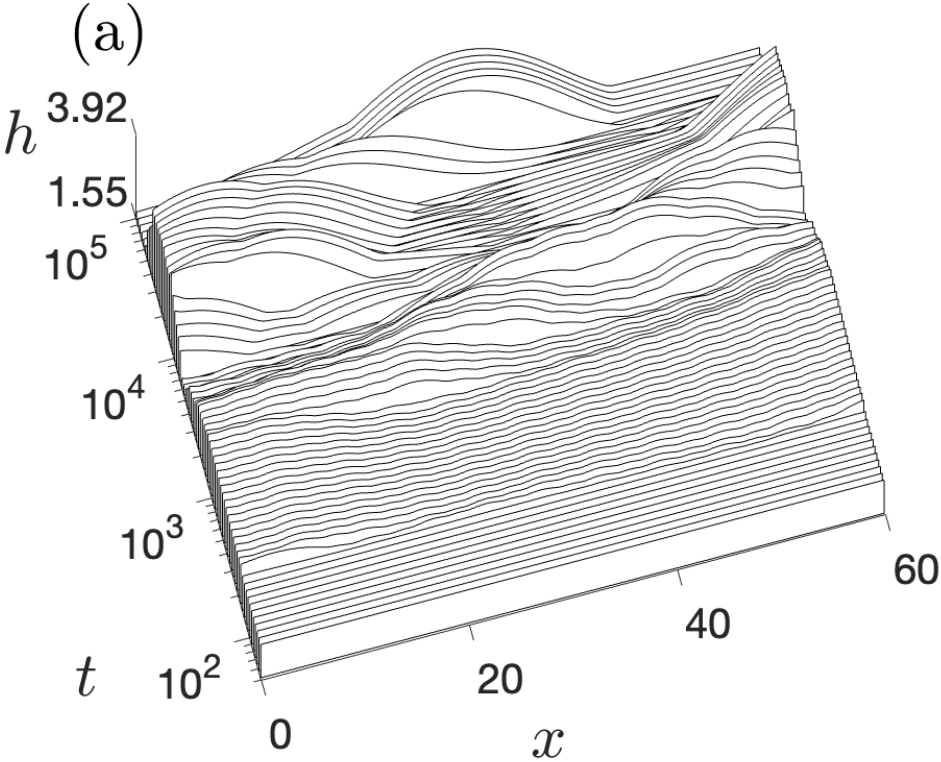}
        \includegraphics[width=0.47\linewidth]{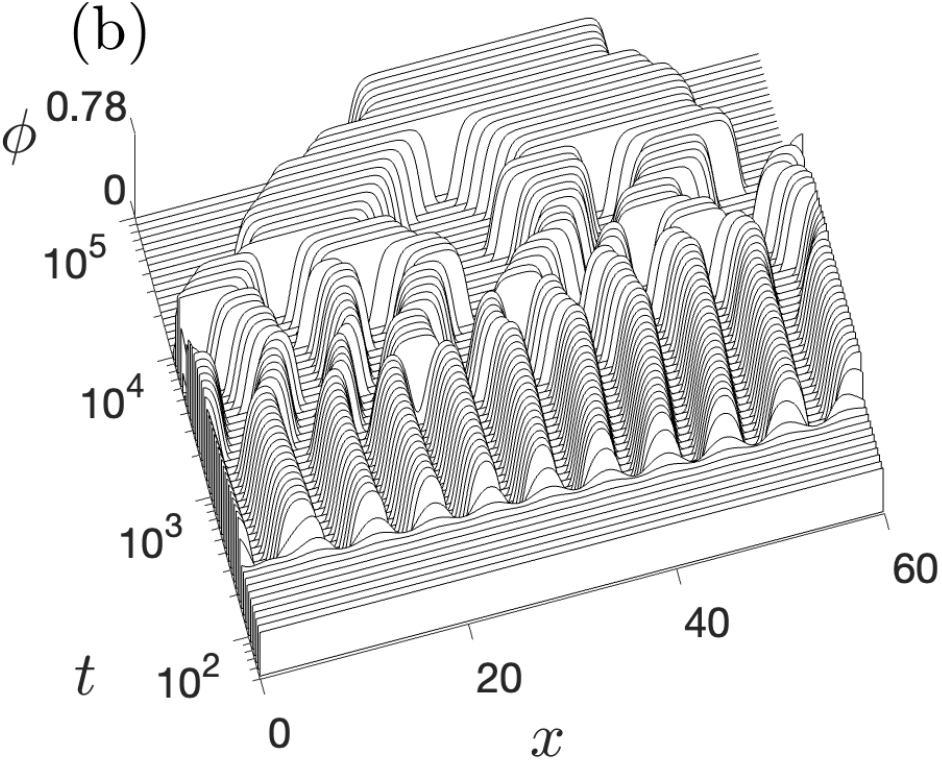}
        
        \includegraphics[width=0.47\linewidth]{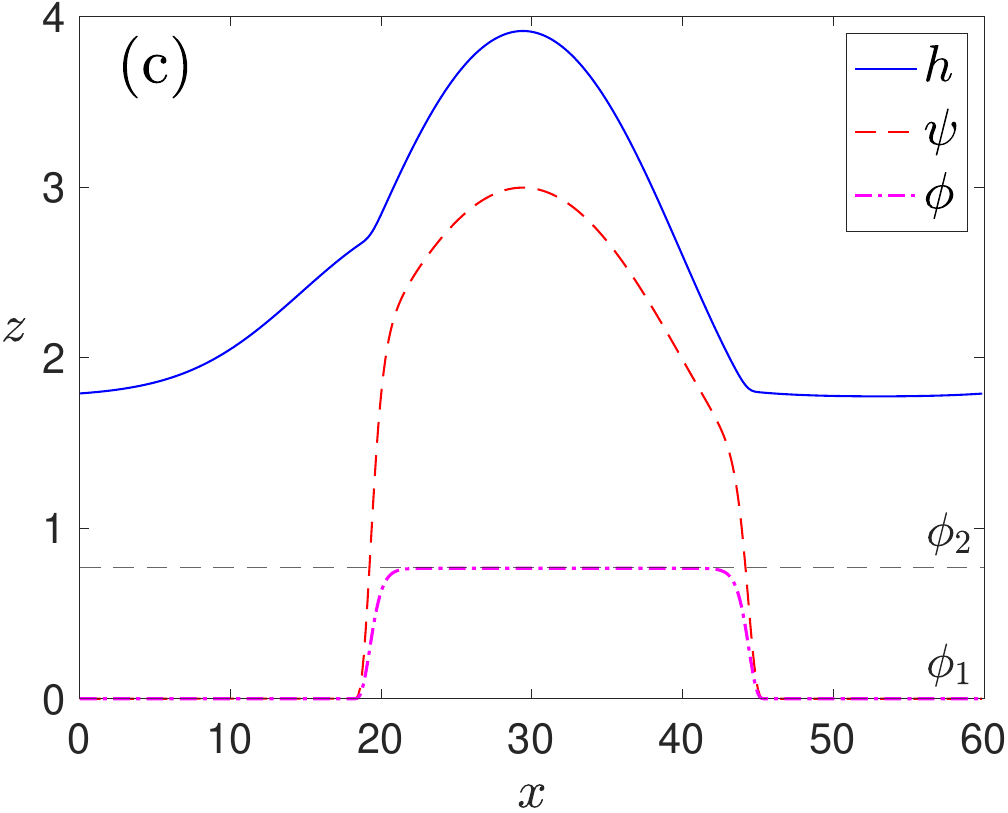}
        \includegraphics[width=0.49\linewidth]{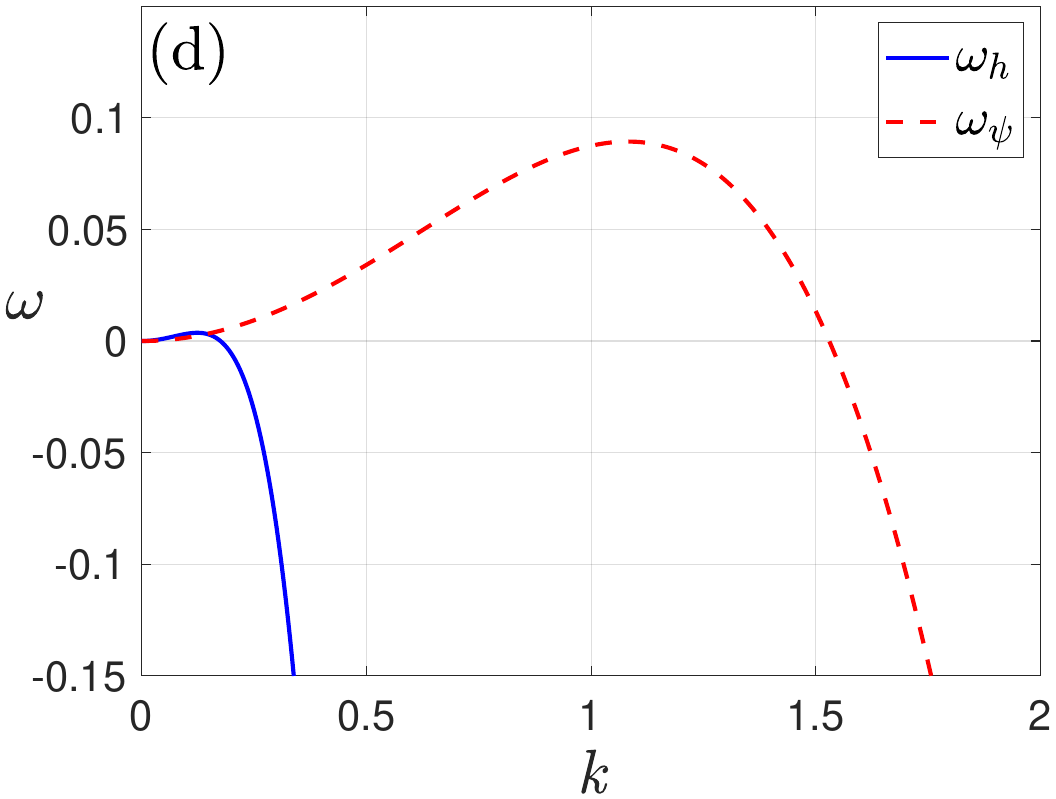}
\caption{Panels (a) and (b) show waterfall plots over time (using $\log t$) of the film height and the local colloid concentration, respectively, for $A' = 1$, $K' = 0.13$, $\alpha' =\beta' = 3$, $\epsilon' = 0.5$, \red{$a^{2} = 2$}, $h_i = 2.5$ and $\phi_i = 0.41$. Panel (c) shows the final equilibrium profiles and (d) shows the dispersion relation for this system.}
\label{case_asymm}
\end{figure}

In our explorations of the model, we have also observed cases where the final profiles are asymmetrical. A typical example is displayed in Fig.~\ref{case_asymm}, which occurs when $A' = 1$, $K' = 0.13$, $\alpha' =\beta' = 3$, $\epsilon' = 0.5$, \red{$a^{2} = 2$}, $h_i = 2.5$ and $\phi_i = 0.41$. This corresponds to the dimensionless temperature $K'/\alpha'=0.043$, which is lower than in the cases considered previously. Here, the coexisting colloid concentrations are $\phi_1 = 0.00055$ and $\phi_2 = 0.77$ (c.f.\ Fig.~\ref{Bulkphasediagram}), so there is a greater difference between these than in the cases considered previously. The time evolution for a system of size $L_x = 60$ is displayed in waterfall plots in Figs.~\ref{case_asymm}(a)--(b), with the corresponding final profiles in panel (c) and dispersion relations in (d). From the dispersion relations we see that the colloids mode is by far the fastest growing in the early stages. Thus, in the dynamics we see this instability dominating the overall dynamics and making the system exit the linear regime relatively quickly. We also see the colloid concentration variations coupling strongly to the film height $h$. Compared to the examples discussed previously, where we observe symmetric final profiles with the colloids distributed evenly outside the droplet [see e.g.\ in Fig.~\ref{case1_evolve}(c)], in the present case the coupling between $h$ and $\phi$ is much stronger. Therefore, the colloidal concentration variations are mirrored in the film height variations over the surface. In Figs.~\ref{case_asymm}(a)--(b) the initial number of peaks that form in $h$ and $\phi$ are accurately predicted by the dispersion relation. Since the system evolves to minimise the free energy, plots of the free energy always decreases over time (not displayed), with each decrease in the total number of colloid agglomerates (bumps) visible in Fig.~\ref{case_asymm}(b) corresponding to a step-like drop in the total free energy. For some asymmetric cases we observe that due to small numerical round-off errors the final droplet can very slowly slide at constant velocity over the smooth surface. Such translations do not change the free energy. The sliding speed depends on choice of grid spacing and the direction can change with a different set of initial randomness. Of course, this cannot be a genuine feature, as confirmed with the aid of bifurcation diagrams in \S~\ref{sec:bif_asymm}.

\section{System equilibrium and bifurcation diagrams}
\label{sec:bif}

\subsection{Bifurcation diagram for a simple case with small coupling}

In the previous sections, we showed examples of the dynamics of our system for a range of different parameters, and we noticed that the final film height and colloid concentration profiles equilibrate to being either in-phase (peaks, or high density regions occur at similar positions) or \red{anti-phase} (where a peak in the film height corresponds to a lower density of colloids). We now investigate systematically the dependence of the equilibrium profiles on the system length $L_x$. We do this by producing bifurcation diagrams for some of the cases previously considered. \red{The bifurcation diagrams are generated using our in-house numerical codes developed in \textsc{Matlab} employing a spectral method, see, for example, \cite{LIN2018291, Blyth_Tseluiko_Lin_Kalliadasis_2018, Blyth_Lin_Tseluiko_2023}. Eqs.~\eqref{eq:21} and \eqref{eq:solsusp-coup-grad} are rewritten as a dynamical system for the Fourier coefficients of $h$ and $\psi$, which is truncated at a sufficiently high wavenumber to ensure accuracy. Steady-state solution profiles correspond to the fixed points of this dynamical system, resulting in a system of nonlinear equations for the nonzero-wavenumber Fourier coefficients of $h$ and $\psi$. Two additional equations are obtained from requiring fixed average values of $h$ and $\psi$. Additionally, we pin the solutions such that $h$ has a local extremum in the middle of the domain, i.e.\ we impose the condition $h_x|_{x=L_x/2}=0$. This necessitates the introduction of an additional unknown parameter, to maintain consistency between the number of imposed equations and unknowns. To achieve this, we introduce a fictitious wave speed, $c$, resulting in the addition of terms $ch_x$ and $c\psi_x$ to the right-hand sides of Eqs.~(\ref{eq:21}) and (\ref{eq:solsusp-coup-grad}), respectively. This speed is trivially found to be zero in all continuation computations. Various solution branches of the resulting system of nonlinear equations for different parameter values are obtained by initially starting from small-amplitude nearly sinusoidal solutions on a domain size close to a cutoff wavelength obtained from the linear stability analysis. Numerical pseudo-arclength continuation is then performed, initially with respect to the domain-size parameter, and subsequently with respect to other relevant parameters, see, for example, \cite{Tseluiko_2013,Lin_etal_2016,Engelnkemper2019,Tseluiko_2020} for more details on numerical continuation techniques for liquid-film problems.}

\begin{figure}
    \includegraphics[width=0.48\linewidth]{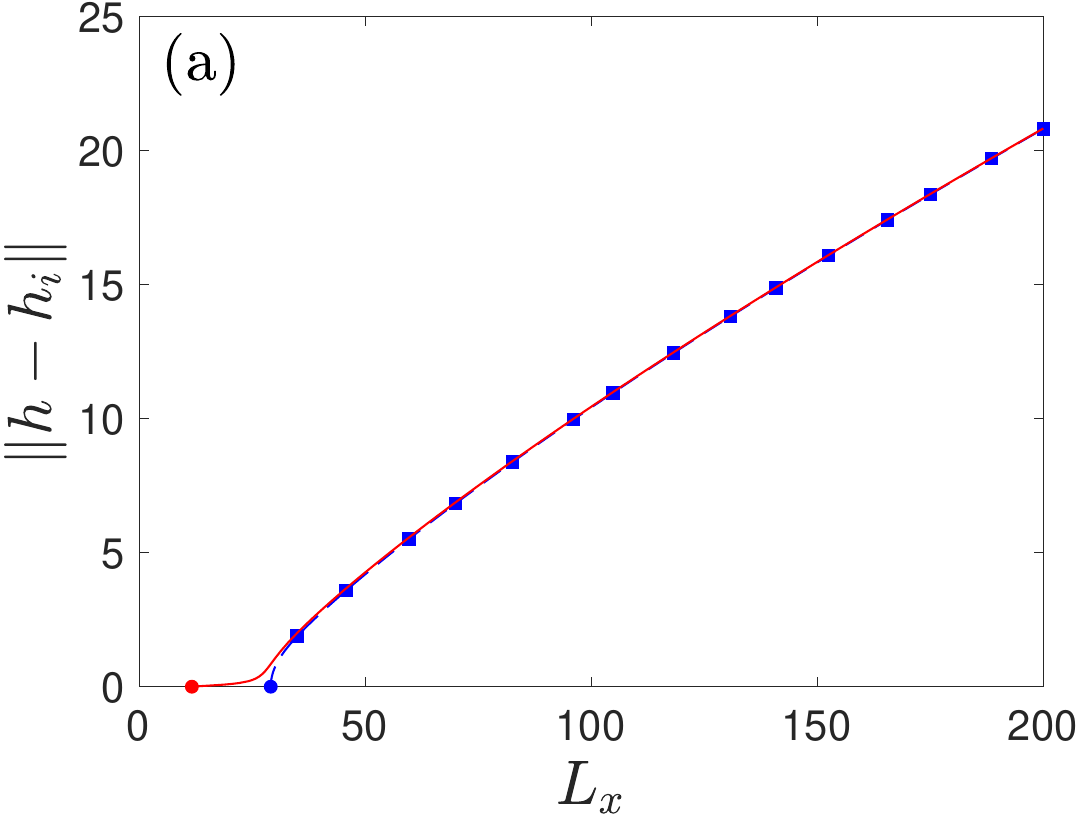}
    \includegraphics[width=0.48\linewidth]{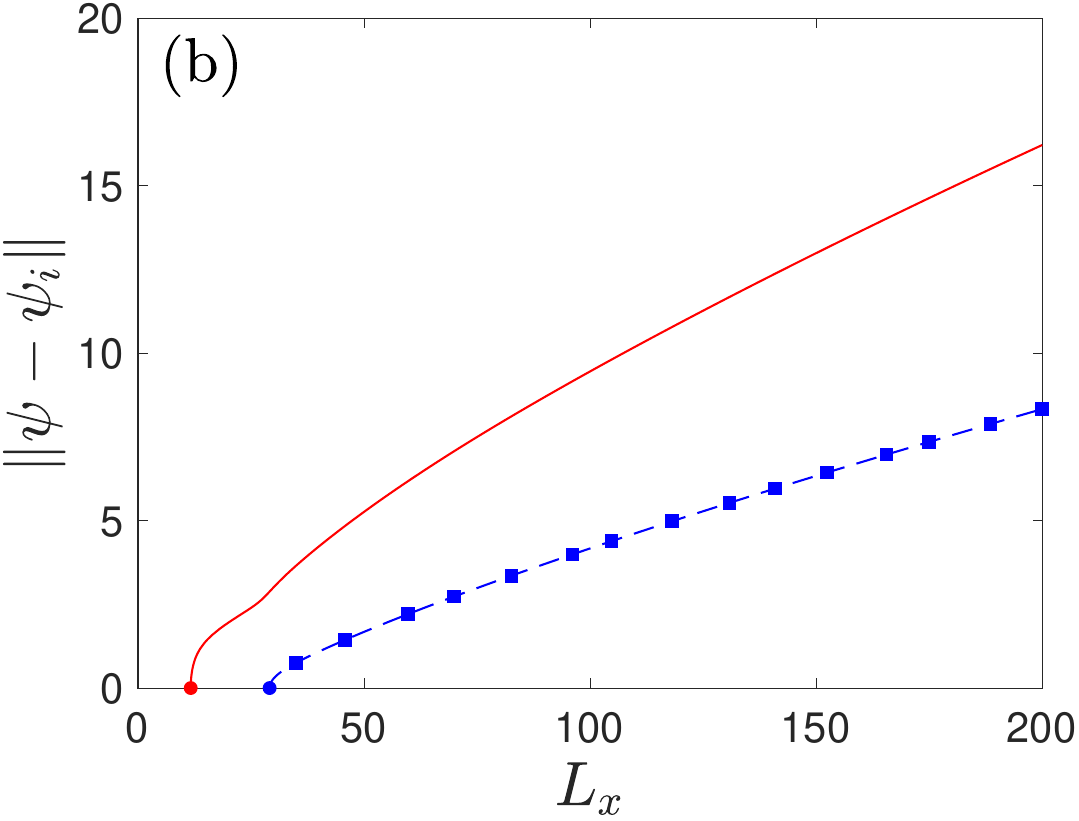}
    \includegraphics[width=0.48\linewidth]{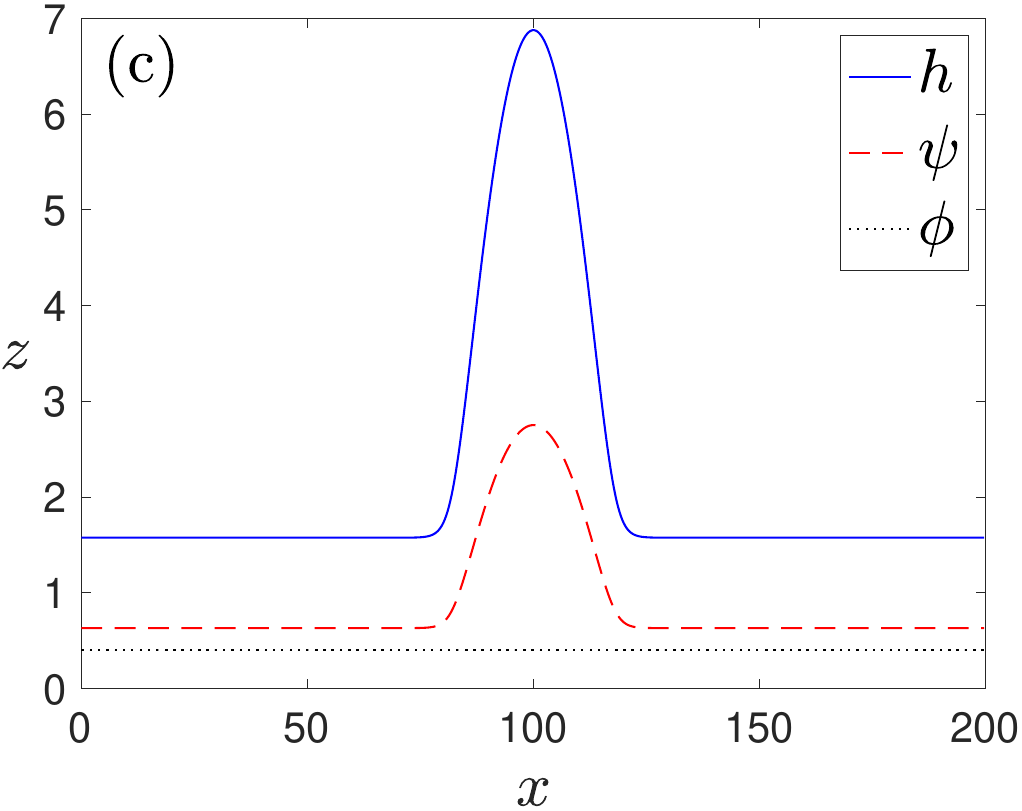}
            \hspace{0.45cm}
    \includegraphics[width=0.48\linewidth]{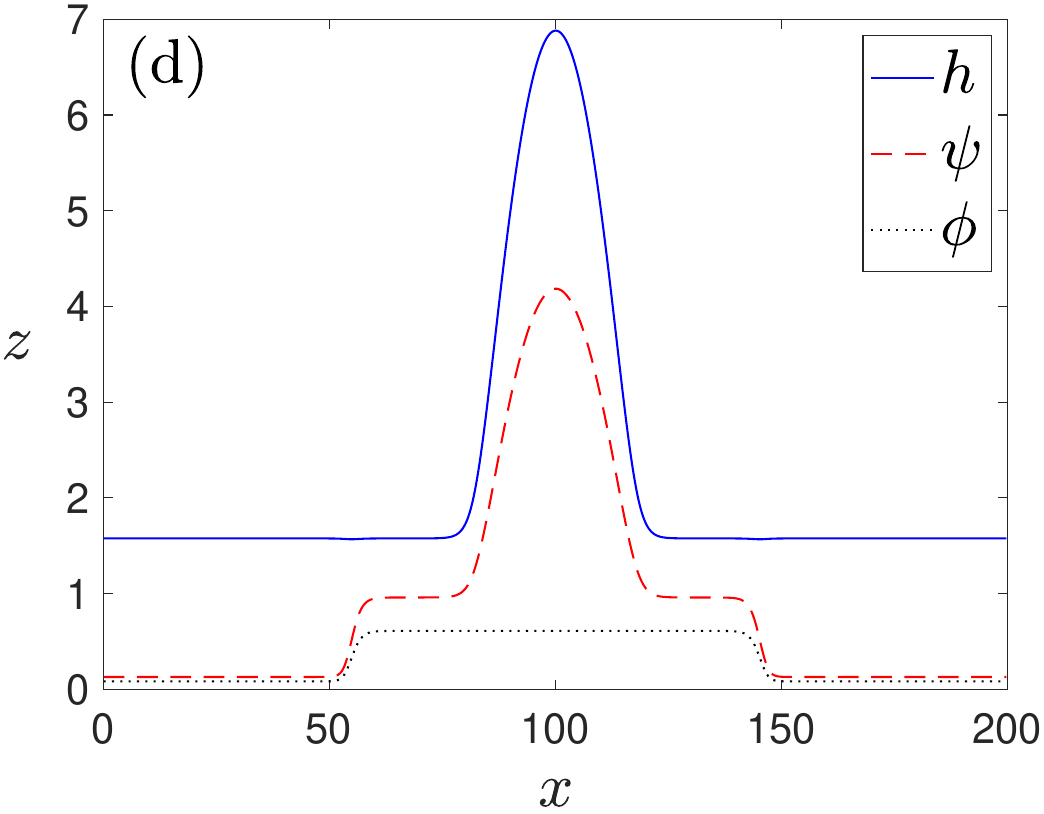}
\caption{Bifurcation diagrams and final states for cases shown in \S~\ref{sec:case1}, where $A' = 2$, $K' = 0.15$, $\alpha'=1$, $\epsilon' = 0.5$, \red{$a^{2} = 2$}, $h_i = 2.2$, and $\phi_i = 0.4$. Panel (a) shows the film height $L^2$-norm corresponding to two main branches of solutions for varying system size $L_x$. These originate from instabilities in either the film height (blue) or in the colloid local concentration (red), shown with circles. Squares represent locations of bifurcation points. Stable and unstable solutions are shown with solid and dashed lines, respectively. Similarly, (b) shows the $L^2$-norm of the corresponding $\psi$ profiles. Panels (c) and (d) show  equilibrium profiles from continuation at the final $L_x=200$ point on diagrams (a) and (b): (c) shows the film height branch [blue lines in (a) and (b)], and (d) the colloid instability branch [red dashed lines in (a) and (b)].}
\label{bif:case1}
\end{figure}

Our bifurcation diagrams depict the equilibria of our system when varying $L_x$ and keeping all other parameters fixed. Each realisation of the equilibrium profiles is represented by a point in each diagram (one for $h$ and one for $\psi$) corresponding to the $L^2$-norm of the profile, subtracting the profile average value $h_i$ or $\psi_i$, i.e.\ 
\begin{equation}
\lvert \lvert h - h_i \rvert \rvert = \sqrt{\int_{0}^{L_x}\lvert h - h_i \rvert ^2 \, \textrm{d} x},
\end{equation}
and
\begin{equation}
\lvert \lvert \psi - \psi_i \rvert \rvert = \sqrt{\int_{0}^{L_x}\lvert \psi - \psi_i \rvert ^2 \, \textrm{d} x}.
\end{equation}
In all the bifurcation diagrams shown here, blue lines represent continuation results starting from the known initial linear instability in the film height and red lines are similarly from the colloid instability (see \S~\ref{sec:LSA}), termed here the film-height mode and the colloid mode. However, due to the coupling between the film height and the local colloid concentration, an instability in either affects both profiles, and hence both red and blue lines appear on the bifurcation diagrams for the norm of the equilibrium profile for both $h$ and $\psi$. Sections of the branches corresponding to stable and unstable solutions are shown with solid and dashed lines, respectively. Bifurcation points to side branches are illustrated using squares. 

The bifurcation diagrams displayed do not show all possible branches, and hence not all possible equilibria of our system, but those included are instructive for understanding the dynamics of our system. Similarly, not all bifurcation points where the stability of solution branches changes are displayed, but some are illustrated with squares. In later cases in this section, some significant points where branches collide are discussed.

We begin by presenting the bifurcation diagrams in Figs.~\ref{bif:case1}(a) and \ref{bif:case1}(b), which correspond to the dynamical simulations discussed in \S~\ref{sec:case1}. The parameter values are $A' = 2$, $K' = 0.15$, $\alpha'=1$, $\epsilon' = 0.5$, \red{$a^{2} = 2$}, $h_i = 2.2$, and $\phi_i = 0.4$. Solution branches corresponding to inhomogeneous profiles begin on the $x$-axis at the points corresponding to where the system is first big enough to fit one unstable wavelength. We can read from Fig.~\ref{case1}(a) the largest wave numbers when the system is unstable (the roots of $\omega_h(k)=0$ and $\omega_\psi(k)=0$) as $k_{h0}=0.215$ and $k_{\psi0}=0.535$. Note that $k_{h0}\equiv\sqrt{2}k_h$ and $k_{\psi0}\equiv\sqrt{2}k_\psi$, which can be obtained from Eqs.~\eqref{eq:k_h} and \eqref{eq:k_psi}, respectively. The corresponding wavelengths are $\lambda_{h0}=2\pi/k_{h0}=29.2$ and $\lambda_{\psi0}=2\pi/k_{\psi0}=11.7$, and are accurately captured in the bifurcation diagrams, Figs.~\ref{bif:case1}(a) and \ref{bif:case1}(b), corresponding to the circular blue and red points on the $x$-axis, where the norms equal zero.

As $L_x$ is increased, the equilibrium profiles change. For $h$ in Fig.~\ref{bif:case1}(a), we see that profiles on the two different branches originating from either the film-height or colloid instability are very similar, since both norm lines lie very close to each other. However, for $\psi$ in Fig.~\ref{bif:case1}(b), the norms are very different and correspond to very different profiles. To illustrate this, we plot representative profiles for the points at $L_x=200$ in Figs.~\ref{bif:case1}(c) and \ref{bif:case1}(d). In Fig.~\ref{bif:case1}(c) the $\psi$ profile is largely flat with a small central bump, whereas in Fig.~\ref{bif:case1}(d) the $\psi$ profile has a larger central bump and shoulders extending into the flat-film region, being significantly different from the flat solution, and thus has a much larger $L^2$-norm. This substantial difference is seen in Fig.~\ref{bif:case1}(b), where the $L^2$-norm at $L_x=200$ on one branch is roughly twice the value on the other. Additionally, we see in the representative profiles in Figs.~\ref{bif:case1}(c) and \ref{bif:case1}(d) that the maxima in both $h$ and $\psi$ occur together, which is the case all along these solution branches. Hence, for this set of parameter values, we conclude that profiles of any system length $L_x$ have the colloids located in-phase with the centre of the droplet. \red{We note that the solutions of the colloid mode branch are linearly stable, whereas the solutions of the thin-film mode branch are all linearly unstable. In addition, there are a number of bifurcation points on the thin-film branch. Although we have not checked the nature of all these bifurcation points, our preliminary analysis shows that most likely they correspond to transcritical bifurcations. The side branches passing through these points are not shown here, as the solutions of these branches are linearly unstable.} We note also that the equilibrium profiles in Fig.~\ref{bif:case1}(d) agree well (up to horizontal translation) with the final profiles after the dynamic evolution of our hydrodynamic model in Fig.~\ref{case1_evolve}, giving confidence to both results.

\subsection{Bifurcation analysis for a case with strong coupling}

\begin{figure}
      \includegraphics[width=0.48\linewidth]{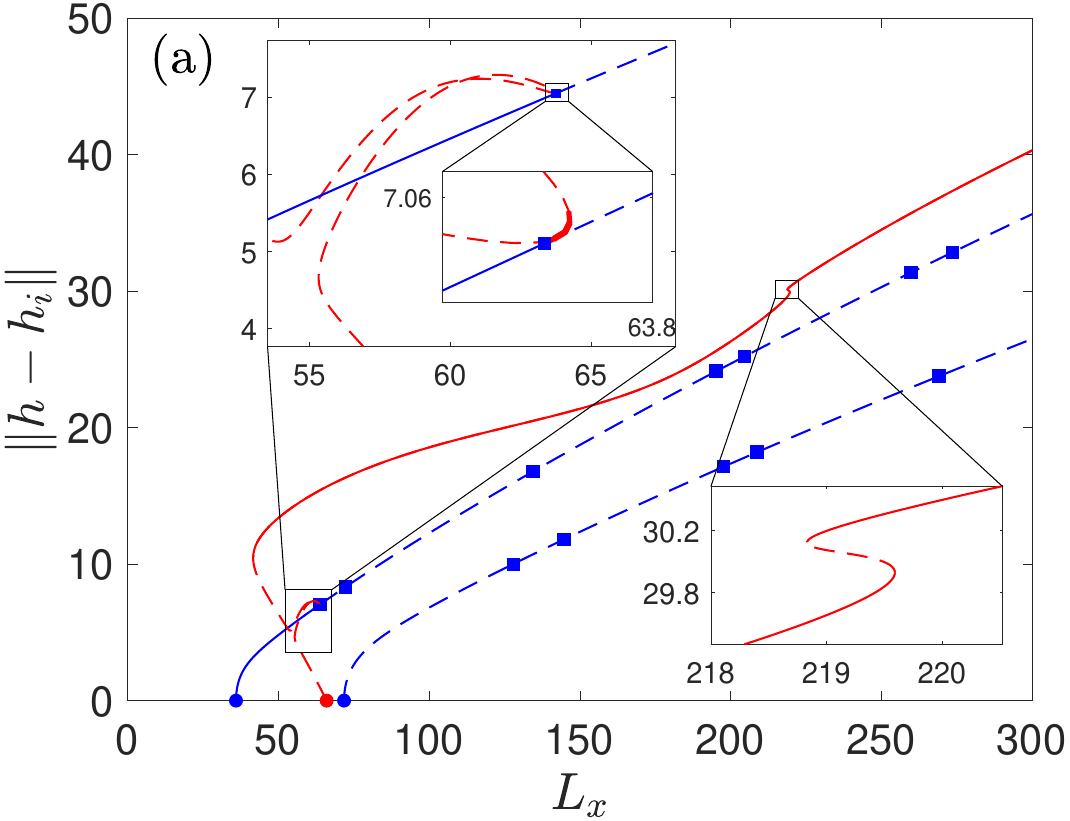}
      \includegraphics[width=0.48\linewidth]{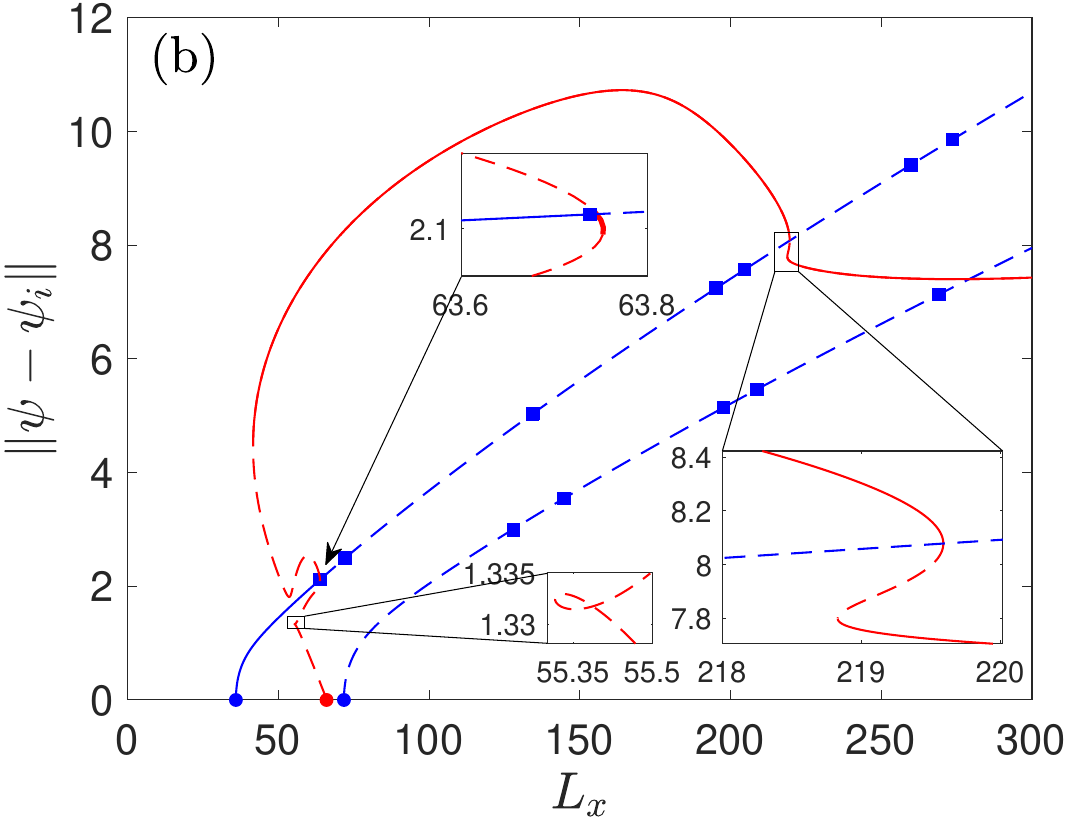}
      \includegraphics[width=0.48\linewidth]{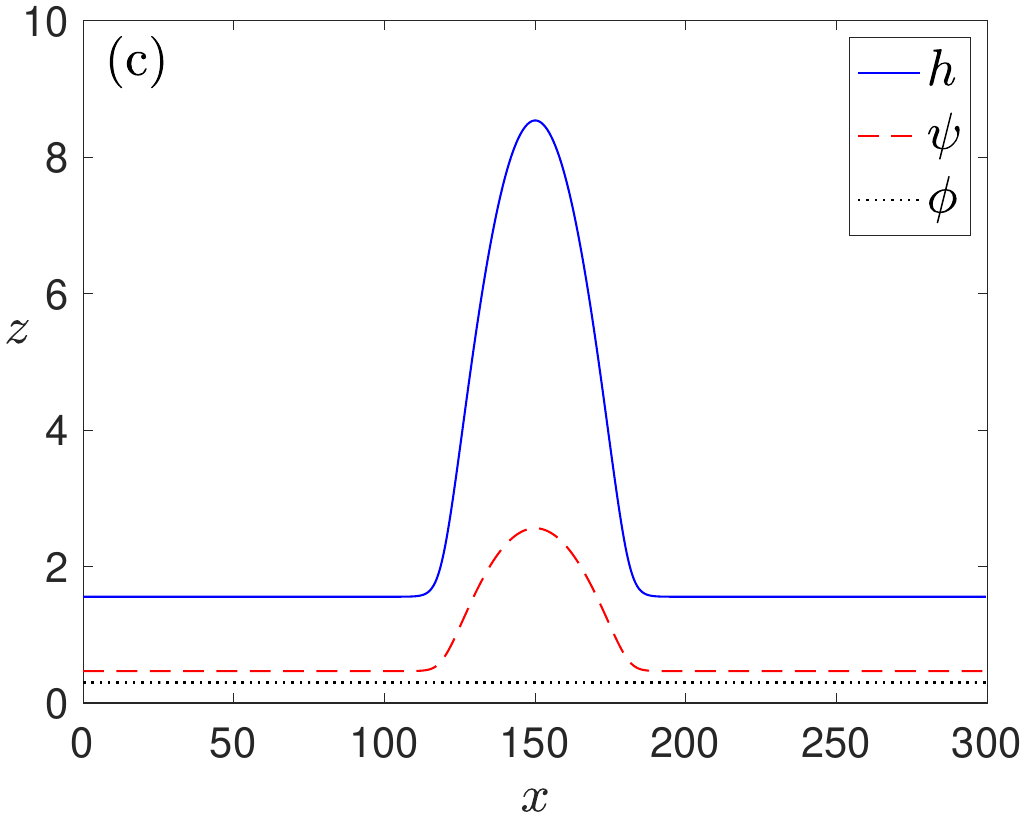}
      \includegraphics[width=0.48\linewidth]{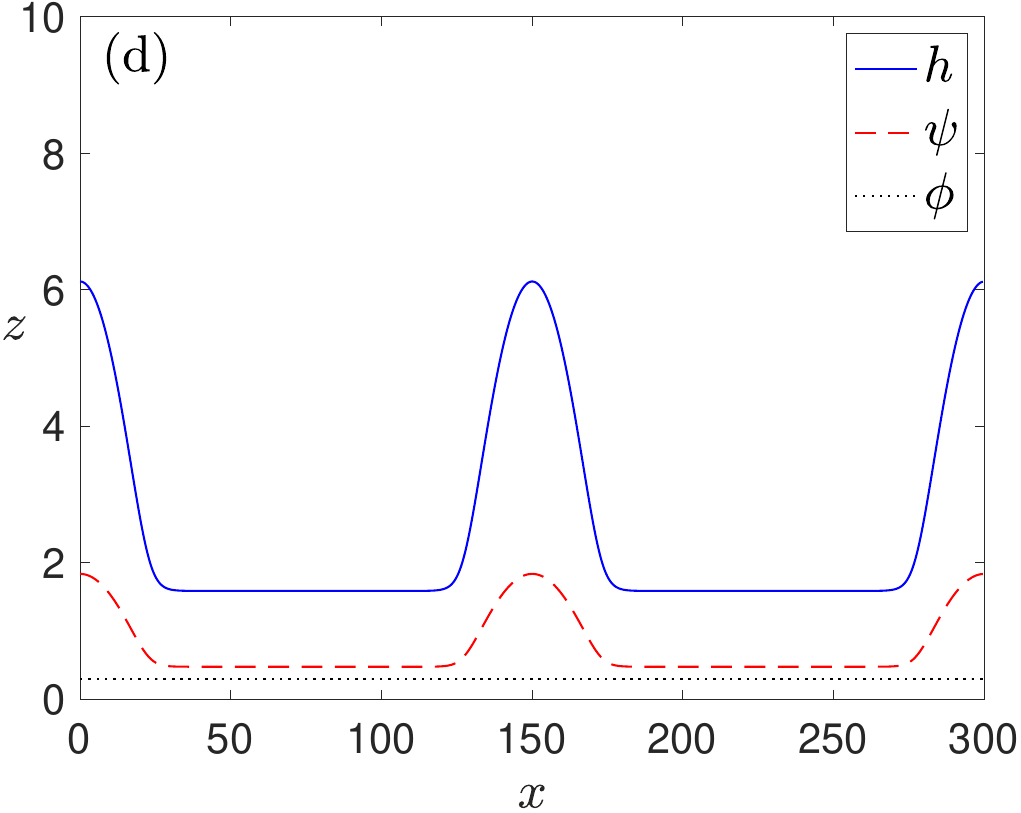}
\centerline{\includegraphics[width=0.48\linewidth]{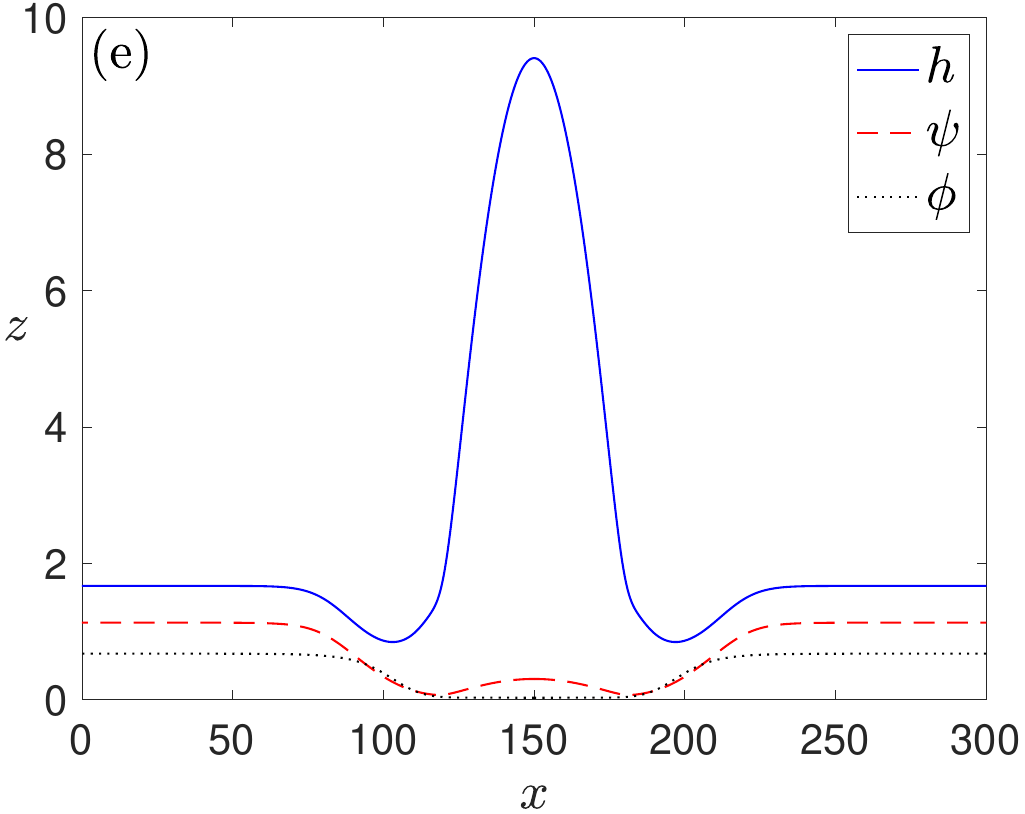}}
\caption{(a)--(b) Bifurcation diagram for parameters as in Figs.~\ref{case3}(d)--(f), where now a second blue line branch corresponding to a double-droplet profile is depicted. \red{The solid lines correspond to stable solutions, whereas the dashed lines correspond to unstable solutions.} (c)--(e) Equilibrium profiles from the film height mode (c) first film height branch and (d) second film height branch; (e) equilibrium profiles from the colloid film mode.}
\label{bif:out}
\end{figure}

In \S~\ref{sec:nandr} we noted that the system can evolve to situations where the colloid local concentration profile can end up being either in- or \red{anti-phase} with the film height. We now investigate how this is manifested in the bifurcation diagrams showing the various possible equilibrium solutions, and how the transition occurs. First, we display in Figs.~\ref{bif:out}(a) and \ref{bif:out}(b) bifurcation diagrams for the (stronger coupling) set of parameters used in Figs.~\ref{case3}(d)--(f), where the final profiles correspond to the film height and the colloid concentration profiles being \red{anti-phase}.

In Figs.~\ref{bif:out}(a) and \ref{bif:out}(b) the two blue lines are branches of solutions corresponding to instabilities in the film height. The first, as in Fig.~\ref{bif:case1}, corresponds to one wavelength (single droplet) in the system and the second, starting at double the wavelength of the first branch, corresponds to profiles of two wavelengths (two droplets). As a result, the starting point of the second film height branch is at twice the value (system length $L_x$) of the starting position of the \red{leftmost} branch. An example of a single wavelength solution on the first branch is shown in Fig.~\ref{bif:out}(c), and an example of a two wavelength solution is shown in Fig.~\ref{bif:out}(d) (recall that the domain is periodic in $x$). In the bifurcation diagram, we also display (red line) the continuation along the colloid mode branch, that \red{crosses} the first film-height branch at the first bifurcation point, \red{which turns out to correspond to a transcritical bifurcation (as can be more clearly seen in the inset, discussed in more detail later). 
} The starting point of this branch is located to the left, at smaller $L_x$, of the second film mode branch and they do not intersect. This corresponds to an \red{anti-phase} profile, with a representative example shown in Fig.~\ref{bif:out}(e). This profile is the same as the one shown in our dynamical simulation in Fig.~\ref{case3}(f) with the droplet at a different location since in dynamical simulation we used a randomly perturbed initial condition.

\begin{figure}
      \includegraphics[width=0.48\linewidth]{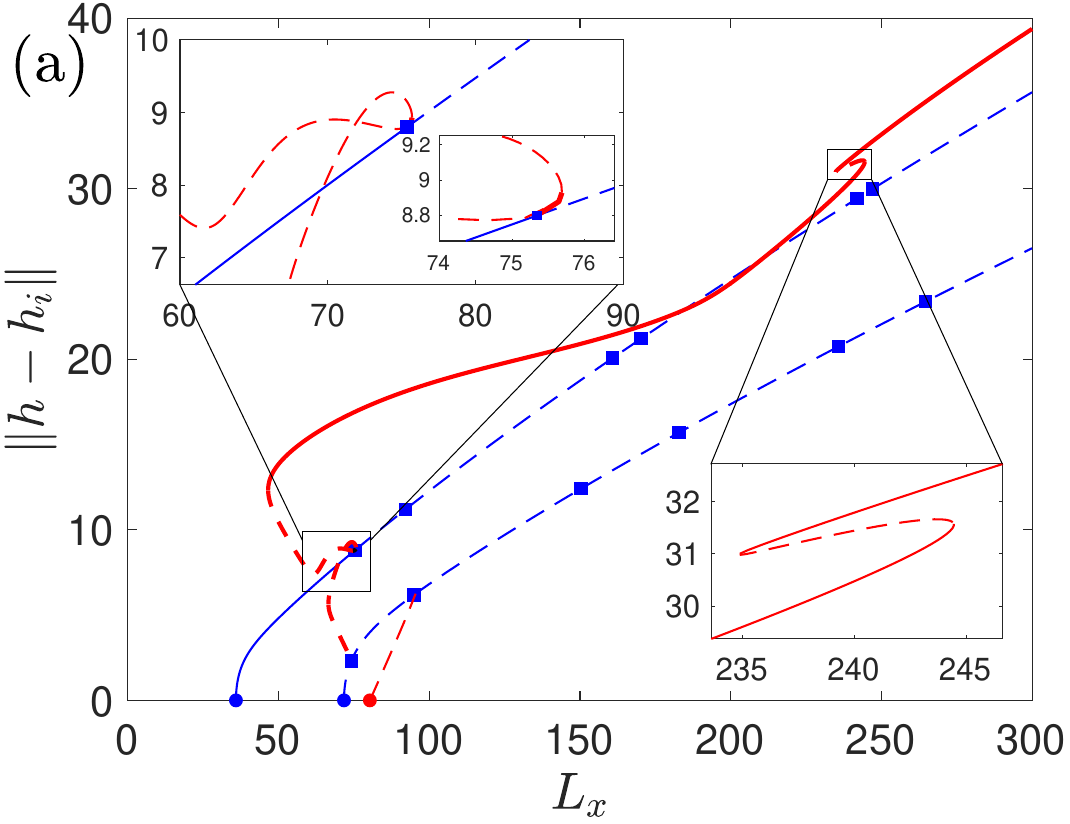}
      \includegraphics[width=0.48\linewidth]{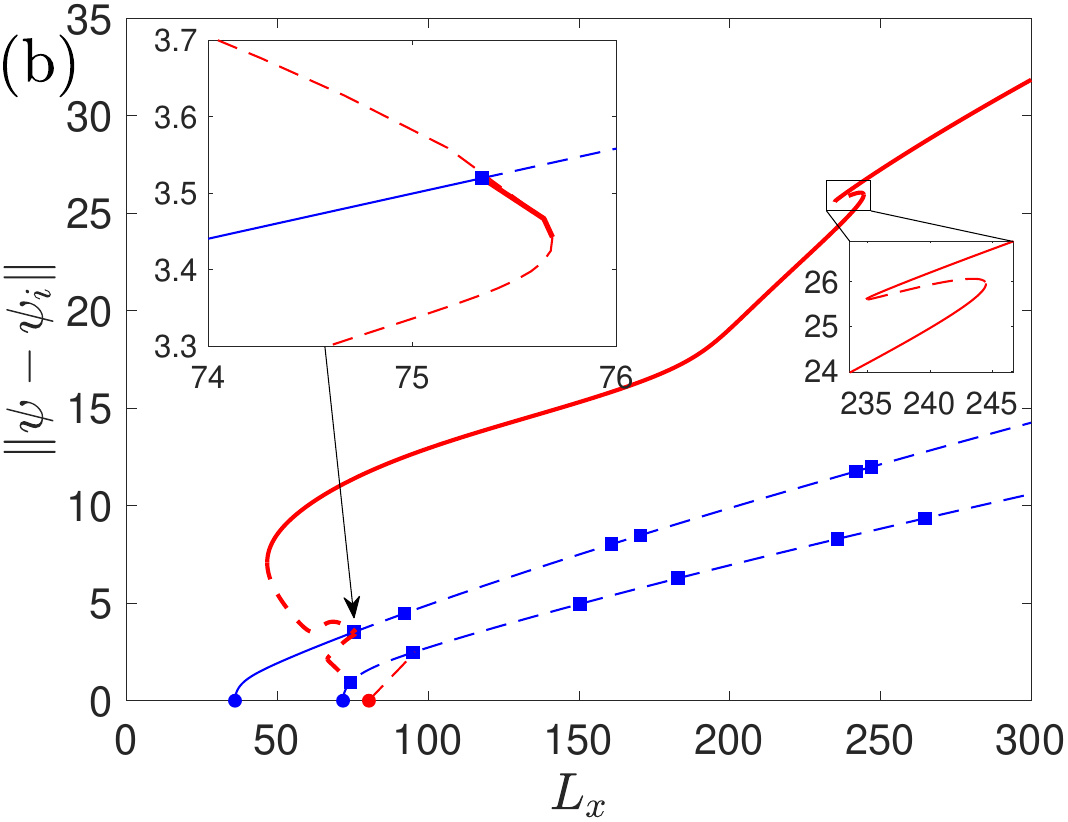}
      \includegraphics[width=0.48\linewidth]{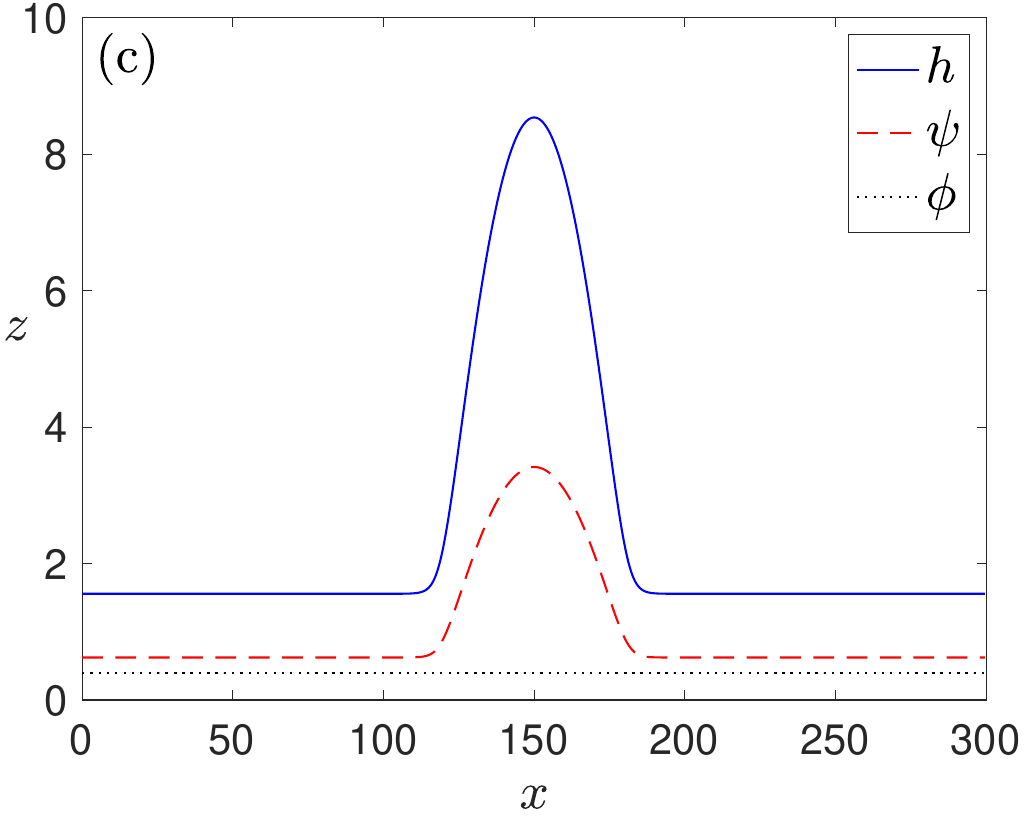}
      \includegraphics[width=0.48\linewidth]{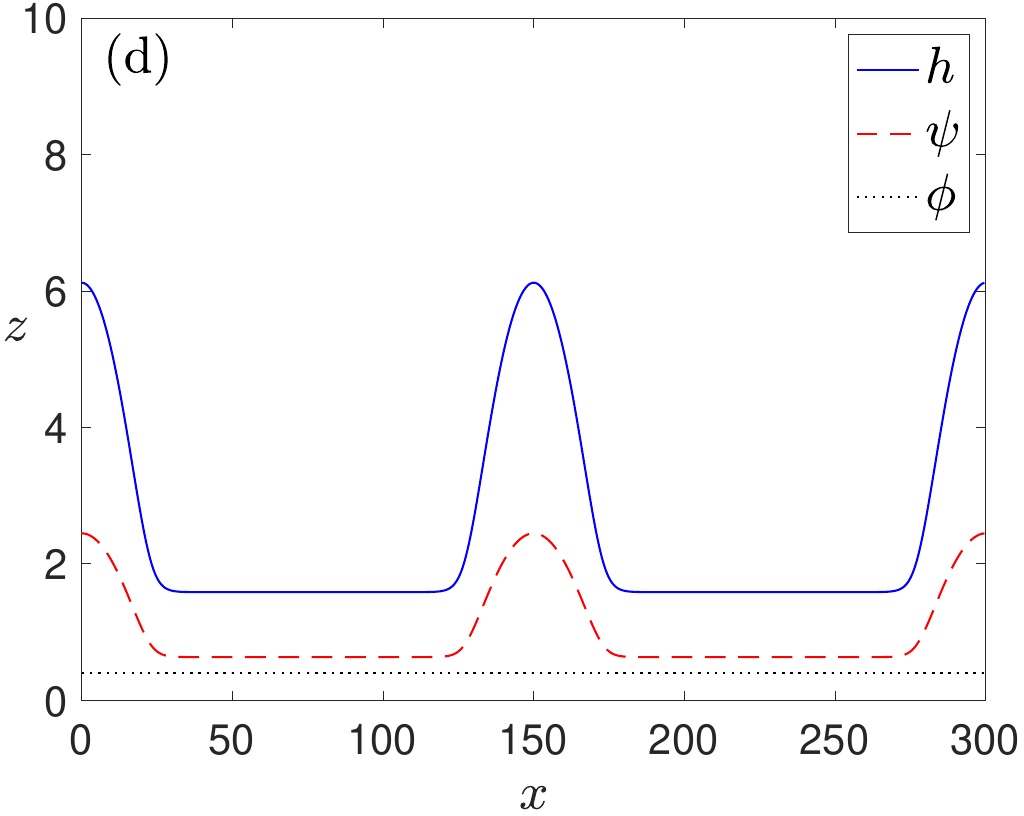}      
      \includegraphics[width=0.48\linewidth]{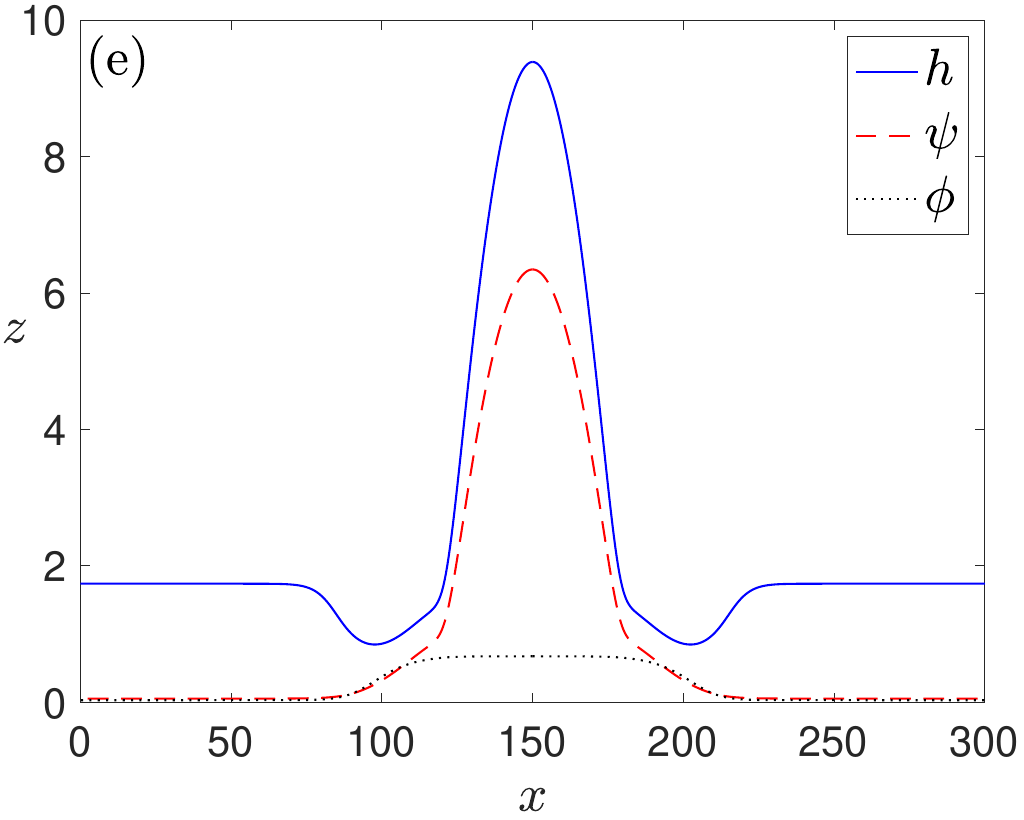}
               \hspace{0.45cm}   
      \includegraphics[width=0.48\linewidth]{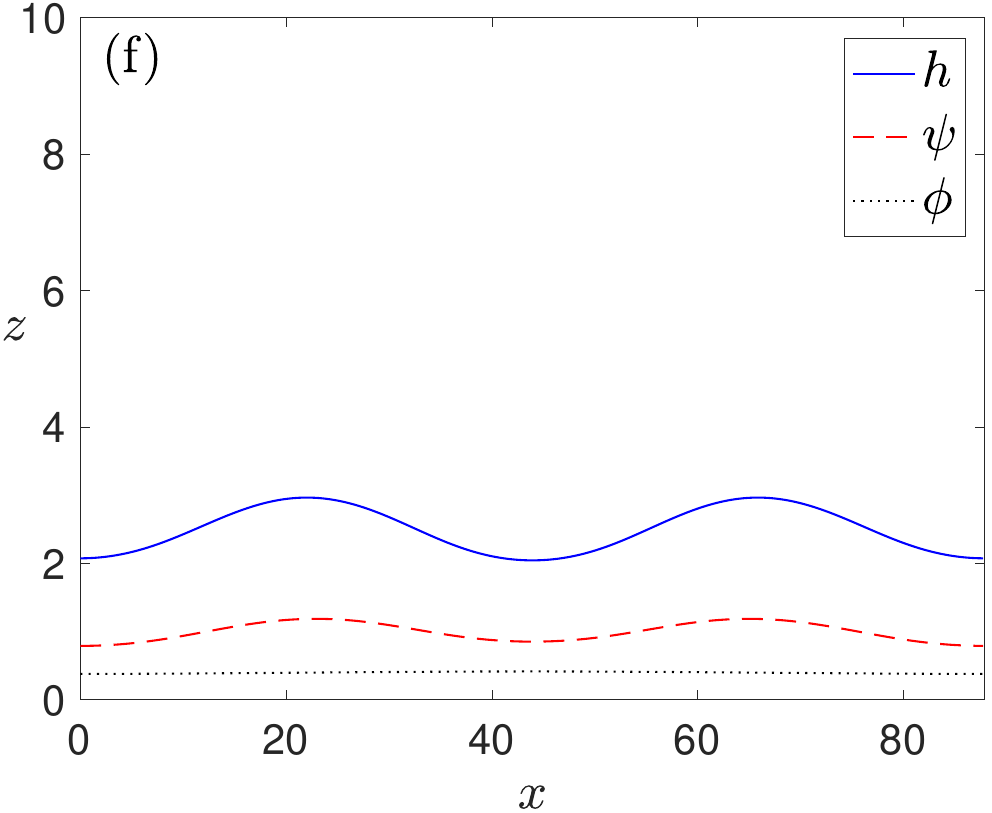}
\caption{Bifurcation diagrams and profiles for an in-phase case, parameters as in Fig.~\ref{bif:out} but with $\phi_i=0.4$. (a)--(b) show bifurcation diagrams for the $L^2$-norm of the film height and colloidal profiles, respectively. Note that there are many other branches in the bifurcation diagram; those displayed correspond to the one- and two-wavelength solutions as predicted by our linear stability analysis. \red{The solid lines correspond to stable solutions, whereas the dashed lines correspond to unstable solutions.} Panels (c)--(e) show profiles on the three displayed branches at $L_x=300$: (c) shows the profile on the first film height mode branch, (d) on the second film height mode branch, and (e) and \red{anti-phase} profile on a colloids branch that terminates on the second film height branch. (f) shows the profile when the colloids mode branch terminates on the second film height branch.}
\label{bif:in}
\end{figure}

The eigenvalues corresponding to solution branches displayed in Figs.~\ref{bif:out}(a)--(b) (not displayed) show that the part of the first blue film height branch before the first bifurcation point is stable, and so is indicated as a solid line, while the rest of the branch is unstable, indicated as a dashed (blue) line. The second blue film height branch is entirely unstable, so is all dashed.

The red colloid branch starts out unstable until reaching the leftmost turning point, except for the small part near where it crosses the film height branch \red{at the transcritial bifurcation point}. This part is magnified in Fig.~\ref{bif:out}(a) \red{in the top inset} and the stable part is highlighted by the thicker red solid line. The branch becomes stable for a very small section between the turning point and the first bifurcation point on the film-height branch (corresponding to a transcritical bifurcation). The \red{colloid branch} is stable from the leftmost turning point until $L_x = 219.5$, where it becomes unstable again for a very short section \red{(see the bottom inset)}. 

When we increase the total average colloid concentration, the zero (neutral wavenumber) in the dispersion relation for the colloids shifts to smaller $k$, making the starting point for the branch of solutions corresponding to the colloid instability to shift to larger $L_x$ (smaller $k$ corresponds to larger wavelengths). A sufficient increase can make the colloid branch start to the right of the second film height branch. For such model parameters (here changing $\phi_i$ from 0.3 to 0.4 and keeping all other values the same), as we increase the system size $L_x$, we find that the colloid-mode branch now \red{crosses} the second film height branch, as shown in the corresponding bifurcation diagrams displayed Figs.~\ref{bif:in}(a)--(b). In such a case, the colloid-mode branch \red{crosses} the second film height branch at a pitchfork bifurcation point. We show the profiles at the intersection point in Fig.~\ref{bif:in}(f). \red{Note that at this point the solution has the period equal to half of the domain size, but this symmetry is broken on the colloid branch away from this pitchfork bifurcation point.}

To obtain the equivalent colloid-mode branch for lower $\phi_i$, we can take an already computed equilibrium solution for $\phi_i=0.3$ (for instance at $L_x=200$), continue the solution in $\phi_i$ to the desired value $\phi_i=0.4$ at the same $L_x$, and use this to then continue forward and backward in $L_x$, to generate the full colloid mode solution branch displayed in Figs.~\ref{bif:in}(a)--(b) (red line). Doing this, on continuing backward in $L_x$ we find that this branch \red{meets} the second branch of the film height bifurcation curve at a pitchfork bifurcation point, rather than the flat state (horizontal axis of our figures). We also use continuation to take the system forwards to $L_x = 300$. Again, the eigenvalue plot (not displayed) suggests that the first branch of the film height mode is stable until the first bifurcation point in Figs.~\ref{bif:in}(a)--(b), and the entire second branch for the thin film is again unstable. 

From Figs.~\ref{bif:in}(a)--(b), we can see that the main branch of the colloids mode (thicker red line) intersects the first branch of the film height curve through the first bifurcation point \red{(which turns out to correspond to a transcritical bifurcation)}, and lands on the first bifurcation point of the second film-height branch \red{at a pitchfork bifurcation}. We show the profiles from each branch at $L_x = 300$ in Figs.~\ref{bif:in}(c)--(e). The profiles from the film height branches (shown in Figs.~\ref{bif:in}(c) and (d)) are similar to the \red{anti-phase} case because the parameters related to the film height have not changed. However, from the colloid branch (Fig.~\ref{bif:in}(e)) we see from the profiles that the film height and colloid concentration profiles are now in-phase. We now explore this transition between \red{anti-phase} to in-phase further.

\subsection{Transition from \red{anti-phase} to in-phase solution profiles}

\begin{figure}
\centering
        \includegraphics[width=0.6\linewidth]{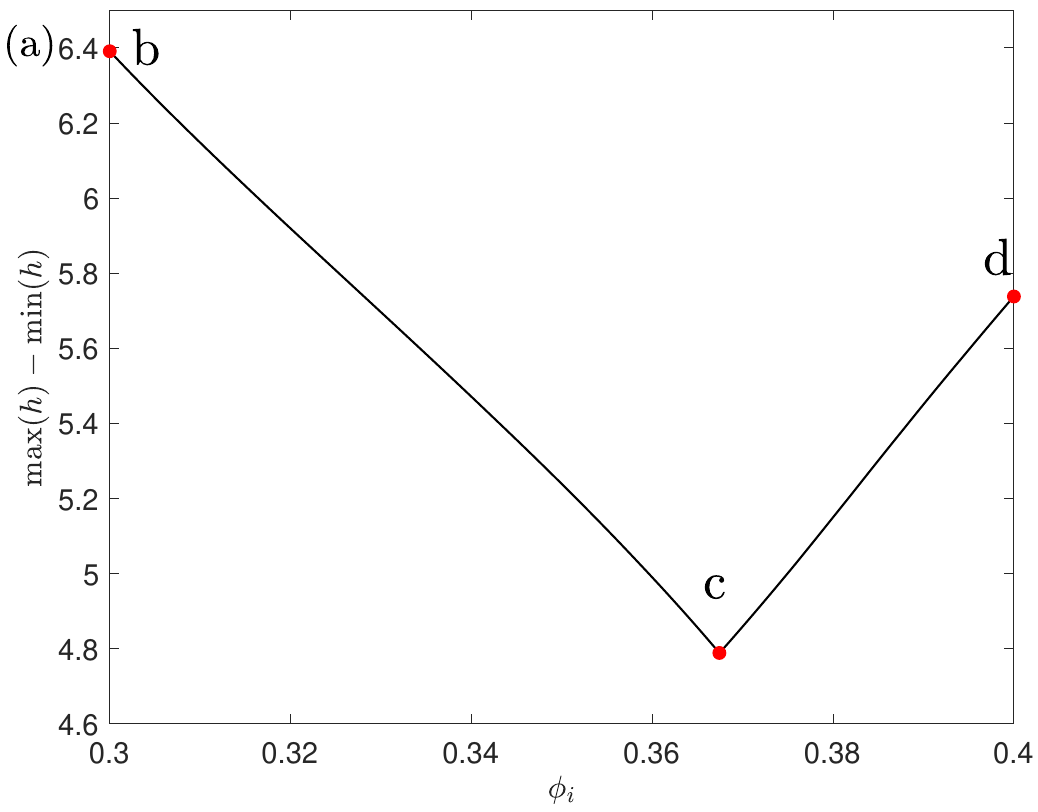}\\
        \includegraphics[width=0.3\linewidth]{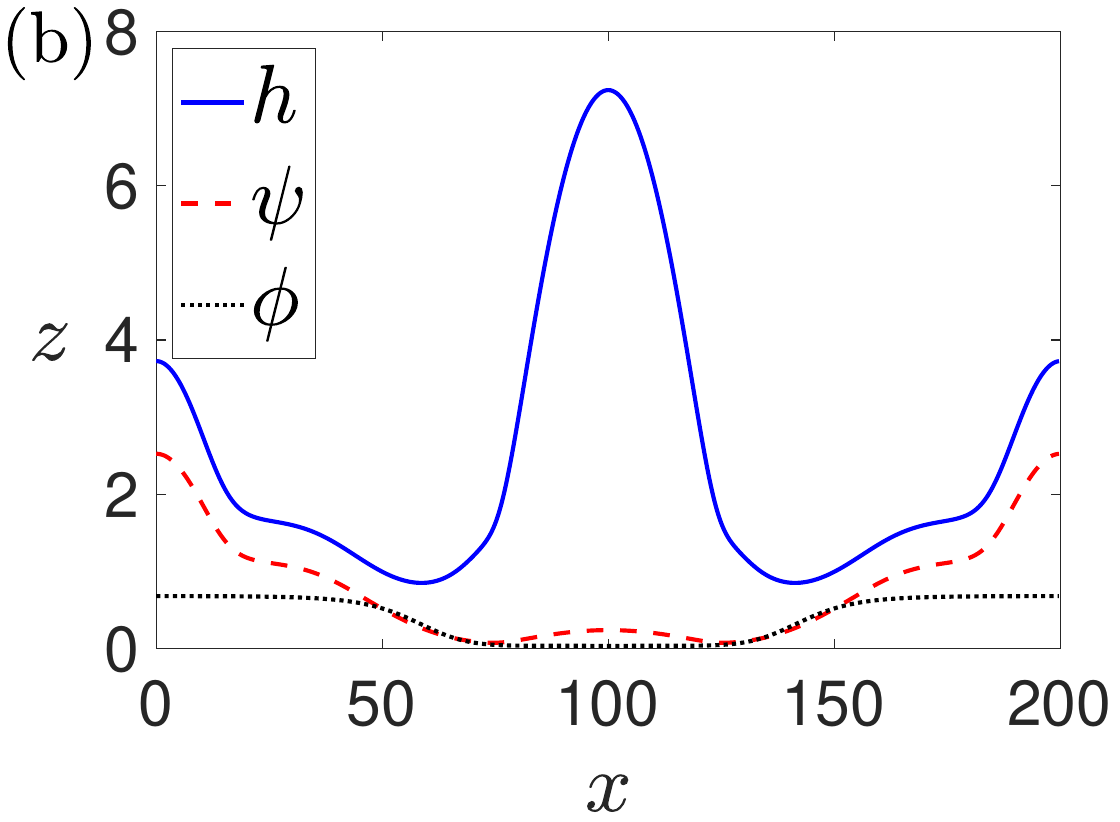}
        \includegraphics[width=0.3\linewidth]{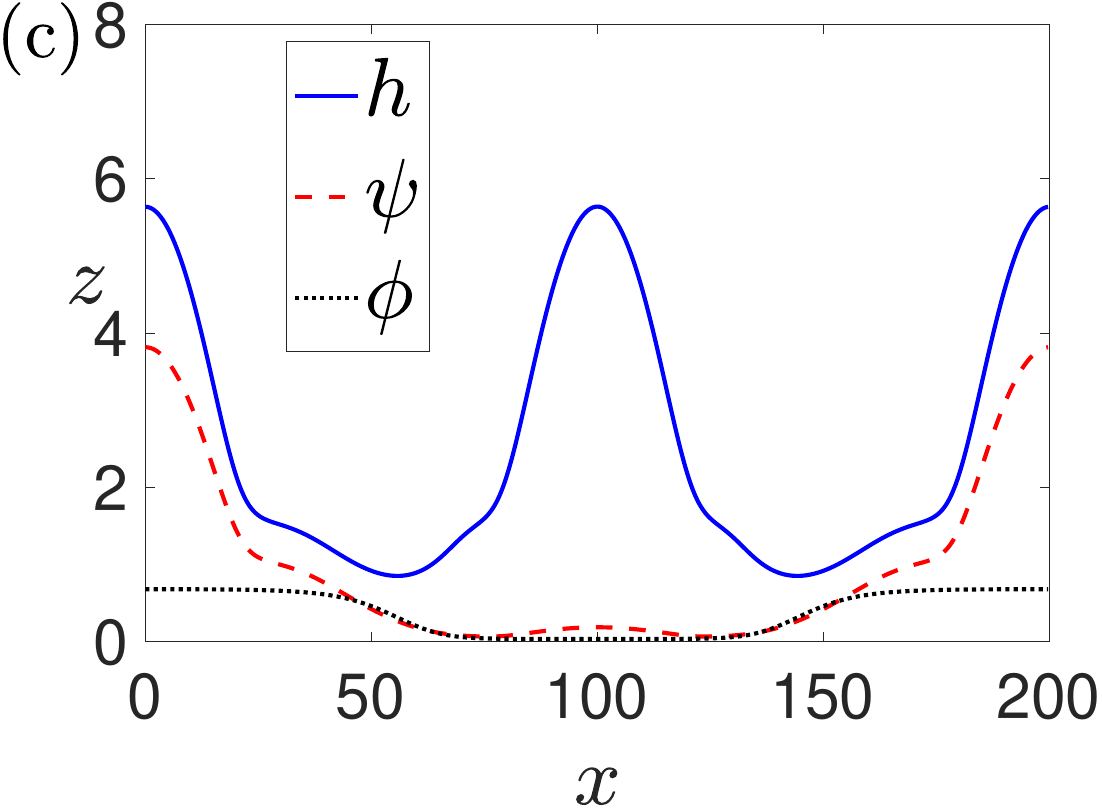}
        \includegraphics[width=0.3\linewidth]{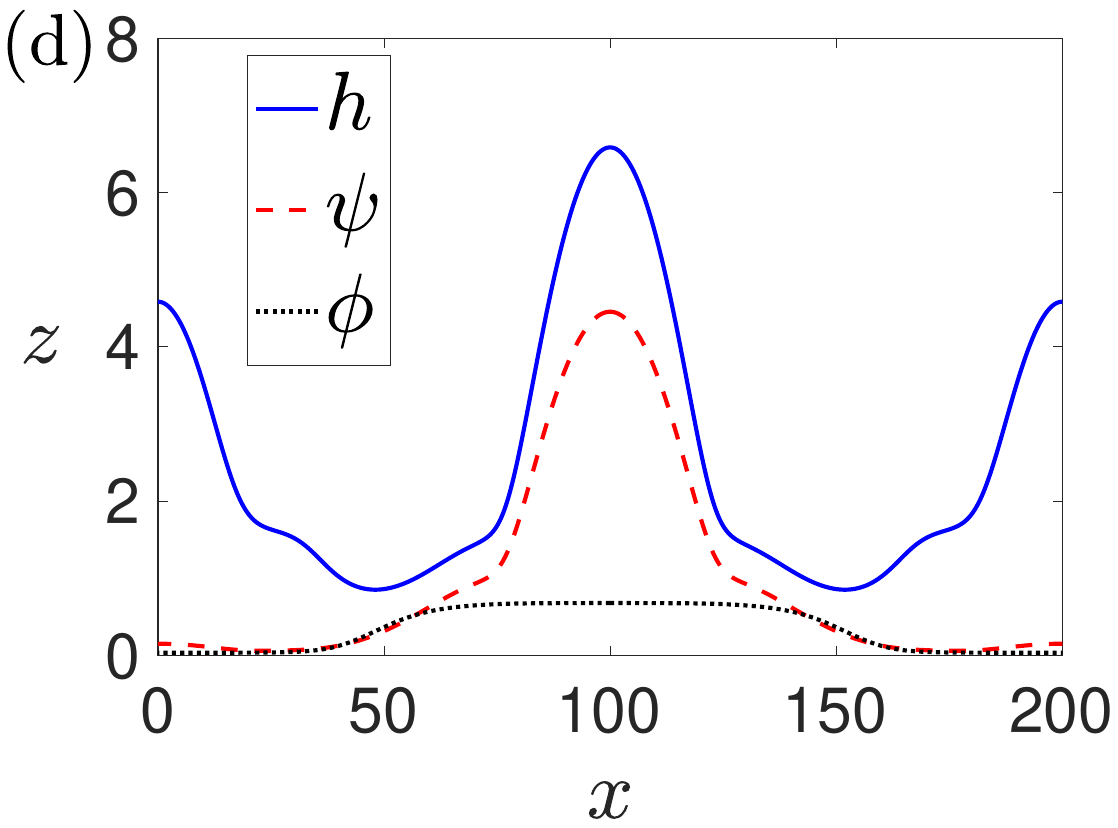}
\caption{Transition between \red{anti-phase} to in-phase at $L_x = 200$. (a) Turning point in the maximum difference in film height. (b)-(d) Profiles for $L_x=200$ at $\phi_i=0.3$, $\phi_i=0.4$ and the critical concentration $\phi_i=0.367$, with other parameters as in Fig.~\ref{bif:out}.}
\label{transition}
\end{figure}

With the two sets of bifurcation diagrams in Figs.~\ref{bif:out} and \ref{bif:in}, we understand that there is a fundamentally different structure between the \red{anti-phase} and in-phase cases. Critically, the transition occurs depending on whether the colloid mode begins at higher or lower $L_x$ than the second film-height branch, or, equivalently, the largest value of $k$ that the colloids are linearly unstable compared to half the value for the film (or twice the wavelength). This change modifies the structure of the bifurcation diagram to enable an intersection with the second film-height branch or not. The transition thus occurs at this point. Starting from a high value and decreasing $\phi_i$, eventually the start point of the colloid branch collides with the start point of the second film-height branch and the connecting branch is annihilated below this critical value.

Another way to visualise the transition is to plot the maximum difference in film height against the initial concentration $\phi_i$; see Fig.~\ref{transition}(a) for $L_x=200$. We observe a clear turning point, located at $\phi_i = 0.367$ (the middle red point) and this is the critical concentration for the system to go from \red{anti-phase} to in-phase. In the $L_x=200$ profiles displayed from the colloids branch, we can see a clear height difference between drops. In Fig.~\ref{transition}(b) (corresponding to the left red point in Fig.~\ref{transition}(a)), the droplet in the middle is taller, with the colloid concentration higher in the second smaller droplet (wrapping around the two sides due to periodicity), making the system \red{anti-phase}. On the other hand, in Fig.~\ref{transition}(d) (corresponding to the right red point in Fig.~\ref{transition}(a)), the drop with the most colloids in is taller, and hence in-phase. In Fig.~\ref{transition}(c) we plot the profile of the system when the concentration is at the turning point of Fig.~\ref{transition}(a) (where $\phi_i = 0.367$), where the heights of both droplets are the same.

\subsection{Bifurcation diagrams for a case with asymmetrical solutions}
\label{sec:bif_asymm}

\begin{figure}
\centering
    \includegraphics[width=0.48\linewidth]{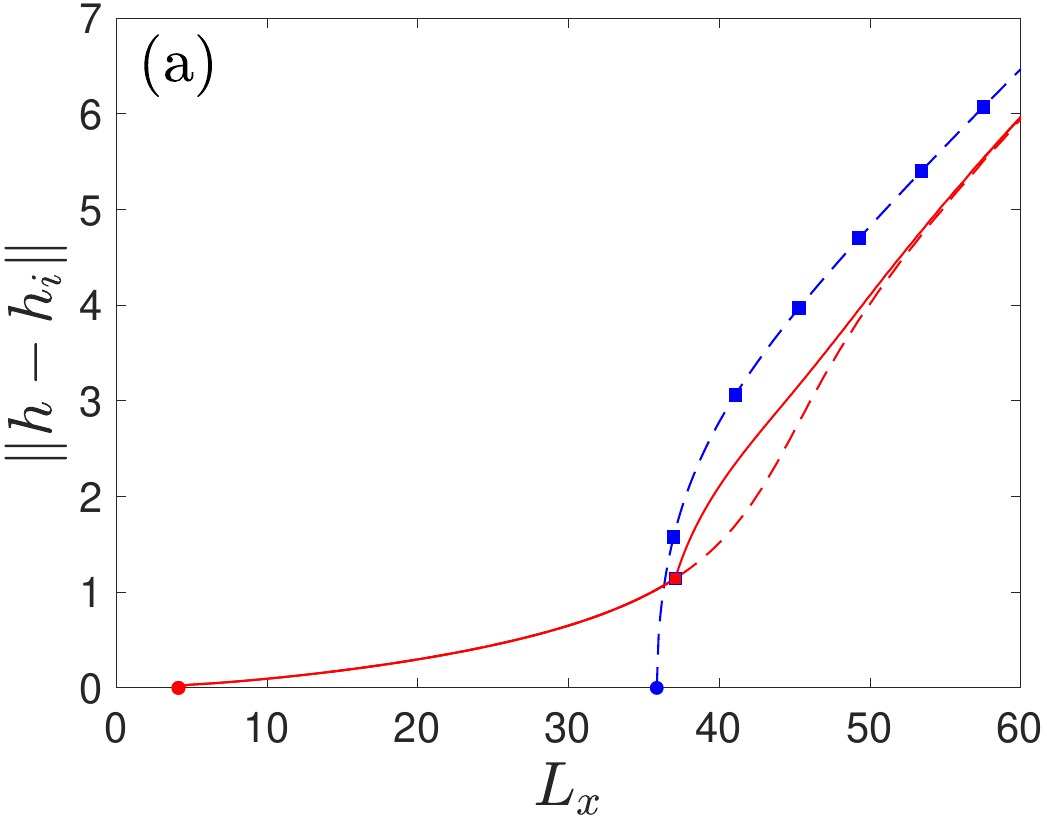}
    \includegraphics[width=0.48\linewidth]{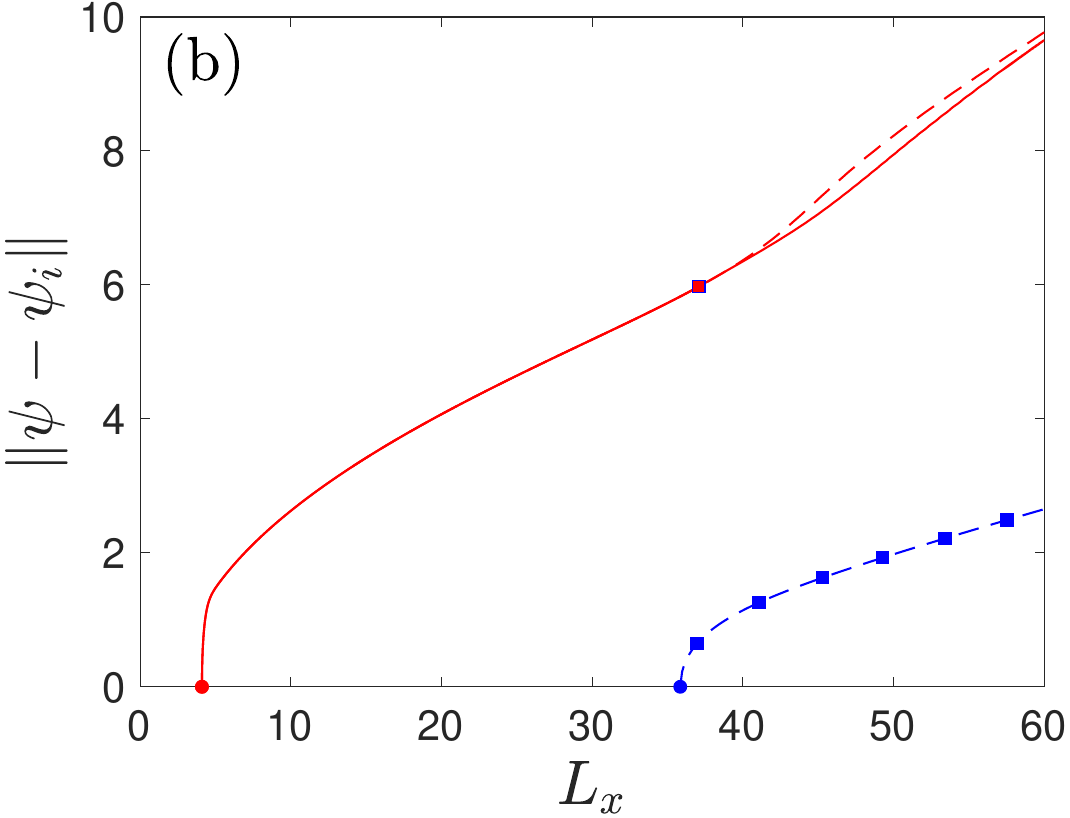}
\caption{Bifurcation diagrams for the same parameters as in Fig.~\ref{case_asymm}. Panel (a) shows the $L^2$-norm for $h$, while (b) shows this for $\psi$. The red dashed line is the solution branch corresponding to unstable equilibria with symmetrical profiles. The red solid line is for the stable equilibria, which to the right of the red square are asymetrical. The blue dashed line corresponds to the unstable thin film branch of solutions.}
\label{bif:asymm}
\end{figure}

\begin{figure}
      \includegraphics[width=0.48\linewidth]{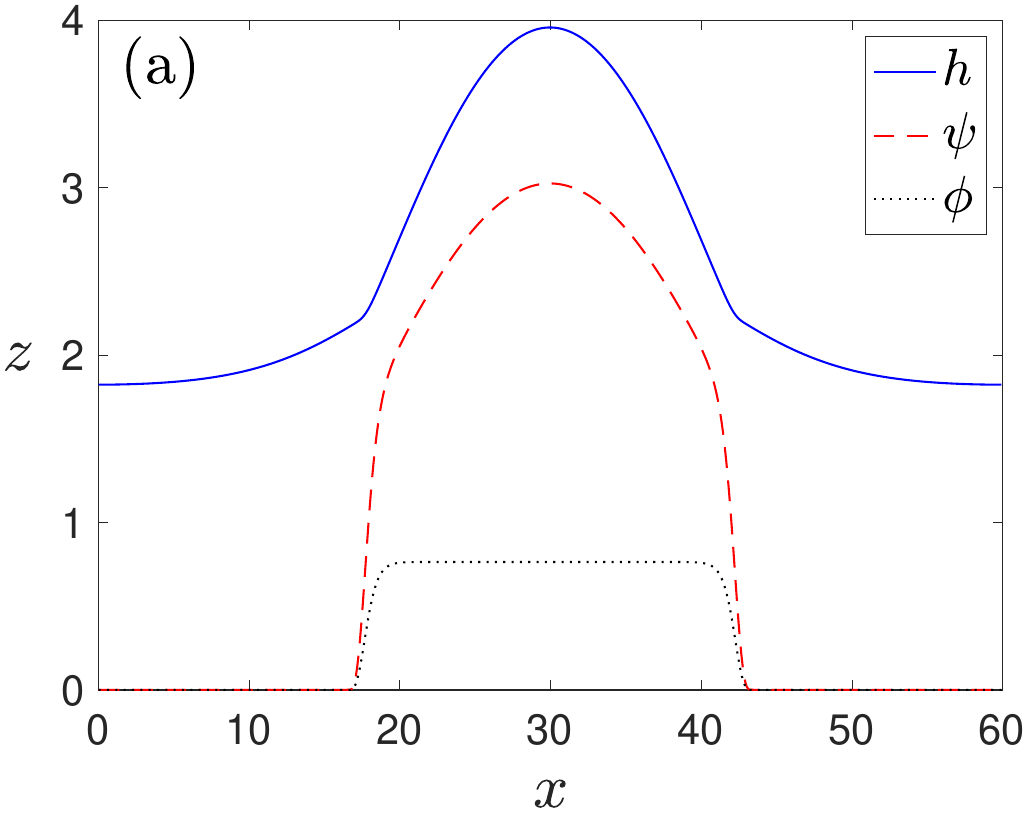}
      \includegraphics[width=0.48\linewidth]{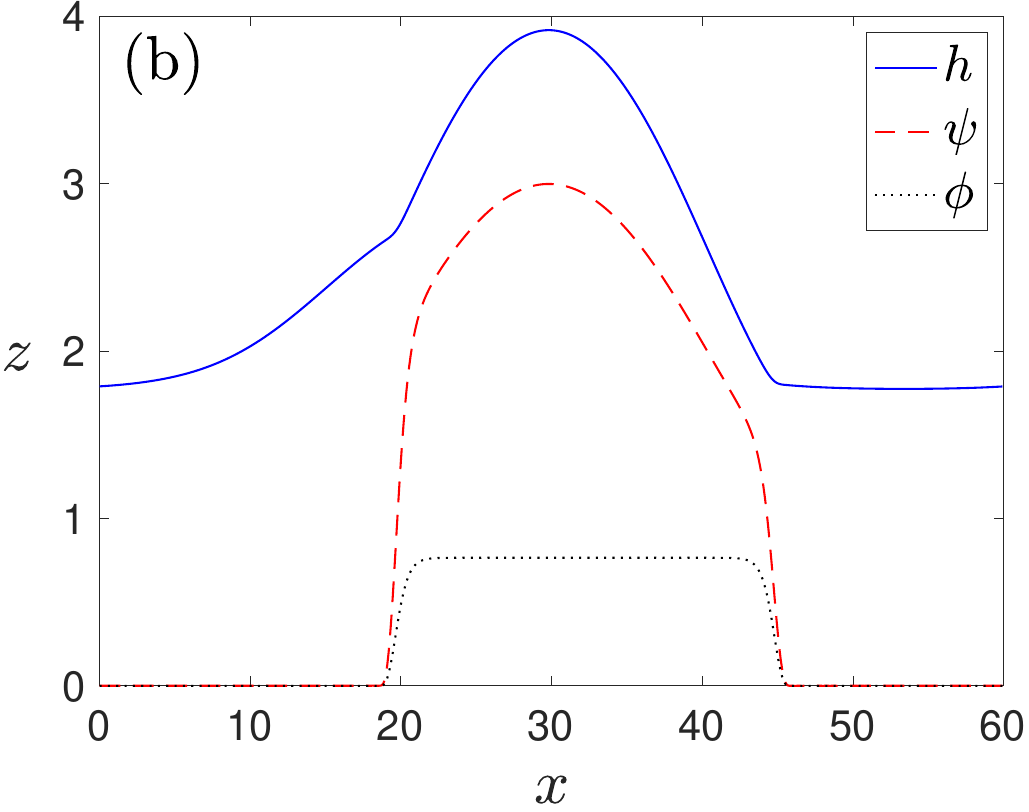}
\caption{Profiles for a system of length $L_x=60$ and corresponding to the bifurcation diagrams in Fig.~\ref{bif:asymm}. Panel (a) shows the unstable symmetrical solution on the first colloid branch and (b) shows the stable asymmetrical solution on the side branch bifurcating from the first colloid branch.}
\label{final_prof:asymm_1}
\end{figure}

In \S~\ref{sec:asymm} we described an example situation with asymmetrical final profiles. Here, we map out the bifurcation diagram for this case, displayed in Fig.~\ref{bif:asymm}. From noting that the neutral wavenumber in the dispersion relation in Fig.~\ref{case_asymm}(d) for the colloids is much \red{bigger} than that for the film height, we can predict that the starting point for the colloid branch is at a much \red{smaller} wavelength (i.e.\ much \red{smaller} system size) than the one for the film-height branch. Arguably the most interesting feature that we observe in Fig.~\ref{bif:asymm} is the pitchfork bifurcation point on the first colloid branch, indicated by a red square. The side-branch that bifurcates from this point is displayed as a red solid line. Interestingly, the solutions on this side-branch are stable and asymmetrical. The profiles for $L_x = 60$ for both the symmetrical and asymmetrical colloid branches are shown in Figs.~\ref{final_prof:asymm_1}(a) and (b), respectively.

In Fig.~\ref{final_prof:asymm_1}(b), we display an asymmetrical profile, corresponding to the colloids being slightly gathered together at one end of the droplet, with not enough of them in the system to span equally to the other end of the droplet. Note that this equilibrium profile is almost the exact replicate of the final equilibrium obtained from our dynamical simulation, shown in Fig.~\ref{case_asymm}(c). Note too, however, that with different random seeds, we may observe equivalent profiles that are either translations or a mirror image of the equilibrium profiles shown in Fig.~\ref{case_asymm}(c), i.e.\ in our dynamical simulations, where the centre of the droplet ends up and on which side of the droplet the colloids gather depends on the random initial conditions. To confirm the stability of the asymmetrical state calculated in the bifurcation diagram for $L_x=60$, we set it as an initial condition for our dynamical code, finding that over time it does not change, i.e.\ it is indeed stable. 

\section{Three-dimensional droplets}
\label{sec:2D}
As discussed at the start of \S~\ref{sec:nandr}, our model equations apply to systems of three-dimensional droplets, i.e.~consist of thin-film equations for the film height and colloid local concentration depending on two spatial variables $x$ and $y$. Our computer code for this case again uses finite differences, but now extended to two spatial variables. This increase in complexity of the code results in the numerical simulations taking considerably longer, but are still manageable on a personal computer for the system sizes discussed here. To benchmark this code, we run simulations with a very narrow domain width $L_y$ in the $y$-direction together with an initial condition that mimics our simulations from the previous section. We find the results agree with high accuracy, up until $L_y$ becomes sufficiently large that the expected effects of the additional dimension begin to develop.

\begin{figure}
\centering
    \includegraphics[width=0.6\linewidth]{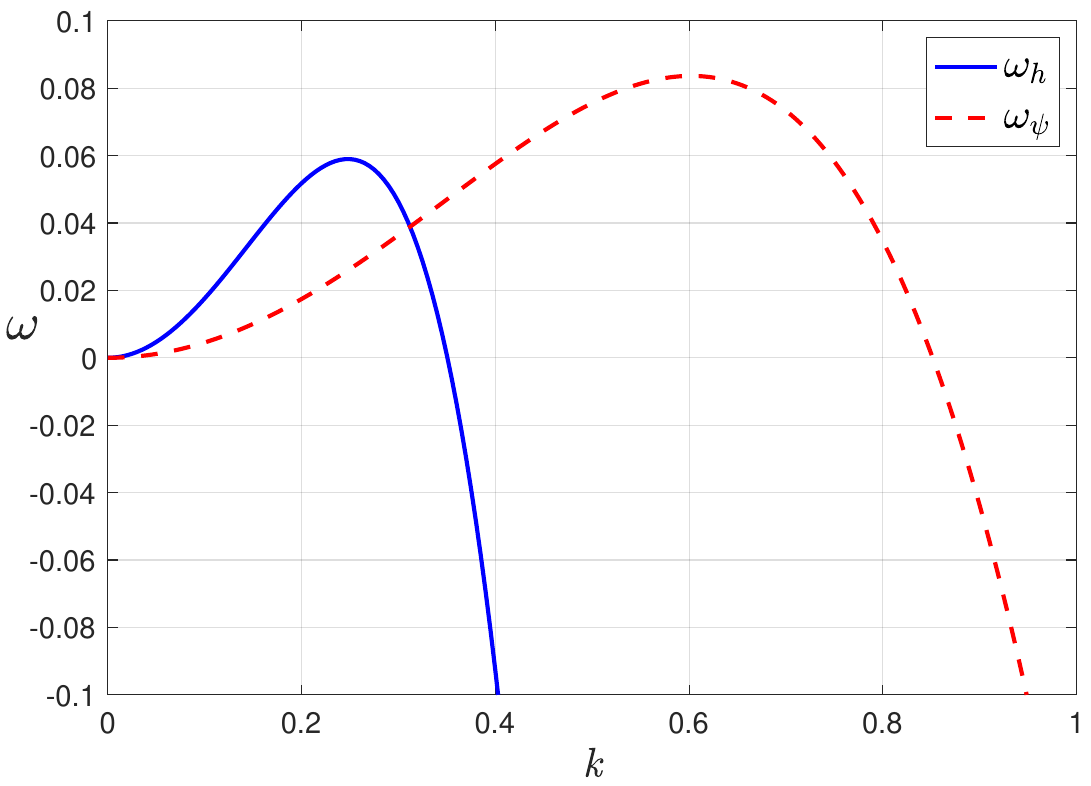}
\caption{Dispersion relation for parameters $A' = 4$, $K' = 0.15$, $\alpha' = 1$, $\epsilon' = 0.2$, \red{$a^{2} = 50$}, $h_i = 2.5$ and $\phi_i = 0.4$, relevant to the simulation presented in Fig.~\ref{twodsimulation}.}
\label{twoddisp}
\end{figure}

\begin{figure}
\centering
        \includegraphics[width=0.4\linewidth]{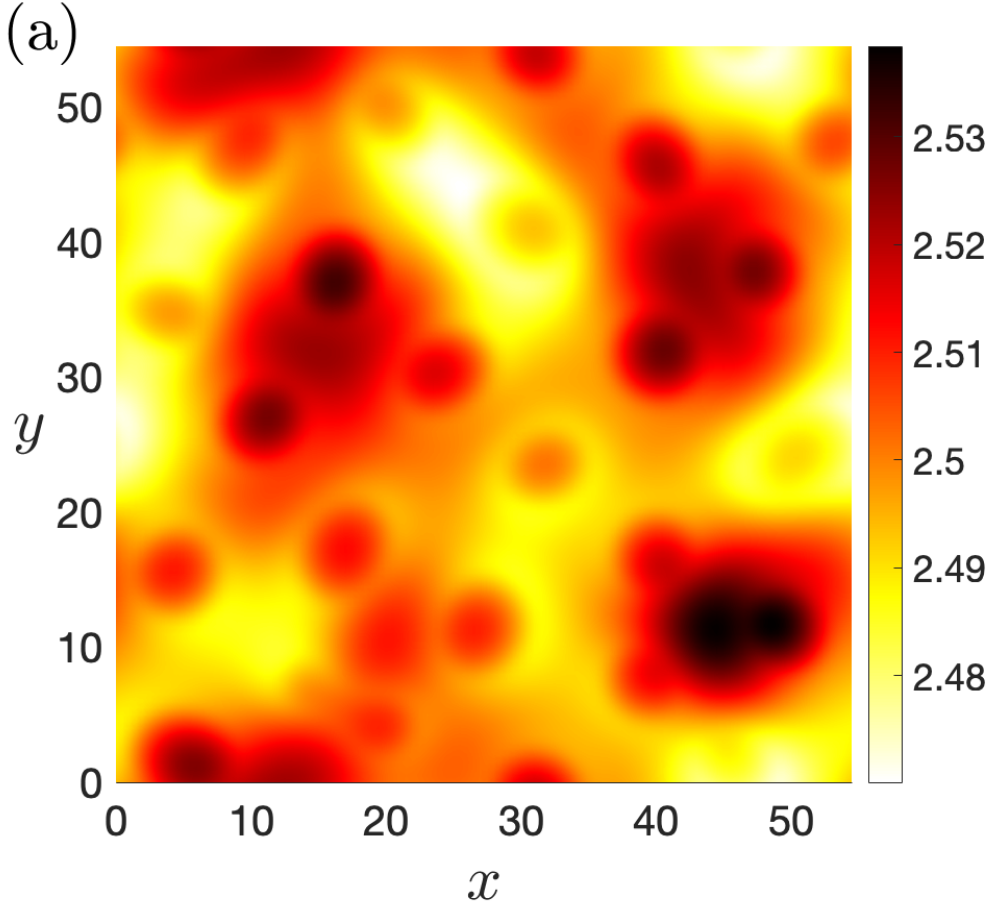}
        \includegraphics[width=0.4\linewidth]{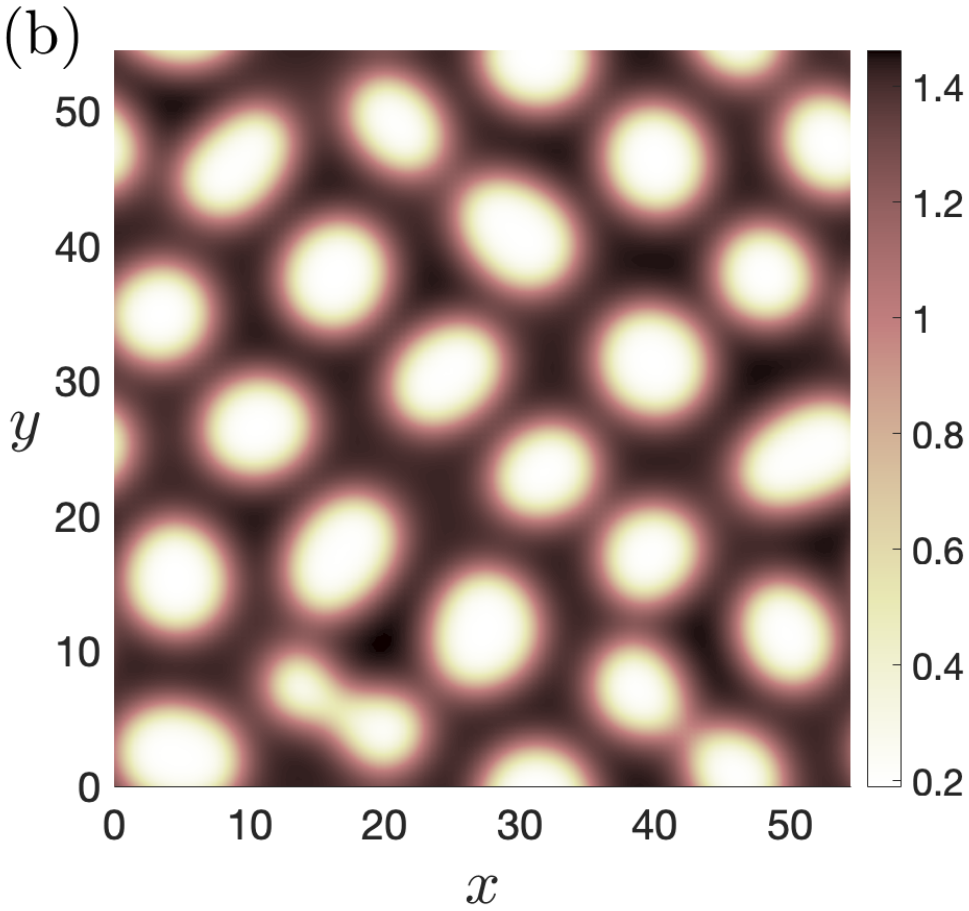}
        \includegraphics[width=0.4\linewidth]{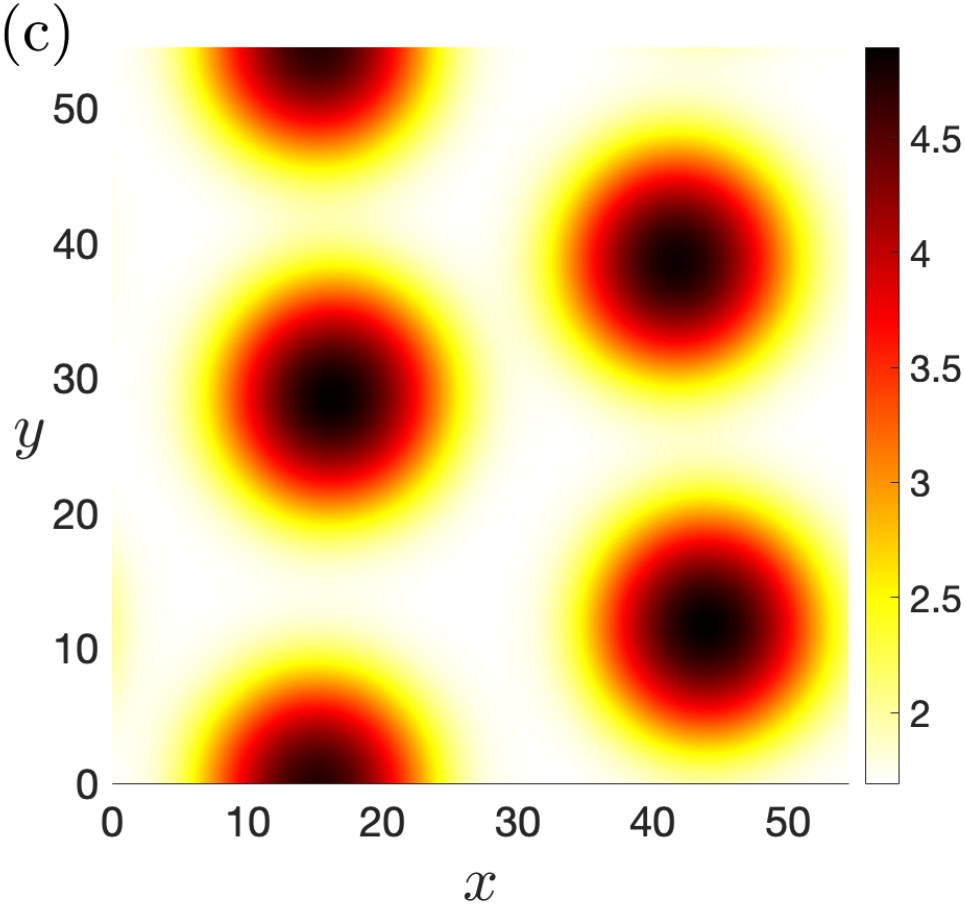}
        \includegraphics[width=0.4\linewidth]{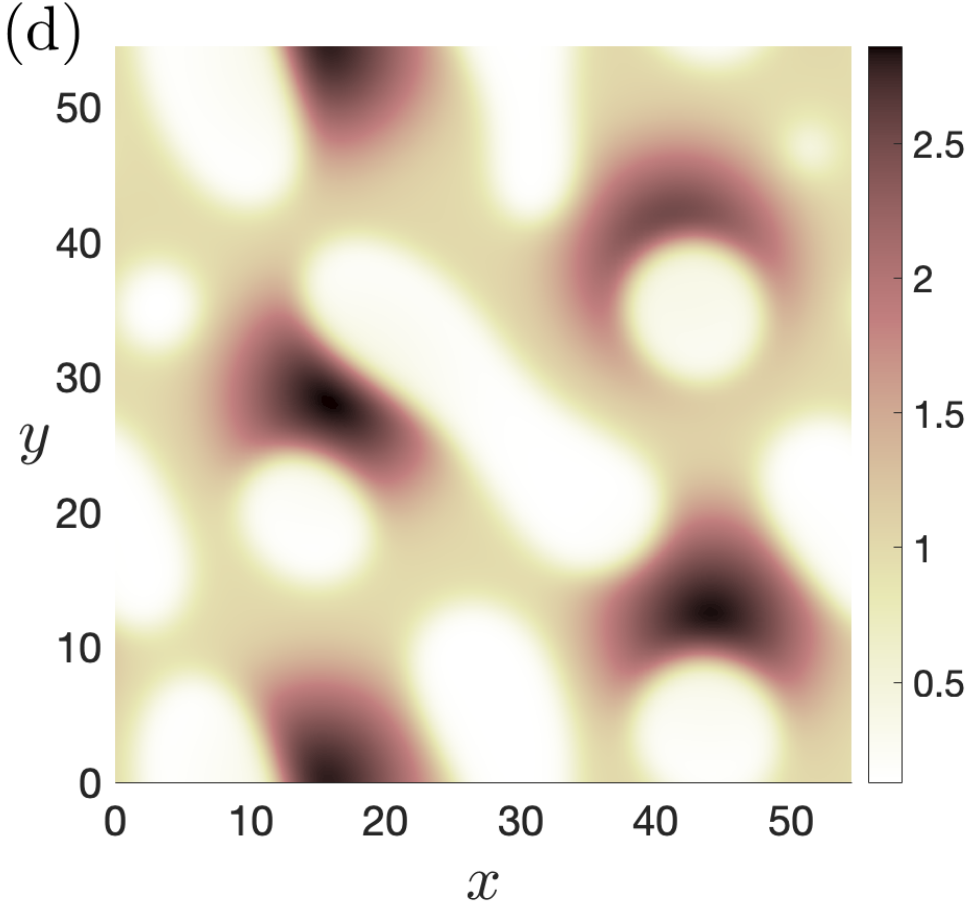}
        \includegraphics[width=0.4\linewidth]{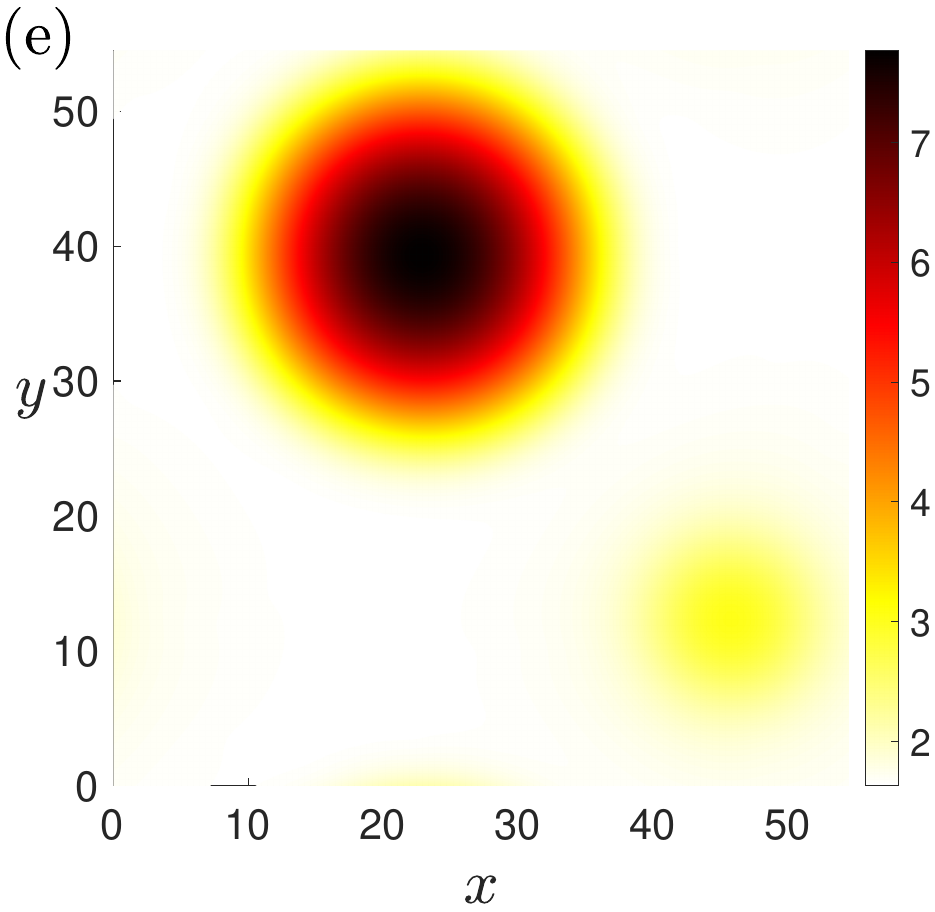}
        \includegraphics[width=0.4\linewidth]{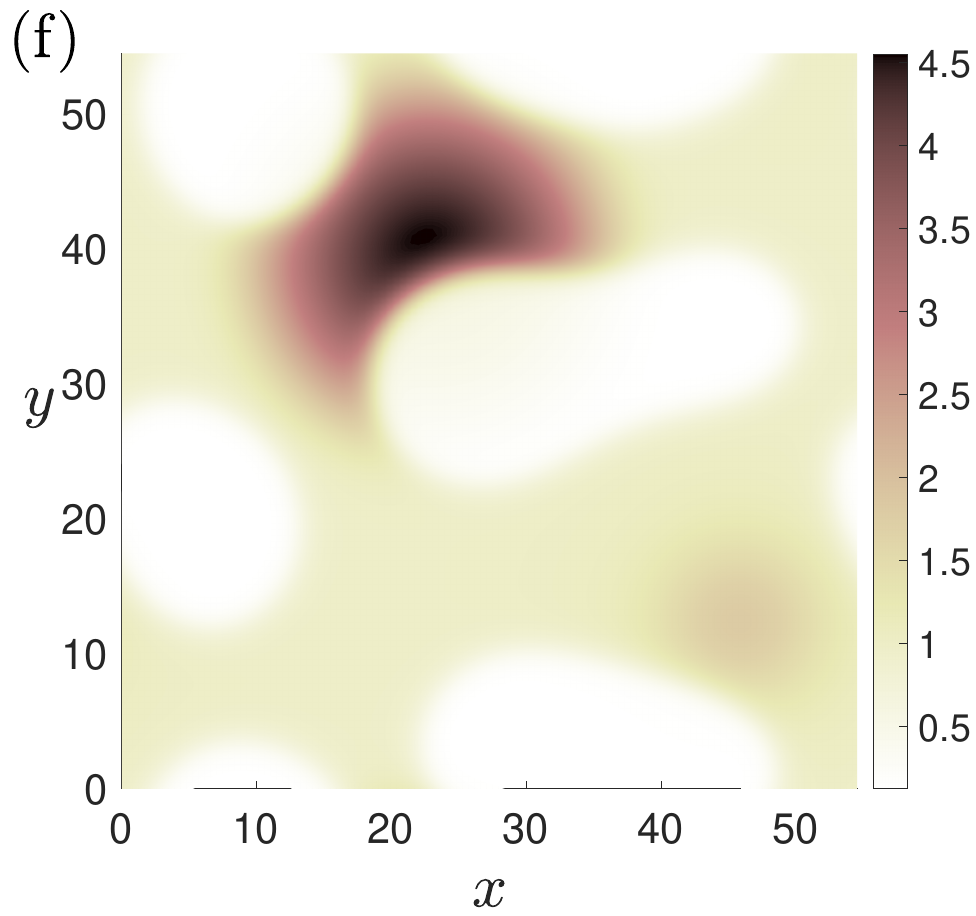}
\caption{Simulation of a square system of size $L_x = L_y = 55$ with periodic boundary conditions and parameters  $A' = 4$, $K' = 0.15$, $\alpha' = 1$, $\epsilon' = 0.2$, $h_i = 2.5$ and $\phi_i = 0.4$. Shown are (left) the film-height profiles $h$ and (right) the local colloid effective height profiles $\psi$ at the times (a,b) $t = 200$, (c,d) $t = 500$, and (e,f) $t = 1100$. The system exhibits coupled dewetting and demixing of the colloids within the film.}
\label{twodsimulation}
\end{figure}

In Fig.~\ref{twoddisp} we show the dispersion relations for a system with parameters $A' = 4$, $K' = 0.15$, $\alpha' = 1$, $\epsilon' = 0.2$, \red{$a^{2} = 50$}, $h_i = 2.5$ and $\phi_i = 0.4$, showing both the film height and the colloids are linearly unstable. Thus, at this state point we should expect both dewetting of the film and also demixing of the colloids within the film. An example of the three-dimensional film height and colloid concentration dynamics corresponding to this set of parameters is shown in Fig.~\ref{twodsimulation}. We set the system size to be $L_x = L_y = 55$, with $110$ discretisation points in either direction and apply periodic boundary conditions. The initial condition is similar to that used in the previous section, being flat profiles with small amplitude random perturbations, consisting of uniformly distributed random numbers for each $(x,y)$ location, for both $h$ and~$\psi$, with amplitudes $\varepsilon_h=\varepsilon_\psi=10^{-5}$.

\begin{figure}
\centering
        \includegraphics[width=0.49\linewidth]{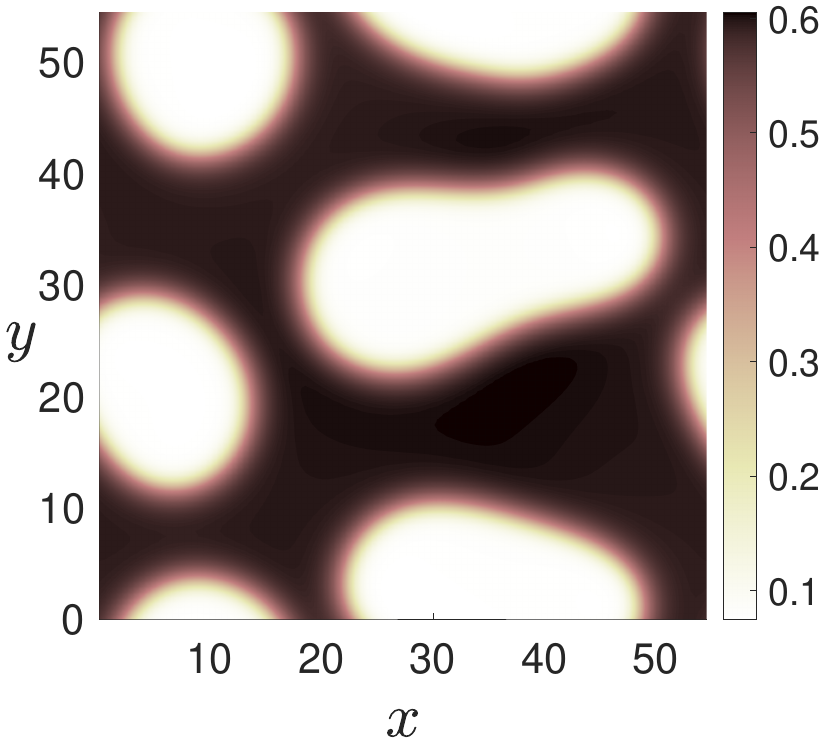}
\caption{The colloidal local concentration profile $\phi$ at the time $t=1100$, corresponding to the coupled dewetting and colloidal demixing (agglomeration) dynamics displayed in~Fig.~\ref{twodsimulation}.}
\label{twodaddinfo}
\end{figure}

From the dispersion relations shown in Fig.~\ref{twoddisp} we can read off that the wavelengths corresponding to most unstable wavenumbers are $\lambda_h = 25$ and $\lambda_\psi = 10.5$. With the length of the domain being $55$, this means that at early times we are likely to observe two wavelengths in the film height, since this is closest to the most unstable liquid height mode, and to observe five wavelengths of the most unstable colloidal demixing mode. This can indeed be seen in Figs.~\ref{twodsimulation}(a) and \ref{twodsimulation}(b), which are for the early time $t=200$, by taking a visual slice in either the $x$- or $y$-direction. In the $\psi$ profile the five wavelengths are obvious, but in $h$ the two wavelengths are slightly harder to see, due to the coupling with the colloids. However, by looking at the large-scale period of red-yellow fluctuations in a given slice, the two wavelengths can be discerned.

After the initial phase corresponding to the linear regime, the profiles begin to coarsen in a similar fashion to the cases where $h$ and $\psi$ depend on one spatial variable discussed previously. By the time $t=500$ the profiles have started to coarsen, forming four separated liquid droplets [Fig.~\ref{twodsimulation}(c)], which eventually coalesce into a single droplet by the final displayed time $t=1100$, in Fig.~\ref{twodsimulation}(e). In a similar fashion, the colloid profiles coarsen in Figs.~\ref{twodsimulation}(d) and (f), but with the coupling to the liquid film height affecting the dynamics and creating obvious regions of colloids located in-phase with the drops.

The time taken for this three-dimensional simulation to reach the final state is considerably shorter than for similar corresponding two-dimensional cases, because the extra spatial dimension gives to the coarsening process an additional degree of freedom in which to occur. The coarsening is the slowest process in the whole dynamics and so speeding this up speeds up the overall dynamics. The local colloidal concentration $\phi$ can be inferred from inspecting the $\psi$ profiles relatively easily. However, to avoid having to do this, we also plot this field $\phi(x,y,t=1100)$ in Fig.~\ref{twodaddinfo}, which shows that at most points in space the local colloid concentration takes the two coexisting bulk values predicted by calculating the phase diagram in Fig.~\ref{Bulkphasediagram}. This also shows that although the dewetting process has been largely completed, the colloidal coarsening still has some way to go. The colloidal dynamics is very slow after this stage because the colloids have to diffuse through the thin precursor film layer to further coarsen.

\section{Concluding remarks}
\label{sec:conc}
Here, we have developed a low-order thin-film model for the dynamics of colloidal fluids on planar surfaces. We have used our model to form an understanding of the manner in which colloidal demixing (i.e.\ agglomeration) can influence and couple to liquid-film dewetting processes. Our model is based on that of \citet{thiele2011note}, which we have extended by adding additional terms to the free energy~\eqref{freeenergyfunc} in order to incorporate the effects of colloidal particle interactions. In particular, to incorporate the influence of attractive interactions between the colloids, which can drive demixing of the colloids within the film.

After non-dimensionalising the model, we have performed a linear stability analysis, uncovering an interesting partially-decoupled dispersion relation that is responsible for the early stage evolution of the system. The analytically determined eigenvalues from the linear stability analysis of initially uniform films demonstrate a rich array of situations depending on whether either the liquid film or the colloids are stable, and the relative wavelength of disturbance that initially grows in either the film height or in the colloid local concentration profile. The decoupling in the dispersion relation does not occur if one assumes (in contrast to our assumption here) that the binding potential $g(h)$ in the free energy~\eqref{freeenergyfunc} is also a function of the local colloid concentration $\phi$. In this case, even when there is no colloidal demixing, the coupling between film height and colloid concentration fluctuations can trigger dewetting \citep{thiele2013gradient}.

We also determined the bulk phase diagram by analysing the thermodynamics of the system. All thermodynamic quantities are determined by the assumed free energy functional~\eqref{freeenergyfunc}, and this also governs the properties of the eventual equilibrium states of the system in the long-time limit. By deriving the model based on a free energy functional which incorporates the necessary physics, ensures that there is a consistency between equilibria of the model and the thermodynamics\red{, which are implicit in the specified} free energy functional. This general approach of starting from the free energy, can be a very fruitful and straight-forward way to develop thin-film models incorporating a wide range of physics \citep{thiele2011note}. Just a few examples of phenomena that can be incorporated into such models in this way include: evaporation leading to pattern formation \citep{frastia2011dynamical, frastia2012modelling}, evaporation in confined spaces \citep{hartmann2023sessile}, surfactant molecules in the bulk and on the surface of films \citep{thiele2016gradient}, freezing and melting \citep{sibley2021ice} and elastic substrates \citep{henkel2021gradient, kap2023nonequilibrium}. \red{An interesting future direction would be to incorporate different couplings between the the film height and the solute concentration, building on the present work and also that of \citet{thiele2013gradient}, to investigate more general situations where $C_2\neq0$ [see Eq.~\eqref{LSAcompare}], so coupling between modes leading to instability could become even richer than the cases investigated here. Additionally, by assuming a constant surface tension and viscosity, our model neglects the possibility of Marangoni forces and the slow dynamics that one should expect when the local colloid concentration becomes high. These would also be interesting areas for future work. For the situations investigated here, replacing the constant viscosity with a function that depends on the local colloid concentration will have no real qualitative effect, other than to change some of the overall timescales for the dynamics. However, when evaporation is present, the local colloid concentration can become sufficiently high that locally the viscosity $\eta\to\infty$, which can be important in determining the drying patterns left by evaporating colloidal suspensions \citep{rabani2003drying, martin2004nanoparticle, frastia2011dynamical, frastia2012modelling, robbins2011modelling}.} It should also be mentioned that {\em dynamical} effects such as slip at the surface or the diffusion of colloids over almost-dry surfaces can also by incorporated easily into this whole class of models \citep{yin2017films}.

\red{A matter worth commenting on is what we might expect to observe if we were to choose parameters and/or the initial conditions so as to have a greater scale separation between the precursor film thickness and the typical height of of droplets.
While it is not possible to comment on all possible scenarios in this regime, in general, we do expect similar features to persist. However, to see for instance the coarsening dynamics, we would need significantly greater computational time. We do envisage some situations that are not captured in our bifurcation diagrams: for instance, with lager volume droplets we expect even more intricate transient states and perhaps also equilibria. Recall that generically, as the system size in such systems is increased, there is space for either more droplets of the size we describe or for fewer larger droplets, so increasing the number of solution branches in the bifurcation diagrams. We leave this for future work on this subject.}

We have used finite difference methods to develop \red{\textsc{Matlab}} codes for solving our model. In order to verify the code, we conducted numerous tests to compare the computational and theoretical results, which all showed good agreement. We then investigated the dynamics of the model in a \red{variety} of situations, including showing bifurcation diagrams for various cases. In particular, we investigated how varying the average concentration of the colloids can change the final state of the system from in-phase to \red{anti-phase}, and we also discovered cases with asymmetrical final solutions. Bifurcation diagrams helped us \red{understand} these situations better, which also cross checked the simulation results. From the dynamics, we can see how the parameters we select and the corresponding theoretical relations that arise \red{affect} the dynamics and final results of the simulation. Moreover, we have developed a code for three-dimensional droplets, again exhibiting excellent agreement to the theoretical predictions for the linear behaviour and the two-dimensional code. As expected, the evolution of the system is much faster in three-dimensions due to the additional dimension, although the computer simulation times are typically significantly longer due to the much larger number of grid points required.

Here, our intention has been to develop a general model for incorporating the influence of colloidal agglomeration and demixing on the dynamics of thin films of liquids. Thus, agreement with specific experiments such as those of \citet{howard2023surfactant}, where colloidal aggregation in drying thin films was observed in experiments on carbon-nanotube suspensions, is at best only qualitative. To match better specific experiments, additional effects need to also be added to the model. For example, to match these particular experiments quantitatively, the evaporation \citep{WilsonRev}, the colloid (non-spherical) shape \citep{Mig}, droplet shape \citep{wray_moore_2023}, surface roughness \citep{SK}, \red{influence on $g(h)$ of molecular ordering at the surface and the nature of the wetting transition \citep{hughes2017},} possibly gravity \citep{moore_wray_2023}, and the surfactants \citep{KalPap} also present in the film need to be included in the model. As future work, such effects could easily be incorporated, though we would suggest systematically incorporating each different one in turn, separately. Finally, the effects of internal flow from droplet orientation \citep{Edwards_etal} and of shielding in multiple droplet systems \citep{Wray_etal} would be interesting to investigate within an augmented model of the type \red{considered} here.

\section*{Acknowledgements}
We gratefully acknowledge stimulating discussions with Ben Goddard and Uwe Thiele. \red{This paper was greatly improved thanks to comments from the three anonymous referees. JZ acknowledges support from EPSRC grant EP/W522569/1.}

\section*{Data availability statement}
Data is available at the Loughborough University Research Repository at\\
http://dx.doi.org/10.17028/rd.lboro.24624306

\appendix

\section{Functional derivatives}\label{appA}

The functional derivatives can be analytically determined, or found using the \emph{Maple} software (we did both), and may be written as
\begin{eqnarray}
\frac{\delta F}{\delta h} & = &-\gamma\nabla^2h+\partial_h g(h) +f\left(\frac{\psi}{h}\right)-\frac{\psi}{h}f'\left(\frac{\psi}{h}\right)-\frac{2\epsilon\psi}{h^3}(\nabla\psi\cdot\nabla h)\nonumber\\
&&+\frac{3\epsilon\psi^2}{2h^4}|\nabla h|^2 + \frac{\epsilon}{2h^2}|\nabla\psi|^2+\frac{\epsilon\psi}{h^2}\nabla^2\psi-\frac{\epsilon\psi^2}{h^3}\nabla^2 h
\end{eqnarray}
and
\begin{eqnarray}
\frac{\delta F}{\delta \psi} & = & f'\left(\frac{\psi}{h}\right)+\frac{\epsilon}{h^2}(\nabla\psi\cdot\nabla h)-\frac{\epsilon}{h}\nabla^2\psi-\frac{\epsilon\psi}{h^3}|\nabla h|^2+\frac{\epsilon\psi}{h^2}\nabla^2 h.
\end{eqnarray}
These have corresponding gradients
\begin{eqnarray}
\nabla\frac{\delta F}{\delta h} & = &-\gamma\nabla^3h+\partial_{hh} g(h)\nabla h +\frac{\psi}{h}\left(\frac{\nabla \psi}{h}-\frac{\psi\nabla h}{h^2}\right)f''\left(\frac{\psi}{h}\right)-\frac{4\epsilon\psi}{h^3}(\nabla\psi\cdot\nabla^2 h)\nonumber\\
&&+\frac{9\epsilon\psi}{h^4}\left(|\nabla h|^2 \cdot \nabla \psi\right) - \frac{3\epsilon}{h^3}\left(|\nabla\psi|^2 \cdot \nabla h\right)-\frac{4\epsilon\psi}{h^3}\left(\nabla^2\psi \cdot \nabla h\right)+\frac{6\epsilon\psi^2}{h^4}\left(\nabla^2 h \cdot \nabla h\right)\nonumber\\
&&-\frac{\epsilon\psi^2}{h^3}\nabla^3 h + \frac{2\epsilon}{h^2}\left(\nabla^2\psi \cdot \nabla \psi\right)-\frac{6\epsilon\psi^2}{h^5}|\nabla h|^3 + \frac{\epsilon\psi}{h^2}\nabla^3\psi
\end{eqnarray}
and
\begin{eqnarray}
\nabla\frac{\delta F}{\delta \psi} & = &\left(\frac{\nabla \psi}{h}-\frac{\psi\nabla h}{h^2}\right)f''\left(\frac{\psi}{h}\right) + \frac{\epsilon}{h}\nabla^3\psi + \frac{2\epsilon}{h^2}\left(\nabla^2\psi \cdot \nabla h\right)+\frac{2\epsilon}{h^2}(\nabla\psi \cdot \nabla^2 h)\nonumber\\
&&+\frac{\epsilon\psi}{h^2}\nabla^3 h - \frac{4\epsilon\psi}{h^3}\left(\nabla^2 h \cdot \nabla h\right)+\frac{3\epsilon}{h^3}\left(|\nabla h|^2 \cdot \nabla \psi\right)+\frac{3\epsilon\psi}{h^4}|\nabla h|^3.
\end{eqnarray}


\end{document}